\documentclass[12pt]{article}
\usepackage{amsmath,epsfig,amssymb, amsfonts,amsthm,epsfig,verbatim, enumitem}
\usepackage[comma,authoryear]{natbib}
\usepackage{color,graphicx,lscape,longtable,mathabx,natbib, float, multirow}
\usepackage{hyperref,subfigure, multirow, setspace,  booktabs}
\graphicspath{{figures/}}%设置图片为当前路径下的graphics文件夹。

\newtheorem{Theorem}{Theorem}

\newtheorem{Rem}{\underline{\bf Remark}}

\newtheorem{Pro}{Proposition}
\newtheorem{Lem}{\underline{\bf Lemma}}

\newcommand{\blind}{1}

\def\bse{\begin{eqnarray*}}
\def\ese{\end{eqnarray*}}
\def\be{\begin{eqnarray}}
\def\ee{\end{eqnarray}}
\def\bsq{\begin{equation*}}
\def\esq{\end{equation*}}
\def\bq{\begin{equation}}
\def\eq{\end{equation}}

\def\var{\hbox{var}}

\def\wh{\widehat}

\def\sumi{\sum_{i=1}^n}

\def\sumj{\sum_{j=1}^p}

\def\bvarepsilon{\boldsymbol\varepsilon}

\def\bzeta{\boldsymbol\zeta}

\def\bmu{\boldsymbol\mu}

\def\bbeta{{\boldsymbol\beta}}

\def\bmu{\boldsymbol\mu}

\def\0{{\bf 0}}
\def\A{{\bf A}}

\def\V{{\bf V}}

\def\R{{\bf R}}

\def\a{{\bf a}}
\def\B{{\bf B}}

\def\V{{\bf V}}
\def\K{{\bf K}}

\def\h{{\bf h}}
\def\b{{\bf b}}
\def\I{{\bf I}}
\def\M{\mbox{ $\mathcal{M}$}}
\def\M{{\bf M}}

\def\H{{\bf H}}

\def\K{{\bf K}}

\def\u{{\bf u}}

\def\v{{\bf v}}
\def\W{{\bf W}}

\def\w{{\bf w}}
\def\X{{\bf X}}

\def\x{{\bf x}}
\def\I{{\bf I}}

\def\Y{{\bf Y}}

\def\z{{\bf z}}

\def\bq{\begin{equation}}
\def\eq{\end{equation}}

\def\wh{\widehat}

\def\trans{^{\rm T}}

\def\squarebox#1{\hbox to #1{\hfill\vbox to #1{\vfill}}}
\def\btheta{{\boldsymbol \theta}}

\def\var{\hbox{var}}

\def\bse{\begin{eqnarray*}}
\def\ese{\end{eqnarray*}}
\def\be{\begin{eqnarray}}
\def\ee{\end{eqnarray}}
\def\bsq{\begin{equation*}}
\def\esq{\end{equation*}}
\def\bq{\begin{equation}}
\def\eq{\end{equation}}
\def\wh{\widehat}

\def\bbo{{\bf 0}}

\def\boxit#1{\vbox{\hrule\hbox{\vrule\kern6pt\vbox{\kern6pt#1\kern6pt}\kern6pt\vrule}\hrule}}

\usepackage{soul, ulem}

\renewcommand{\baselinestretch}{2}
\def\spacingset#1{\renewcommand{\baselinestretch}%
{#1}\small\normalsize} \spacingset{1}

\begin{document}
\baselineskip 17pt
\renewcommand {\thepage}{}
\include{titre}
\pagenumbering{arabic}
\begin{center}
{\Large\bf Two-directional simultaneous inference for high-dimensional models}
\end{center}

\baselineskip 15pt

\if1\blind
\begin{center}
Wei Liu$^{1}$, %\email{weiliu@smail.swufe.edu.cn},
Huazhen Lin$^{1*}$, Jin Liu$^{2}$ and  Shurong Zheng$^{3}$\\
	   $^{1}$Center of Statistical Research and  School of Statistics,\\
Southwestern University of
Finance and Economics, Chengdu, China \\
	    $^{2}$Centre for Quantitative Medicine, Program in Health Services \& Systems Research,\\
Duke-NUS Medical School\\
	    $^{3}$School of Mathematics and Statistics,\\
Northeast Normal University, Changchun, China
	
\end{center}
\footnotetext{*Corresponding author. Email: linhz@swufe.edu.cn.
}
\fi

\begin{abstract}
\baselineskip 18pt
This paper proposes a general  two-directional simultaneous inference (TOSI) framework  for high-dimensional  models with a manifest variable or latent variable structure, for example, high-dimensional mean models, high-dimensional sparse regression models, and high-dimensional latent factors models. TOSI  performs simultaneous inference on a set of  parameters from two directions,  one to test whether  the assumed zero parameters indeed are zeros and one to test whether  exist zeros in the parameter set of nonzeros. As a result, we can better identify whether the parameters are zeros,  thereby keeping the data structure fully and parsimoniously expressed.  We theoretically prove that the single-split TOSI is asymptotically unbiased and the multi-split version of TOSI can control the Type I error below the prespecified  significance level.     Simulations  are conducted to examine the performance of the proposed method in finite sample situations and two real datasets are analyzed. The results show that the TOSI method can provide  more predictive and more interpretable estimators  than existing methods.
\end{abstract}


\par

\baselineskip 20pt
\spacingset{1.8} % DON'T change the spacing!
\section{Introduction}
\label{sec:intro}

Over the past {two decades}, great progress has been made in the field of high-dimensional data, where  the
number of parameters can be much larger than the sample size.  The most popular and powerful methods for handling high-dimensional data are  regularization methods \citep{tibshirani1996regression,fan_variable_2001} %,zou2005addendum, zhang2010nearly
and screening methods \citep{fan_sure_2008,  ma_concordance_2017}, % fan_nonparametric_2011, li_feature_2012,
which can be used to separate the set of parameters into  a set
 $G^{o}$ and its complement set $G^{no}$, which are inactive (or zero) and active (or non-zero) sets, respectively. %After obtaining sets $G^o$ and $G^{no}$, we regard them as deterministic sets instead of random sets.
Given {these} two sets, %\footnote{\red What we require to emphasize is that these two sets are regarded as deterministic quantities instead of stochastic quantities in this paper. It is reasonable since we can always obtain the sets by a small independent subsample. We will further introduce this point in identifying the structure of a model.},
two natural problems arise: (1) whether all the {elements} in $G^o$ are insignificant; and (2) whether all the {elements} in $G^{no}$ are significant. {Clearly, we can fully and parsimoniously express the data structure once we address these two problems; % identify  all the {elements} in $G^o$  are insignificant and  all the {elements} in $G^{no}$  are significant;
thus, it is important to perform statistical inference %The statistical inference
that  quantifies the uncertainty associated with the two problems.} %{ does}  matter and  has not been fully addressed.
{We formally formulate the problems (1) and (2)  as} %it is of interest to formulate the hypothesis testing problems as  }
\begin{eqnarray}
\label{eq:H1}
H_{0,G^{o}}: \btheta_j = \bbo, \forall j \in G^o  ~~~~~~~~~ \mbox{vs}  ~~~~~~~~ H_{1,G^{o}}: \btheta_j \neq \bbo, \exists j \in G^o,\\
\label{eq:H2}
\tilde H_{0,G^{no}}: \btheta_j = \bbo, \exists j \in G^{no}  ~~~~~~ \mbox{vs} ~~~~~~ \tilde H_{1,G^{no}}: \btheta_j \neq \bbo, \forall j \in G^{no},
\end{eqnarray}
where  $G^o$ and $G^{no}$ are the previously specified  inactive and active  sets, respectively, {and}
$\{\btheta_j \in R^{q}, j=1, \cdots, p\}$ is a set of  parameters of interest {with large $p$ and  fixed $q$.}
For example, $\btheta_j$ is the $j$th regression coefficient  in
a high-dimensional sparse regression model,  $\btheta_j$  is the mean of the $j$th variable  in a high-dimensional mean model, and $\btheta_j$ is the $j$th loading vector in a latent factor model.

Existing methods  focus on assessing problem (1), while statistical  inference  to quantify the  uncertainty associated with the identification of a group of important variables (problem (2)) is totally ignored in the literature. In fact, with  the inference  in the active set to   eliminate the possibility of including true zeros as nonzeros in the set,
we can obtain  interpretable and simpler  models. However,
the {assessment of} problem (2)   is more difficult than that of  problem (1) since  the explicit form % for {\red each test of $\btheta_j= \bbo$ ?? confusing with multiple testing??}
 is available for  problem (1) under the null hypothesis, but not for problem (2), where each $\btheta_j$ can take any value under the null hypothesis.
In {this} paper, under a general framework for high-dimensional data,
we propose {a two-directional simultaneous  test (TOSI)
to test $H_{0,G^{o}}$ and $\tilde H_{0,G^{no}}$ , i.e., whether all {elements} in $G^o$ are insignificant {(problem (1))} and
whether all {elements} in $G^{no}$ are significant {(problem (2))}.}

Existing methods for assessing problem (1) can be roughly divided into two categories:  $p$-value adjustment methods (PAMs) and  simultaneous inference methods (SIMs). PAMs are proposed for  testing  a single parameter  (TSP)  in  high-dimensional sparse regression models \citep{zhang2014confidence,degeer2014on}. % lsst16,buhlmann2013statistical, meinshausen2009p,
{Specifically, by} performing TSP on $H_0: \btheta_j = \bbo,  \ \mbox{vs}  \ \ H_{1}: \btheta_j \neq \bbo $ for each parameter $\btheta_j$ in $G^o$, we can obtain a set of $p$-values.
For these nominal $p$-values,  the PAMs for problem (1) are proposed to control the family-wise error rate (FWER) \citep{holm1979simple} or false discovery rate \citep{benjamini2001control}.
{Recently,  the sample-splitting techniques are  commonly used  under PAM framework.
For example, \cite{rinaldo2019bootstrapping} and \cite{barber2019knockoff} used single-split method for the post-selection inference so that the uncertainty from the estimation and {variable selection} can be ignored.
To achieve multi-adjustment of each $p$-value of regression coefficients,   \cite{wasserman2009high} proposed a single-split method,
\cite{meinshausen2009p} proposed a multi-split method,
and \cite{mandozzi2016hierarchical} proposed a hierarchical version of the
multi-split method in \cite{meinshausen2009p}.}
However, the  PAMs lack power due to {the strictness of FWER methods.}
Several SIMs  are  developed to improve the power. The max-type test statistic is a typical SIM.
\cite{jiang2004asymptotic} proposed {a} max-type test for a high-dimensional correlation matrix with a restriction of $p\leq n$.
For a set of mean parameters $\{\mu_j\}_{j=1}^p$ in a high-dimensional mean model  with   $p = o(\exp(n^c))$($0<c<1$), \cite{chernozhukov2013gaussian} and \cite{lou2017simultaneous} proposed a multiplier bootstrap method to {conduct} simultaneous testing based on $\max_j|\hat \mu_j - \mu_j|$.
\cite{zhang2017simultaneous}  and \cite{dezeure2017high} considered $\max_{j}|\hat\beta_j - \beta_j|$ to make simultaneous inference for  high-dimensional sparse regression {models} under {homogeneous}  and {heterogeneous} errors, respectively. {To enhance the power of the max-type test,  \cite{zhang2017simultaneous} also proposed a single-split method that divides a sample into two parts, called a three-step procedure, performing  variable selection using the first part and  simultaneous inference using the second part. Because  only part of the data are used for inference using the single-split method, the testing power is also restricted.} Since the max-type statistic used in the above works does not have a known {asymptotic} distribution, the bootstrap method is often {used} to determine the critical value of the test statistic \citep{chernozhukov2013gaussian, zhang2017simultaneous, dezeure2017high}.

However, these max-type  methods suffer from limitations. First, the max-type test can only  quantify the statistical uncertainty associated with the identification of  a group of insignificant  features, that is,  problem (1). Second, they are proposed to make simultaneous inference about problem (1) for specific models, for example, high-dimensional sparse regression models or  high dimensional mean model, the corresponding  test and inference differ by case.
Third, bootstrap methods {are computationally intensive} %requires intensive computation
and  may fail when the assumption of  independent observations does not hold \citep{chernozhukov2013gaussian,zhang2017simultaneous}, such as  in latent models where  the latent factors are estimated and, hence,  correlated with each other.

To overcome the aforementioned problems, we propose a  general framework based on sample splitting. Our contributions are as follows.

\textbf{Generality}:  Existing inference methods focus on  such as high-dimensional mean or variance models and high-dimensional sparse regression analysis,
in which the corresponding test and inference may differ case by case.
In the paper, we  provide a generalized framework for {two-directional} inference  for the models mentioned above, as well as  models which have not been considered in the literature, such as latent variable models.

\textbf{Interpretability}: By better identifying the sets of zeros and nonzeros,
we can explicitly  explore the latent structure in data, thereby achieving interpretability.  Furthermore, when we identify the sets of zeros $G^{o}(\lambda)$  and nonzeros   $G^{no}(\lambda)$ based on  extra samples for any given tuning parameter $\lambda$, the TOSI {method} can  choose a  tuning parameter $\lambda$  for which  $H_{0, G^{o}(\lambda)}$  is accepted and $\tilde H_{0, G^{no}(\lambda)}$ is rejected {if such $\lambda$ exists}, that is, the TOSI method  can  select  $\lambda$ so that the resulting sets of zeros ($G^{o}(\lambda)$) and nonzeros ($G^{no}(\lambda)$) are  statistically insignificant and significant, respectively.
Hence,  the $\lambda$ selected by TOSI
is meaningful to identify  important and unimportant variables. This is also observed in our simulation studies and two motivating data of  liquor sales and criminal data.
For example,  in  the high-dimensional sparse regression model of Experiment 1  in Section \ref{sec:simu},  {the LASSO with $\lambda$ chosen using cross-validation  exactly selects important variables at a frequency
(CS) of $10\%$ on average, but the LASSO with $\lambda$ chosen using TOSI can achieve a frequency of $95\%$ on average (Table \ref{tab:SelectLambda}).} In our motivating data of a liquor sales  data that $\lambda=0.1866$ selected using  ten-fold cross-validation identifies 9 important among  249  variables, while  $\lambda=0.3172$ selected using TOSI further identifies three unimportant  variables from the 9 variables, as presented in Table \ref{tab:winetest}. Further checking  via existing testing methods  for $H_{0, G^{o}(\lambda)}$ with $\lambda=0.3172$  show that  the  243 variables indeed were unimportant. Thus, the analysis results obtained via TOSI were more interpretable than those from CV LASSO. Similarly, for the criminal data, $\lambda=0.1788$ selected using  ten-fold cross-validation identifies 9 important variables and 91 unimportant variables, while  $\lambda=0.1828$ selected using TOSI further identified three unimportant  variables from the 9 variables, as presented in Table \ref{tab:criminaltest}.

\textbf{Computation}: Denote  $|G^o|$ and $|G^{no}|$ as the  size of $G^o$ and $G^{no}$, respectively. Using the two-stage test, we {convert} the simultaneous test for  $H_{0,G^{o}}$ and  $\tilde H_{0,G^{no}}$  into  $|G^o|$ and $|G^{no}|$ TSPs on  $H_0: \btheta_j = \bbo,  \ \mbox{vs}  \ \ H_{1}: \btheta_j \neq \bbo $ for $j\in G$ and $G^{no}$, respectively.
Any TSP method can be used. {Hence,} the computation and programming  are very simple.

\textbf{Asymptotic theory}:
We establish the validity of {the TOSI methods of two versions}, including a single-split version and a multi-split version. We prove that the single-split TOSI is asymptotically unbiased and the multi-split version of TOSI can control the Type I error below the prespecified  significance level.  The study in this paper is  the first attempt to discuss a two-directional  simultaneous  test.

The rest of {this} paper is organized as follows. In Section \ref{est1}, we present the general TOSI framework. The theoretical properties are investigated in Section \ref{asymp}. {We introduce the inference for sparse latent factor models and the applications for selection of penalty parameters  in Sections \ref{sec:lat} and \ref{sec:tune}, respectively.}  The performance of the proposed testing
procedure is {evaluated} via simulation studies in Section
\ref{sec:simu}.  In Section \ref{sec:real}, we apply {the TOSI method} to analyze two real
datasets with a sparse linear regression model.
A brief discussion about further research along  this direction is provided in
Section \ref{sec:dis}.  Technical proofs are relegated to the
Supplementary Materials.  In addition, we implement our proposed method in an efficient and user-friendly R package, {which is available
at \url{https://github.com/LinhzLab/TOSI}.}

\section{General framework}\label{est1}

\subsection{TOSI inference with $L=1$}
\label{sec:tosi}
Let $L$ denote the sample splitting times. We start with  $L=1$.
Consider $\{\z_i, i = 1,\cdots,n\}$  i.i.d. samples from a population $\z \in R^d$, where $d$ can exceed $n$. We are interested in testing a set of  parameters $\{\btheta_j \in R^{q}, j=1, \cdots, p\}$ with  fixed integer $q$.
Denote  $G^o\subseteq [p]$ as any subset of interest, where $[p]\hat=\{1,\cdots, p\}$. In practice, $G^o$ can be the indices of parameters that are penalized to zeros by using an extra sample independent of $\z_i$. We consider hypotheses (\ref{eq:H1}) and (\ref{eq:H2}).  Throughout this article, we allow the size $|G^o|$ and $|G^{no}|$ to grow as fast as $p$ which
can be the exponential order of $n$.

Suppose there exists an  estimator $\hat \btheta_j$ of $\btheta_j$ satisfying $\sqrt{n}(\hat\btheta_j - \btheta_j)\stackrel{d} \rightarrow  N(0, \Sigma_j)$, where $\Sigma_j$ can  be consistently estimated using $\hat\Sigma_j$.  We  randomly split the data $\{\z_i\}_{i=1}^{n}$ into two parts, $\mathcal{D}_1$ and $\mathcal{D}_2$. Without loss of generality, we take $|\mathcal{D}_1|=|\mathcal{D}_2|=n/2\hat= \bar n$ by {assuming} $n$ to be even.
{To test the null hypothesis in problem (1) that  all $\btheta_j$s with $j\in G^o$ are zeros, we propose a two-stage maximum (ToMax) test as below.}
\begin{description}
  \item[Stage I]: Use $\mathcal{D}_1$ to obtain estimator $\hat  \btheta^{(1)}_{j}$ of $\btheta_j$ and estimator $\wh \Sigma_{1j}$  of $\Sigma_{j}$,  then find $j_{\max} \in G^o$ such that $\|\wh\Sigma_{1 j_{\max}}^{-1/2}\hat  \btheta^{(1)}_{1j_{\max}} \| \geq \|\wh\Sigma_{1j}^{-1/2}\hat  \btheta^{(1)}_{j}\|$ for any $j \in G^o$.
  \item[Stage II]: Use $\mathcal{D}_2$ to obtain estimator $\hat  \btheta^{(2)}_{j_{\max}}$ of $\btheta_{j_{\max}}$, estimator $\wh \Sigma_{2j_{\max}}$ of $\Sigma_{j_{\max}}$ and calculate the $p$-value as $\wh p_{\max}=P\{\wh T_{{\max}} > \chi^2_{1-\alpha}(q)\}$, where $\wh T_{{\max}}=\bar n \hat  \btheta^{(2)\trans}_{j_{\max}} \wh\Sigma_{2j_{\max}}^{-1}\hat  \btheta^{(2)}_{j_{\max}}$.
\end{description}
Note that $j_{max}$ is a random variable determined by sample $\mathcal{D}_1$. Intuitively, in Stage I, we select an index with the most extreme statistics in group $G^o$, and we subsequently conduct  hypothesis {testing} for this index at Stage II.

\begin{Rem}
If the null hypothesis  that all parameters in $G^o$ are zeros holds, then {ToMax is equally likely to choose any index as}
$j_{\max}$. Thus,  the FWER can be controlled at the prespecified level.
If the null hypothesis is not true {and we denote} $G^o_{1} = \{j\in G^o: \btheta_j \neq \bbo\}$ and  $G^o_{0} = \{j\in G^o: \btheta_j = \bbo\}$,  then we have $\|\wh\Sigma_{1j}^{-1/2}\hat  \btheta^{(1)}_{j}\|\stackrel{p} \rightarrow \|\Sigma_j^{-1/2}\btheta_j\|=O(1)$ for $j\in G^o_{1}$ and $\|\wh\Sigma_{1j}^{-1/2}\hat  \btheta^{(1)}_{j}\|=o_p(1)$ for $j \in G^o_{0}$.  Under some conditions, we prove that the ToMax test is  asymptotically  unbiased.
\end{Rem}

\begin{Rem}
The two stages of ToMax  {convert}  the simultaneous test into $|G^o|+1$  TSP tests for $\btheta_j=\bbo, j\in G^o$, where  Stage I is equivalent to {conducting} $|G^o|$ TSP tests  for each $j \in G^o$ because $\|\wh\Sigma_{1j}^{-1/2}\hat  \btheta^{(1)}_{j}\|$ is an equivalent expression of the $p$-value for $\btheta_j=\bbo $, and Stage II  conducts  a TSP test   for  the selected index $j_{\max}$. Any  {TSP method} can be used in Stages I and II.
\end{Rem}

{ To test the null hypothesis in problem (2) that there exists $j\in G^{no}$ being zero, we propose a two-stage minimum (ToMin) test as below.}
\begin{description}
  \item[Stage I]: Use $\mathcal{D}_1$ to find $j_{\min} \in G^{no}$ such that $\|\wh\Sigma_{1j_{\min}}^{-1/2}\hat  \btheta^{(1)}_{j_{\min}}\| \leq \|\wh\Sigma_{1j}^{-1/2}\hat  \btheta^{(1)}_{j}\|$ for any $j \in G^{no}$, where $\hat  \btheta^{(1)}_{j}$ and $\wh\Sigma_{1j}$ are obtained based on $\mathcal{D}_1$.
  \item[Stage II]: Use $\mathcal{D}_2$ to estimate $\hat  \btheta^{(2)}_{j_{\min}}$ and calculate the p-value as $\wh p_{\min}=P\{\wh R_{{\min}} > \chi^2_{1-\alpha}(q)\}$, where $\wh R_{{\min}}=\bar n \hat  \btheta^{(2)\trans}_{j_{\min}} \wh\Sigma_{2j_{\min}}^{-1}\hat  \btheta^{(2)}_{j_{\min}}$, and {$\wh\Sigma_{2j_{\min}}$} is obtained based on $\mathcal{D}_2$.
\end{description}

\begin{Rem}
In the case that the null hypothesis in problem (2) is true, that is,  there exists a parameter in $G^{no}$ that is zero {and denoting} $G^{no}_{1} = \{j\in G^{no}: \btheta_j \neq \bbo\}$ and  $G^{no}_{0} = \{j\in G^{no}: \btheta_j = \bbo\}$,  we have $\|\wh\Sigma_{1j}^{-1/2}\hat  \btheta^{(1)}_{j}\|\stackrel{p} \rightarrow \|\Sigma_j^{-1/2}\btheta_j\|=O(1)$ for $j\in G^{no}_{1}$ and $\|\wh\Sigma_{1j}^{-1/2}\hat  \btheta^{(1)}_{j}\|=o_p(1)$ for $j \in G^{no}_{0}$. Under some conditions, we can select the  index $j_{\min} \in G^{no}_{0}$ {with probability of one} such that $\|\wh\Sigma_{1j_{\min}}^{-1/2}\hat  \btheta^{(1)}_{j_{\min}}\|=o_p(1)$. Thus,  the FWER can be asymptotically controlled at the prespecified level.
{In the case that the alternative hypothesis is true,}  we prove that the ToMin test is asymptotically unbiased and the power converges to one under some conditions.
\end{Rem}

By applying the ToMax and ToMin tests, we can simultaneously perform hypothesis testing for problems (1) and (2), termed  as TwO directional Simultaneous Inference (TOSI).

\subsection{TOSI inference with $L>1$}
When $L=1$, we randomly split the data $\{\z_i\}_{i=1}^{n}$ into two parts for once, which may cause a loss of efficiency of inference.
We consider an improvement  that uses  a multi-split method, i.e. $L >1$,  to make full use of data. Since the multi-split testing  method   for $H_{0,G^{o}}$ is the same as that of testing $\tilde H_{0,G^{no}}$, we only introduce the method for $H_{0,G^{o}}$. Specifically, we repeat the data splitting $L$ times
 and  obtain $p$-values $\{\wh p_{l,\max}, l=1,\cdots, L\}$  based on $\{\wh T_{l,\max}, l=1,\cdots, L\}$ for problem (\ref{eq:H1}).
Through Theorem \ref{th:maxtest}, it asymptotically holds
\begin{equation}\label{eq:pmax}
P(\wh p_{l,\max} \leq u) = u \mbox{ under $H_{0,G^{o}}$}.
\end{equation}

Illustrated by the idea in \cite{romano2019multiple}, we propose a testing procedure by
defining a rule that we reject $H_{0,G^{o}}$ if at least $k_{\max}$ out of the $L$ p-values are less than or equal to $\gamma$, where $0<k_{\max} \leq L$ and $\gamma \in (0,1)$. According to Markov's inequality, we have
\begin{equation}\label{eq:pmaxM}
P(\mbox{reject }H_{0,G^{o}})=P(\sum_{l=1}^{L}1_{\{\wh p_{l,\max} \leq \gamma\}} \geq k_{\max})\leq \frac{E\{\sum_{l=1}^{L}1_{\{\wh p_{l,\max} \leq \gamma\}} \}}{k_{\max}},
\end{equation}
where $1_{\{\wh p_{l,\max} \leq \gamma\}}$ is the indicator function.
Then, combing \eqref{eq:pmax} and \eqref{eq:pmaxM}, it asymptotically holds that
$P(\mbox{reject }H_{0,G^{o}}) \leq \frac{\gamma}{r}$
under $H_{0,G^{o}}$, where $r=\frac{k_{\max}}{L}$.
Therefore, if we choose $\gamma$ and $r$ such that $\frac{ \gamma}{r}=\alpha$, then the Type I error is asymptotically controlled below level $\alpha$.
Interestingly, we can regard the L times test to be a multiple test on the same null hypothesis $H_{0, G^{o}}$. When $H_{0, G^{o}}$ is true, the family wise error rate (FWER) is equal to $P(k_{\max} \geq 1)$ { which is  the Type I error when $r=1/L$. This is similar with the  Bonferroni  correction, which leads us to consider the following  more powerful method called ToMax$(L)$ by using Bonferroni-Holm (BH) procedure:}

 {\it Let $k_{\max}$ be the number of BH-adjusted p-values that are less than $\alpha$. We  reject $H_{0,G^{o}}$ if the
number of rejections $k_{\max} \geq 1$.}

 The validity of  ToMax$(L)$ is ensured by   Theorem \ref{th:multitestBH}. Clearly, ToMax is a special case of ToMax$(L)$ with $L=1$.

\section{Theoretical properties}\label{asymp}
We now investigate the statistical properties
of the TOSI test.  Recall
$G^o_{1} = \{j\in G^o: \btheta_j \neq \bbo\}$, $G^o_{0} = \{j\in G^o: \btheta_j = \bbo\}$,  $G^{no}_{1} = \{j\in G^{no}: \btheta_j \neq \bbo\}$, $G^{no}_{0} = \{j\in G^{no}: \btheta_j = \bbo \}$ and {denote $G_1= G^o_{1} \cup G^{no}_{1}$, the  nonzero index set, and $s=|G_1|$.}  $a_n \gg b_n$ implies that  $a_n$ dominates $b_n$ in order. We use $c$ to represent general positive constant which  may be different in different places.

\subsection{Conditions and explanation}
 We require some  conditions for the theoretical
properties displayed in Theorems \ref{th:maxtest}--\ref{th:multitestBH}.
\begin{itemize}
\item[\underline{{\bf (A1)}}] For each $j$, $\sqrt{n}(\hat\btheta_j - \btheta_j)\stackrel{d} \rightarrow  N(0, \Sigma_j)$, where $\Sigma_j$ can  be estimated consistently by $\hat \Sigma_j$.
\item[\underline{{\bf (A2)}}] $\max_{j\in G^o_{1}} \|\hat\btheta_j\| \gg \max_{j\in G^o_{0}}\|\hat\btheta_j\|$ if $G^o_{1} \neq \emptyset$.
\item[\underline{{\bf (A3)}}]  $\min_{j\in G^{no}_{1}} \|\hat\btheta_j\| \gg \min_{j\in G^{no}_{0}}\|\hat\btheta_j\|$ if $G^{no}_{1} \neq \emptyset$.
\item[\underline{{\bf (A4)}}] $\lim\limits_{n\rightarrow \infty} \inf\limits_{j\in G_{1}}\|\sqrt{n} \Sigma_j^{-1/2}\btheta_j\| > c>0$.
\end{itemize}

Conditions (A1)--(A4) are  weak and easily satisfied. %If $p$ is fixed, Conditions (A1)--(A3) hold naturally.
Condition (A1) ensures each population parameter $\btheta_j$ has asymptotically normal estimator $\hat\btheta_j$. In fact, this condition can be relaxed to  any known asymptotic distribution.
The relationship between $p$ and $n$ is implicitly contained in Conditions  (A2)--(A3). For example, in the  case with  the cardinalities of $G^o$ and $G^{no}$ being the same order of $p$, i.e., $|G^o|=O(p)$ and $|G^{no}|=O(p)$, it can be shown that  $\max_{j\in G^o_{0}}\|\hat\btheta_j\| = O_p(\sqrt{\frac{\ln (p)}{n}})$ and $\max_{j\in G^o_{1}} \|\hat\btheta_j-\btheta_j\|=O_p(\sqrt{\frac{\ln (p)}{n}})$ under some tail  probability restrictions, this coupling with $\max_{j\in G^o_{1}} \|\btheta_j\|\gg \sqrt{\frac{\ln (p)}{n}}$ due to $\max_{j\in G^o_{1}} \|\hat\btheta_j\|\geq \max_{j\in G^o_{1}} \|\btheta_j\| - \max_{j\in G^o_{1}} \|\hat\btheta_j-\btheta_j\|$, Condition (A2)  holds if $n \gg \ln (p)$. Similarly, Condition (A3) holds  {if $n \gg \ln (s)$  and $\min_{j\in G^{no}_{1}} \|\btheta_j\| \gg \sqrt{\frac{\ln (s)}{n}}$.}  Condition (A4) is a requirement for the lower bound of signals that is used to prove the unbiasedness of TOSI test.  We  give two  examples to explain Conditions (A1)--(A4), especially for (A2) and (A3).

{\bf Example 1 (High-dimensional mean models): } In high-dimensional mean models, $\btheta_j= E(z_{ij})$ with $q=1$, where $z_{ij}$ is the $j$-th component of $\z_i$ and $\inf_j var(z_{ij}^2) > c >0$. We can choose $\hat\btheta_j \hat= n^{-1}\sumi z_{ij}$.   Then Conditions (A1)--(A4) are satisfied if:  \underline{{\bf (B1)}}: there exist $r_1>0$ and $r_2>0$ such that  $P(|z_{ij}|>t)\leq \exp(-(t/r_2)^{r_1})$ for any $t>0$ and $j$, and \underline{{\bf (B2)}}: $\inf_{j\in G_1} |\btheta_j| \gg \sqrt{\frac{ln(p)}{n}}=o(1)$.

{Specifically}, Condition (B1) ensures the existence of a moment { at} any order, which leads to Condition  (A1)  by the central limit theorem. Furthermore, since $\max_{j\in G^o_{1}}|\hat\btheta_j|  \gg \sqrt{\frac{ln(p)}{n}}$ and $\max_{j\in G^o_{0}}|\hat\btheta_j|= O_p(\sqrt{\frac{ln(p)}{n}})$,  Condition  (A2) holds. Then, note that
$\min_{j\in G^{no}_{1}} |\hat\btheta_j|\geq \min_{j\in G_1} |\btheta_j|  - \max_{j\in G^{no}_{1}}|\hat\btheta_j -\btheta_j|$, by {applying} Conditions (B1) and (B2), we have $\min_{j\in G^{no}_{1}} |\hat\btheta_j| \gg \sqrt{\frac{ln(p)}{n}}$. Thus, Condition (A3) holds because $ \min_{j\in G^{no}_{0}}|\hat\btheta_j|$ cannot exceed the order $\sqrt{\frac{1}{n}}$. Finally, $\inf_j var(z_{ij}^2) > c >0$ implies (A4).

{The minimum signal assumption (B2) seems stringent. Actually, existing sample-splitting testing methods, by primarily focusing on testing $H_{0,G^{o}}$, all required a similar assumption to ensure that the error rate in the variable selection step is ignorable. Otherwise, the type I error in the inference step cannot be well controlled.
For example, \cite{wasserman2009high}  first proposed a single-split method to obtain the $p$-values of regression coefficients in high-dimensional sparse linear regression models, assuming a similar condition.
\cite{meinshausen2009p} improved the single-split method by proposing a multi-split method, and  assumed a sure screening property: $\lim_{n\rightarrow \infty}P(\tilde S \supseteq S)\rightarrow 1$,  which is more stringent than the minimum signal assumption~\citep{fan_sure_2008}.
\cite{zhang2017simultaneous}  proposed a single-split method, and also required the sure screening property for valid inference.
}

{\bf Example 2 (High-dimensional sparse linear regression models):} {There exist many classical inference methods for the sparse linear model, including the sample-splitting-based testing methods introduced in Section  \ref{sec:intro} and  non-sample-splitting-based testing methods, where the most popular method is the  bias-correction-based method, i.e., LASSO-type correction \citep{zhang2014confidence,degeer2014on} and ridge-type correction \citep{buhlmann2013statistical}, that obtains the $p$-value for each regression coefficient {and may  lack power due to the strictness of FWER methods.}
In addition,  \cite{meinshausen2015group} tested a specified group of regression coefficients based on  $l_1$-norm  but required a constraint Gaussian error assumption. In contrast to TOSI, these methods are limited  to handle the problem (1). {As another research line, \cite{lsst16} proposed post-selection inference approaches focusing on the confidence interval of each
coefficient in the best linear approximation to $E(y|\x)$ given a subset of selected covariates and served a different purpose from TOSI and other aforementioned methods.
}}

In the high-dimensional sparse linear regression model,
${\Y} = \X \btheta+ {\bvarepsilon},$
where   $\Y= (y_1, \cdots, y_n)\trans$ is a response vector, $\X=(\x_1, \cdots, \x_n)\trans$ is a $(n\times p)$-dimensional covariate matrix,  error ${\bvarepsilon}=(\varepsilon_1,\cdots, \varepsilon_n)\trans$ with $E(\varepsilon_i)=0$ and $var(\varepsilon_i)=\sigma^2$ is independent of $\X$, and an unknown regression vector $\btheta=(\btheta_{1}, \cdots, \btheta_{p})\trans$.  Suppose $\X$ has i.i.d. rows with mean zero and covariance matrix $\Omega= (\omega_{ij})$.  In this case, we have data $\z_i=(y_i, \x_i\trans)\trans, i=1,\cdots, n$ with $q=1$. We can choose $\hat \btheta$ to be the de-biased LASSO estimator in \cite{zhang2014confidence} and \cite{degeer2014on}.

Denote  $\Omega^{-1} \hat = \Theta=(\theta_{jk})_{j,k=1}^p$. Recall $s=|\{j:\btheta_j\neq 0\}|$ and let $s_j= |\{1\leq k\leq p:  \theta_{jk} \neq 0, k\neq j\}|$.  For Example 2, Conditions (A1)-(A4) hold if:
\underline{{\bf (C1)}} $\x_i$ is a sub-Gaussian random vector;
\underline{{\bf (C2)}} the minimum eigenvalue $\lambda_{\min}$ of $\Omega$ satisfies that $\lambda_{\min}> c$ and $\max_j \omega_{jj} <C$;
\underline{{\bf (C3)}} $\frac{s\ln (p)}{\sqrt{n}}=o(1)$ and $\max_j \frac{s_j \sqrt{\ln (p)}}{\sqrt{n}}=o(1)$;
\underline{{\bf (C4)}} $\varepsilon_i$ is a sub-Gaussian random variable; and
\underline{{\bf (C5)}} $\min_{j\in G_1}|\btheta_{j}| \gg \sqrt{\frac{\ln (p)}{n}}$.

{Condition (C1) is  similar to Assumption 2.1 in \cite{zhang2017simultaneous} and (B1) in \cite{degeer2014on} to
control the tail behavior of covariates.  Condition (C2) is the same as Assumption 2.2 in \cite{zhang2017simultaneous}, {which is used to upper bound the spectral norm of the precision matrix $\Theta$}.  Condition (C3) is a standard sparsity assumption for {regression coefficients $\btheta$} and the precision matrix $\Theta$, which is also assumed in Theorem 2.4 in \cite{degeer2014on}. Condition (C4) constrains  the tail behavior  of error term, which  is also used in  the Assumption 2.3(i) in \cite{zhang2017simultaneous}. Condition (C5) is a minimum signal strength assumption to  ensure the error rate in the variable selection step is ignorable and similar condition can be  found in assumptions (A2) and (A3) in \cite{wasserman2009high} and assumption (A1) in \cite{meinshausen2009p} since they also adopted the sample-splitting strategy for inference.}

\subsection{Theoretical results}

Denote $\btheta_{G^o}=\{\btheta_j: j\in G^o\}$ and $\btheta_{G^{no}}=\{\btheta_j: j\in G^{no}\}$. Let $\beta_{\hat T_{\max}}(\btheta_{G^o})=P(\hat T_{\max} > \chi^2_{1-\alpha}(q))$ and $\beta_{\hat R_{\min}}(\btheta_{G^{no}})=P(\hat R_{\min} > \chi^2_{1-\alpha}(q))$ be the power functions of ToMax and ToMin tests, respectively, then we present two theorems that ensure the validity of ToMax and ToMin tests.

\begin{Theorem}\label{th:maxtest}
Suppose that Conditions  (A1), (A2) and (A4) are satisfied, we have,
\begin{itemize}
\item[(i)]  Under $H_{0,G^{o}}: \btheta_j = \bbo, \forall j \in G^o $,  $\wh T_{{\max}}$ is asymptotically distributed as $\chi^2(q)$.
\item[(ii)] Under $H_{1,G^{o}}:\btheta_j \neq 0, \exists j\in G^o$,  then for a prefixed significance level $\alpha$,
$\beta_{\hat T_{\max}}(\btheta_{G^o})$ $\geq \alpha,$
when $n\rightarrow \infty$. In particular, \ if \ $\inf_{j\in G^o_{1}}\|\sqrt{n}\Sigma_j^{-1/2}\btheta_j\| \rightarrow \infty$, then $\beta_{\hat T_{\max}}(\btheta_{G^o}) \rightarrow 1$.
\end{itemize}
\end{Theorem}

\begin{Theorem}\label{th:mintest}
Suppose that Conditions  (A1), (A3) and (A4) hold, we have
\begin{itemize}
\item[(i)]  Under $\tilde H_{0,G^{no}}: \btheta_j = \bbo, \exists j \in G^{no}$,  $\wh R_{{\min}}$ is also asymptotically distributed as $\chi^2(q)$.
\item[(ii)] Under $ \tilde H_{1,G^{no}}: \btheta_j \neq \bbo, \forall j \in G^{no}$,  then for a prefixed significance level $\alpha$,
$\beta_{\hat R_{\min}}(\btheta_{G^{no}})$ $\geq \alpha,$
when $n\rightarrow \infty$. In particular, if $\inf_{j\in G^{no}}\|\sqrt{n}\Sigma_j^{-1/2}\btheta_j\| \rightarrow \infty$, then $\beta_{\hat R_{\min}}(\btheta_{G^{no}}) \rightarrow 1$.
\end{itemize}
\end{Theorem}

\begin{figure}[H]
  \centering
  \subfigure[ToMax]{
  \includegraphics[width=7.7cm, height=4.4cm]{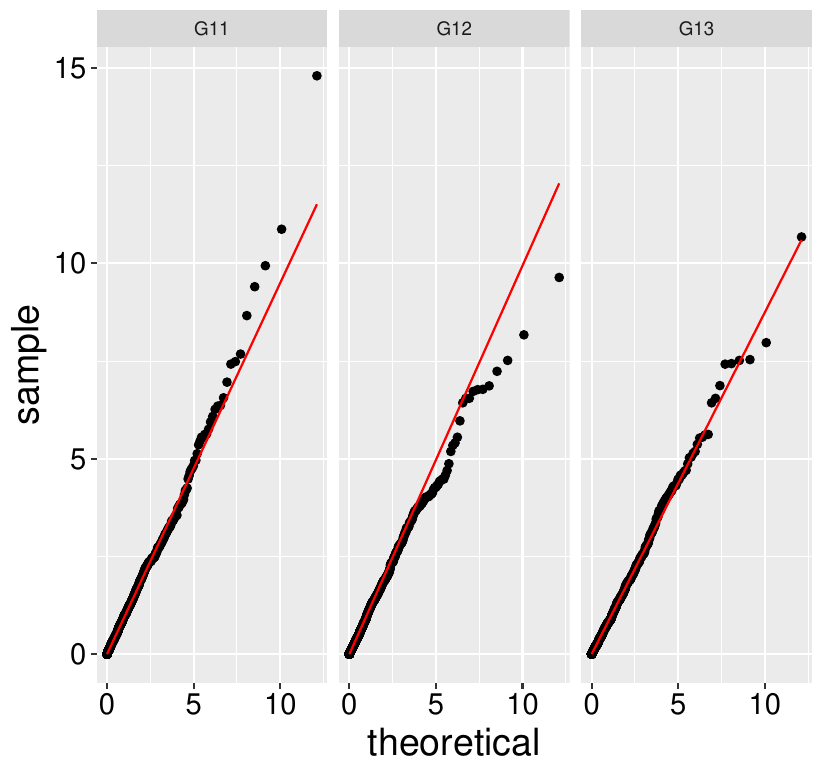}
	
}
\subfigure[ToMin]{
  \includegraphics[width=7.7cm, height=4.4cm]{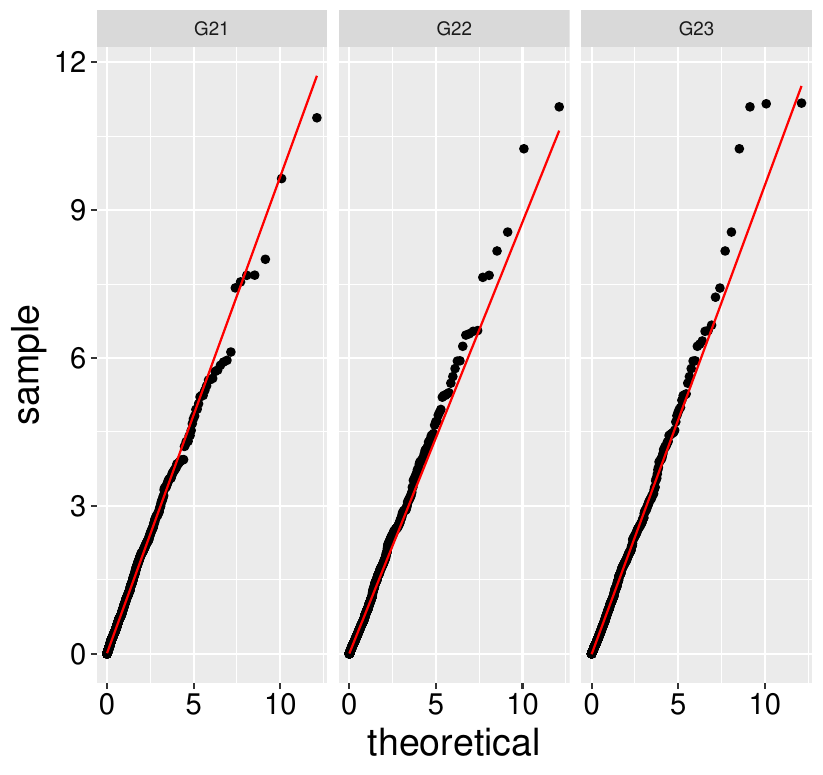}
}
\caption{(a)\&(b): QQ plots  from 2000 repeats for ToMax (with respect to three given sets $G_{11}$, $G_{12}$ and $G_{13}$) and ToMin (with respect to three given sets $G_{21}$, $G_{22}$, $G_{23}$)  with $\chi^2_{(1)}$ distribution under high-dimensional sparse linear regression  models ($n=100$) in Experiment 1. }\label{fig:RegM}
\end{figure}

We also perform  simulation studies to verify the asymptotically distributions of $\wh T_{{\max}}$ and $\wh R_{{\min}}$ in Theorems \ref{th:maxtest} and \ref{th:mintest}. Figure \ref{fig:RegM}(a)\&(b) show the QQ plots of the empirical distribution of ToMax and ToMin  vs the $\chi^2_{(1)}$ distribution under the corresponding null hypothesis for high-dimensional sparse linear regression  models, see  Experiment 1 in Section \ref{sec:reg}, which confirms the conclusion in Theorems \ref{th:maxtest} and \ref{th:mintest}.

We formally present the validity of ToMax$(L)$ through the following theorem.

\begin{Theorem}\label{th:multitestBH}
Under Conditions (A1), (A2) and (A4), for testing procedure ToMax$(L)$, it asymptotically holds that
$P(\mbox{reject }H_{0,G^{o}}|H_{0,G^{o}} \mbox{is true}) \leq \alpha.$
\end{Theorem}

\begin{Rem}
It is worth to be noted that Theorem \ref{th:multitestBH} automatically produces an combined p-values except of giving a decision of rejection or acceptance. In particular, we denote the BH-correction p-values to $\{\wh p^{adj}_{l,\max}, l=1,\cdots, L\}$, then the final combined p-value is $\wh p^{adj}_{\max}=\min_l \wh p^{adj}_{l,\max}$. If $\wh p^{adj}_{\max}< \alpha$, then we reject the null hypothesis.
\end{Rem}

Similarly, we can obtain  multi-split version and related theoretical properties for testing $\tilde H_{0,G^{no}}$, named by ToMin$(L)$. {ToMax(L)/ToMin(L) with $L > 1$ is a conservative method  in the sense that it controls the Type I error to not exceed the nominal level $\alpha$ rather than equal to $\alpha$. However, with multiple splits, each individual can be used for both Stages I and II if $L$ is sufficiently large; thus, the data are utilized more efficiently and  the power of  ToMax$(L)$/ToMin$(L)$ increases with increasing $L$, which  is confirmed by  our extensive simulation studies for high-dimensional sparse regression models, sparse latent factor models, and high-dimensional mean models in Section 6 and Supplementary Materials. However, when $L$ is sufficiently large, sufficient information has been used such that continually increasing $L$ cannot improve the power. Hence, in practice, we can choose a larger $L$ so that a stable conclusion can be obtained based on the resulting $p$-value.}

\section{Example 3. Sparse latent factor models}\label{sec:lat}
In this section, we introduce another example to illustrate the application of TOSI method to latent factor models. {The simultaneous inference in latent factor model can be used to select the important variables contributing to latent factors, such as cell-type-relevant genes in the area of genomics. Taking the single cell RNA sequencing (scRNA-seq) data as an example, the scRNA-seq data are measured on tens/hundreds of thousands of cells and tens of thousands of genes. The normalized data can be modelled by a linear factor model since cells often occupy a limited number of cell types~\citep{hou2020systematic}, where $\h_i$ is interpreted as the cell-type-related latent factors and $\b_j={\bf 0}$ means gene $j$ has no expression in all considered cell types. By selecting the genes with $\b_j\neq {\bf 0}$, we achieved the  variable (genes) selection for the downstream analyses.} % mongia2019mcimpute,

Suppose that the observations $\x_i= (x_{i1}, \cdots,
x_{ip})^{\trans}$  are correlated because they share a  latent factor $\h_i=(h_{i1},\cdots, h_{iq})^T$ with $q\ll p$. We  consider the  model,
\begin{eqnarray}
\label{eq:model1}
&&x_{ij} = \left\{\begin{array}{ll}\b_{j}^{\trans} \h_i+u_{ij},& j\in J\\
u_{ij},& j \in J^c,\end{array}\right.
\end{eqnarray}
where $J=\{j: \b_j \neq {\bf 0}, j=1,\cdots,p\}$, $J^c$ is the complement of $J$,
$\B = (\b_1,\cdots, \b_p)^{\trans}$ is a $p \times q$
deterministic matrix  ($\b_j={\bf 0}$ if $j\in J^c$),
$\u_i = (u_{i1},\cdots, u_{ip})^{\trans}$ is an error term independent of $\h_i$, $E(\u_i) = \bf 0$ and $\var(\u_i)=\mathrm{diag}(\sigma_1^2, \cdots, \sigma_p^2)$. By identifying the set $J$ in (\ref{eq:model1}), we can investigate which features contribute  to the latent factors $\h_i$. Denote $\b_j=(b_{j1},\cdots, b_{jq})^{\trans}$;  we further allow some of $b_{jk}, k=1,\cdots, q$ for $j \in J$ to be zero. By {identifying}  $b_{jk}$ to be {either} zero or nonzero, we can investigate  whether variable $j$ is associated with the $k$th component of the latent factor. Thus, we can explicitly  explore the  path between the high-dimensional {observed variables} and the latent factor to achieve the interpretability of the latent factor.
The high-dimensional sparse latent factor model  (\ref{eq:model1}) is substantially different from the manifest models in Examples 1--2 because of the {unobserved} latent factors $\h_i$.

Denote $\H=(\h_1,\cdots, \h_n)^{\trans}$. Since $\b_{j}^{\trans} \h_i=(\M^{\trans}\b_j)^{\trans} (\M^{-1}\h_i)$ for any invertible matrix $\M \in R^{q\times q}$, model (\ref{eq:model1}) is not
identifiable. To make it identifiable, similar to \cite{bai2013principal} and \cite{asz010}, we assume
 \underline{{\bf (E1)}} $n^{-1}\H^{\trans}\H=\I_q$; \underline{{\bf (E2)}} $\B^{\trans}\B$ is diagonal with
decreasing
diagonal elements {and} the first nonzero  element in each column of $\B$
is positive. Furthermore, for simplicity  we assume that the means of $x_{ij}$s
and $\h_i$s  have already been removed, namely,
$E(x_{ij})=0$ and $E(\h_i)=0$.

We propose {a} new Non-Iterative Two-Step  estimation (NITS)  in Appendix C.1 of Supplementary Materials for estimating $\H$ and $\B$, which obtains a sparse solution of $\B$. The resulting  estimators $\wh\H$ and $\wh\B$ have  closed forms; hence,  the computation and implementation are simple.
The large sample properties, including the identifiability of models and the
oracle property of the NITS estimators,  and their proofs are deferred to {the Supplementary Materials}.

\subsection{Inference on $\B$} \label{sec:tosifac}
We are interested in  making inferences on two aspects of two problems: (1) whether $\b_j$ identified as zero is indeed zero; whether $\b_j$ identified as nonzero  are significantly  different from zero; (2)  the same problems for  $b_{jk}$.
Therefore,  we can fully identify which  variables  are  associated with the latent factor $\h_i$ to improve the interpretability.  The inference on  entries $b_{jk}$ is similar with that for $\b_j$ and is omitted here.

Using  the estimation and variable selection described in {the Supplementary Materials}, we can partition all $p$ $\b_j$s into two groups: $G^o$ , the all unimportant index set and $G^{no}$, the active index set.  We are interested in the following  hypotheses,
\begin{eqnarray}\label{eq:H01}
&  H_{0,G^{o}}: \b_j = \bbo, \forall j \in G^o \ \mbox{vs}  \ \ H_{1, G^{o}}: \b_j \neq \bbo, \exists j \in G^o,\\
\label{eq:H02}
&\tilde H_{0,G^{no}}: \b_j = \bbo, \exists j \in G^{no}  \ \mbox{vs}  \ \  \tilde H_{1,G^{no}}: \b_j \neq \bbo, \forall j \in G^{no}.
\end{eqnarray}

To obtain the {test statistics for} (\ref{eq:H01}) and (\ref{eq:H02}),  we consider  $\widetilde\B=(\widetilde\b_1,\cdots, \widetilde\b_p)^{\trans}$ from
\begin{eqnarray} \label{eq:step1}
&&({\widetilde\H}, \widetilde\B)=\mbox{argmin}_{\B,\H}\parallel \X-\H\B^{\trans}\parallel_F^2.
\end{eqnarray}
Under Conditions (D1)--(D5) in Section \ref{sec:asy}, we {can show that}  $\widetilde\b_j$ satisfies
$\sqrt{n}(\widetilde\b_j - \b_{j}) \sim N(0, \sigma_j^2\I_q),$
where $\sigma_j^2$ is the variance of $u_{ij}$ {and} can be estimated by $\hat\sigma_j^2= n^{-1} \sumi (x_{ij} - \widetilde\h_i^{\trans}\widetilde\b_j)^2$.
See  Lemma 2 in {the Supplementary Materials}. The asymptotic normality of $\widetilde\b_j$ is also given in \cite{bai2013principal}  {in} the context of high-dimensional panel data. %  and \cite{bai2016maximum}
With $\btheta_j$ and $\hat\btheta_j$ replaced by $\b_j$ and $\widetilde\b_j$, respectively, the TOSI test described in Section \ref{sec:tosi}  can also be obtained for problems (\ref{eq:H01}) and (\ref{eq:H02}).

\subsection{Asymptotic Properties}
\label{sec:asy}

We now establish  the validity of
the proposed TOSI procedure.
   To establish the asymptotic properties, we need the regularity Conditions (D1)-(D6) given in  {Appendix C.1.3 of Supplementary Materials.}
Conditions (D1)--(D5) yields the asymptotical normality of $\widetilde\b_j$, which implies Condition (A1).
Conditions (A2) and (A3) can be proved by considering Condition (D6). Finally, Condition (D6.3) implies (A4). Then we give the following theorem whose proofs are deferred to Appendix C.2 in Supplementary Materials.

\begin{Theorem}\label{th:rowchiqmaxtest}
Under Conditions (D1)-(D6),  the same conclusions in Theorems \ref{th:maxtest} -- \ref{th:multitestBH} can be obtained.
\end{Theorem}

{
\section{Application to the  penalty parameter selection}\label{sec:tune}

As mentioned in Section \ref{sec:intro}, the TOSI method  can  select the penalty parameter $\lambda$ so that the resulting sets of zeros ($G^{o}(\lambda)$) and nonzeros ($G^{no}(\lambda)$) are  statistically insignificant and significant, respectively. Hence,  the $\lambda$ selected using TOSI is meaningful to identify  important and unimportant variables.
Here, we take the sparse linear regression model as an example to illustrate the selection of penalty parameter based on TOSI.   We first randomly split the data $\{\z_i\}_{i=1}^{n}$ into two parts $\mathcal{D}$ and $\mathcal{D}_{s}$ for inference and variable selection, respectively.  We design a bisection method for searching  $\lambda$ according to the inference results from TOSI. To start the searching, we  set an initial searching domain $\mathcal{H}_0=[\lambda^{(1)}_{lower}, \lambda^{(1)}_{upper}]$, and  use the K-fold cross validation (CV) to select an initial penalty parameter $\lambda^{(1)}\in \mathcal{H}_0$,  the estimated sets of zeros $G^{o}(\lambda^{(1)})$ and non-zeros $G^{no}(\lambda^{(1)})$ using LASSO based on the sample $\mathcal{D}_{s}$.
In the $l$th iteration, we update the inference and  estimators as follows.
  \begin{description}
  \item[Step a.] Based on sample $\mathcal{D}$, we use TOSI to test $H_{0,G^{o}(\lambda^{(l)})}: \btheta_j = 0, \forall j\in G^{o}(\lambda^{(l)})$ and $\tilde H_{0,G^{no}(\lambda^{(l)})}: \btheta_j = \bbo, \exists j \in G^{no}(\lambda^{(l)})$,
      resulting in four cases: (a) both are rejected; (b) both are accepted;  (c) the former is accepted and the latter is rejected;  and  (d)  the former is rejected and the latter is accepted. {Due to} the variable selection consistency based on LASSO, (d) is  rare and  ignored here.
      \begin{itemize}
        \item Case (a) implies there exists nonzero elements  in  $G^{o}(\lambda^{(l)})$, thus, we move $\lambda^{(l)}$ to a smaller one $\lambda^{(l+1)}=(\lambda^{(l)}+\lambda^{(l)}_{lower})/2$, and set $\lambda^{(l+1)}_{upper}=\lambda^{(l)}$ and $\lambda^{(l+1)}_{lower}=\lambda^{(l)}_{lower}$.
        \item Case (b) indicates there exists zero index in  $G^{no}(\lambda^{(l)})$, then  we move $\lambda^{(l)}$ to a larger one with $\lambda^{(l+1)}=(\lambda^{(l)}+\lambda^{(l)}_{upper})/2$, and set $\lambda^{(l+1)}_{lower}=\lambda^{(l)}$ and $\lambda^{(l+1)}_{upper}=\lambda^{(l)}_{upper}$.
        \item  In the case (c), the searching process is stopped and $\lambda^{(l)}$ is regarded as optimal.
      \end{itemize}
  \item[Step b.] Based on sample $\mathcal{D}_{s}$, we apply LASSO with the penalty parameter $\lambda^{(l+1)}$ to determine sets $G^{o}(\lambda^{(l+1)})$ and $G^{no}(\lambda^{(l+1)})$.
  \end{description}
Repeat the iteration until the searching  is stopped.
The simulation studies in Section \ref{sec:reg} show that the  bisection method based on TOSI performs better than the existing methods for  tuning penalty parameter, including K-fold CV, AIC, BIC, and scaled LASSO, as shown in  Table \ref{tab:SelectLambda}.
}

\section{Numerical studies}\label{sec:simu}
In this section, we conduct simulation studies to assess
the finite-sample performance of the proposed TOSI method {in}  {comparison  with the existing simultaneous inference methods and $p$-value-adjusted  methods.} For testing $H_{0,G^o}$, if there exists an adjusted $p$-value less than $\alpha$, then $p$-value-adjusted  methods reject it.
{We use testing size and power to evaluate the performance of the inference methods.}  The resulting size and power are obtained based on the empirical average from 500 repeats.  We also investigate the performance of the  TOSI in  guiding the selection of penalty parameters.

We considered three experiments corresponding to three models which
are high-dimensional mean models, sparse linear regression models and sparse factor models, with  $G^o$ and $G^{no}$ set as follows. For evaluating the testing size, we set $G^o$ to $G_{11}\hat=\{p-1,p\}, G_{12}\hat=\{(p/2), \cdots, p\}$ and $G_{13}\hat= \{s+1,\cdots,p\}$  and $G^{no}$ to $G_{21}\hat=\{p-1,p\}, G_{22}\hat=\{s+1, \cdots,p\}$ and $G_{23}\hat =\{1,\cdots,p\}$. For  evaluating the testing power, we set $G^o$ to $G_{14}\hat=\{2,s+1\}, G_{15}\hat=\{3, (s+1),\cdots,p\}$ and $G_{16}\hat= \{3,4, s+1,\cdots, p\}$   and $G^{no}$  to $G_{24}\hat=\{1,2\}$, $ G_{25}\hat=\{1,\cdots,4\}$ and $G_{26}\hat =\{1,\cdots, s\}$.
To save space, the results of high-dimensional mean models are deferred to Appendix D in Supplementary Materials.

\subsection{Experiment 1: High-dimensional sparse regression models}\label{sec:reg}

{We consider a high-dimensional regression model with the same setting as \cite{zhang2017simultaneous}.
In detail,} $y_i = \x_i^{\trans}\bbeta + \bvarepsilon_i, \bbeta=(\beta_1, \cdots, \beta_{50}),  i=1,\cdots, n=50 \mbox{ or } 100,$
where $\x_i \sim N(0, \Sigma^x)$ with $\sigma^x_{jk} = 0.8^{|j-k|}$, $\bvarepsilon_i \sim t(4)/\sqrt{2}$, $t(4)$ is Student's t-distribution with {degrees of freedom equal to four}. {We set $\beta_{j} = \rho z $ for $j\leq s$ and $\beta_{j} = 0$ for $j>s$ with $s=5$, where $z$ is a random variable following uniform distribution $U[0,2]$.  We consider three  signal-noise-ratio settings by taking $\rho=0.3, 0.5$ and $0.8$, respectively.} $\bbeta$ is fixed after being generated.

{TOSI adopts the debiased estimator   \citep{zhang2014confidence}  for high-dimensional sparse regression models in Stages I and II, % to construct the testing procedure,
where ten-fold cross validation is used to select the tuning parameters in  the nodewise LASSO; see \cite{zhang2014confidence} for details.

To benchmark the testing performance of TOSI in testing $H_{0, G^{o}}$, we compare it with five methods: (1) one-step procedure using test statistic $\max_j|\hat \beta_j - \beta_{j}|$ proposed in \cite{zhang2017simultaneous}, denoted as ZC1-17; (2) three-step procedure based single sample splitting in \cite{zhang2017simultaneous}, denoted as ZC3-17; (3) Benjamini-Yekutieli $p$-value-adjusted method~\citep{benjamini2001control} based on the $p$-values obtained from \cite{zhang2014confidence}, denoted as ZZ-14; (4) $p$-values corrected method based on the sample multi-splitting approach in \cite{meinshausen2009p}, denoted as MMB-09; (5) Holm $p$-value-adjusted method~\citep{holm1979simple} based on the $p$-values obtained from \cite{buhlmann2013statistical}, denoted as B-13.}
The results for different {settings}  from 500 replicates are presented in Table \ref{tab:regsplit}.
We conclude  (1) our ToMax/ToMin can asymptotically control the Type I error at the nominal level if sample size is adequate. (2) {ToMax$(L)$/ToMin$(L)$ has  a conservative size  below the significance level $0.05$, which is consistent with Theorem \ref{th:multitestBH}, and  has higher power than ToMax/ToMin by increasing $L$.} (3) The  difference in power between ToMax$(L)$ and {ZC1-17/ZC3-17} decreases as $L$  increases, and the ToMax$(8)$ outperforms {ZC1-17 and ZC3-17}.
 (4) {ToMax$(L)$ outperforms the three $p$-values-adjusted methods (ZZ-14, B-13 and MMB-09) in terms of testing size and power, especially for testing {sets} with smaller cardinalities.
 The $p$-values-adjusted methods control Type I error {too conservatively}
 and hence have lower testing powers.
 ZC1-17 somewhat fails to control the Type I error (0.07$\sim$ 0.13)  when sample size {is} not sufficiently large.}   (5) Larger $n$  {or signal-noise-ratio} improves the size and power due to the stronger signal. {The results for different signal-noise-ratio settings are referred to Table S2 in Supplementary Materials.}

{In addition to linear regression, we also  showcase the application of TOSI on a non-linear sparse logistic regression model, see Table S3 for results in Supplementary Materials,  which suggests  that TOSI outperforms the existing methods and  similar conclusions {as} those for sparse linear regression models can be obtained.  }

TOSI successfully guided the selection of the penalty parameters. To illustrate this result, we compare TOSI$(L=1)$ with  the cross-validation based on LASSO regression (CV LASSO), BIC based on LASSO (BIC), AIC based on LASSO (AIC) and scaled LASSO for Experiment 1 with $s=3$ and $\rho=0.3$.  First, we generate an independent sample $\mathcal{D}_{s}$ with sample size $50$. {Then, based on the bisection method given in Section \ref{sec:tune}, we sequentially conduct TOSI test based on $\mathcal{D}$ to select $\lambda$}, while
CV LASSO, BIC, AIC and scaled LASSO are based on sample $\mathcal{D}_{s} \cup \mathcal{D}$; see Appendix D for details in Supplementary Materials.
Table \ref{tab:SelectLambda}  shows that the average number of variables being selected (NV), the  percentage of  occasions {when} the important
variables are included in the selected model (IN),  and the percentage of occasions {when} exactly  select important variables (CS) over 500
replications. We observe that both methods {can} select important variables, however, CV LASSO, AIC and scaled LASSO usually over-selects the  variables and exactly selects important variables (CS) at a frequency  of $10\%, 36\%$ and $8\%$ on average, respectively. By contrast, TOSI achieves the highest frequency of $95\%$ on average. Thus, TOSI can more accurately identify the model structure. {Finally, by setting three different  nominal levels ($\alpha=0.1, 0.05$ and $0.01$), we verify that TOSI is robust to  the prespecified nominal level in identifying the model structure; see Table S4 in Supplementary Materials.}

\subsection{Experiment 2:  High-dimensional latent factor models}
Let $\b_{.k}$ be the $k$th column of $\B=(\b_1,\cdots,\b_p)^{\trans}, p=150$. To construct the sparse matrix $\B$, for $k<q$, we set the $j$-th component of $\b_{.k}$ to be nonzeros for $j\in A_k=\{(k - 1) \bar s_0 +1, \cdots, k \bar s_0\}$; and  the components of $\b_{.q}$ in locations $A_q=\{(q - 1) \bar s_0 +1, \cdots, s\}$ to be nonzeros, where  $\bar s_0 = \lfloor s/q \rfloor$ and $s$ {are} the number of $\b_{j}$s such that $\b_{j}\neq 0$, $\lfloor x \rfloor$ {is} the largest integer less than $x$. Hence, $J=\cup_{k=1}^q A_k=[s]$ and
the $k$th factor is contributed by the variables $x_{ij}$'s with $ j \in A_k$. We set $s=[3p/4]$
and randomly generate the nonzeros of $\b_{.k}$ from $\rho(1.5 - 0.24(k-1) + z)$, where $z\sim U[0,1]$ {and} $\rho=0.3$ controls the strength of the signal. %is, the stronger the signal is.
{Clearly}, $\B$ satisfies the identifiability condition (E2). We independently generate $\h_{i}, i=1,\cdots, n$  from a multivariate normal distribution
with mean zero and covariance matrix $(\sigma_{ij})_{q\times q}$ with
$\sigma_{ij}=0.5^{|i-j|}$. Then, we center and normalize $\H=(\h_{1},\cdots,\h_{n})\trans$ so that
$\H$ satisfies the identifiability condition (E1). We consider $q=1$. The results for the NITS estimator are deferred to   {Appendix C.3} of Supplementary Materials.
Table \ref{tab:Facsplit} shows testing size and power of TOSI under various settings, the similar conclusion with those for Table \ref{tab:regsplit} can be drawn. {In Table 2, we also compare ToMax$(L)$ with  a newly developed $p$-values-adjusted method (UY-21) by \cite{uematsu2021inference}. From Table \ref{tab:Facsplit}, we observe UY-21, with FDR 0.05, controls Type I error too conservative and has  a lower power than ToMax$(L)$.}

\section{Real data analysis}\label{sec:real}
{In this section, we apply TOSI to a  liquor sales dataset and a criminal dataset by using  high-dimensional linear regression models, where  the debiased estimator  in \cite{zhang2014confidence} is used to construct the testing procedure.}

\begin{figure}[H]
  \centering
  \subfigure[]{
	 \includegraphics[width=5.1cm, height=5.0cm]{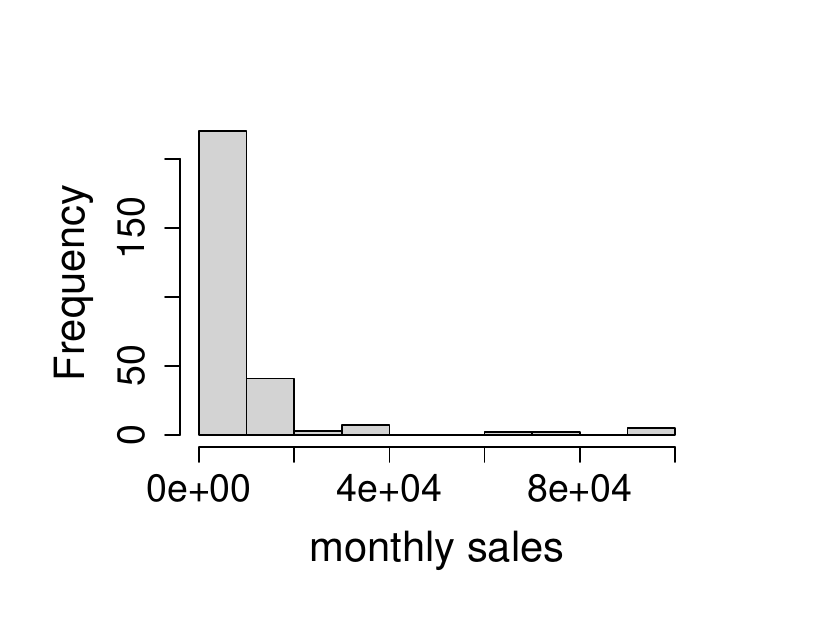}
	\label{fig:DBindex}}
\subfigure[]{
	 \includegraphics[width=5.1cm, height=5.0cm]{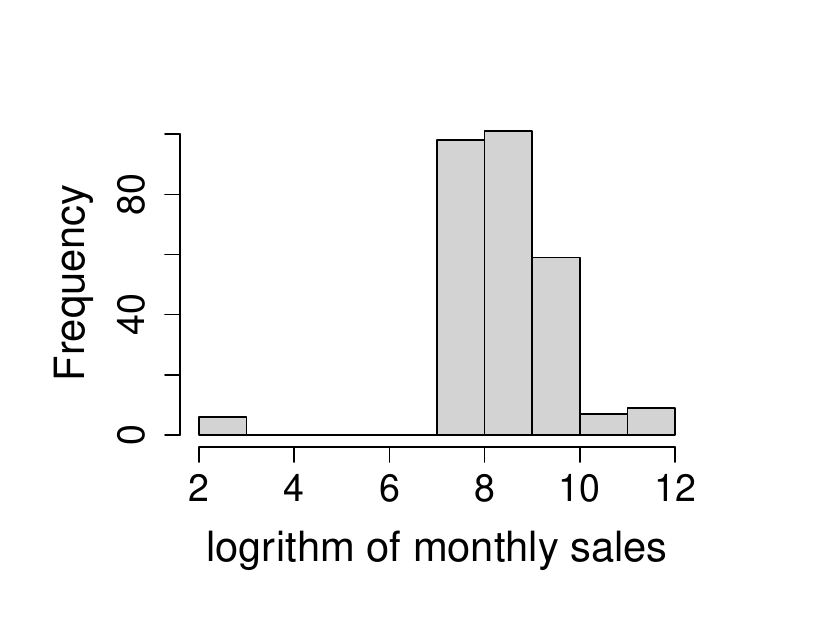}
	\label{fig:DBindex}}
\subfigure[]{
	 \includegraphics[width=5.1cm, height=5.0cm]{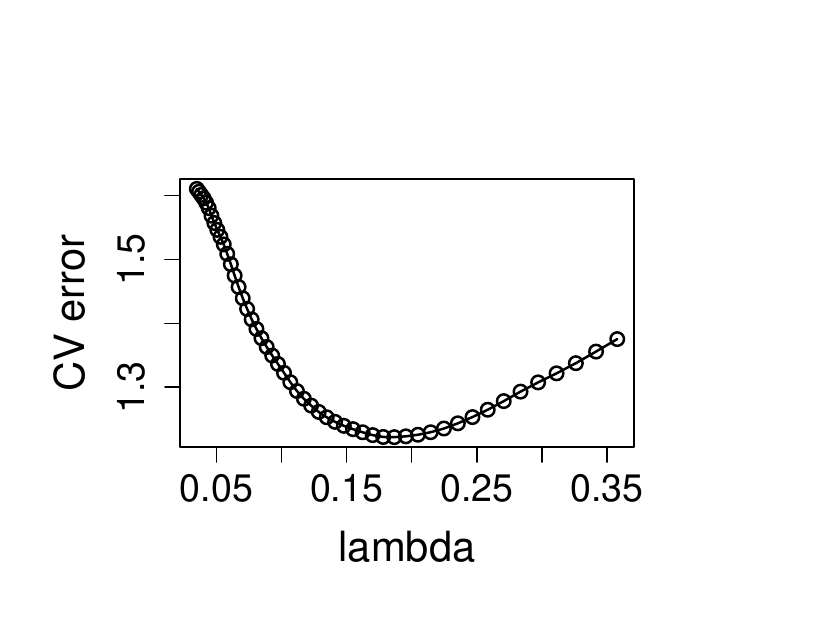}
	\label{fig:DBindex}
}
  \caption{(a)\&(b) Histograms of the target variable {\it monthly sales} and its log-transformation in liquor sales data; (c) Ten-fold cross validation error VS lambda in liquor sales data.}\label{fig:melambda}
\end{figure}

{
\subsection{Liquor sales data}

TOSI is now applied to analyze  a liquor sales dataset of Jiangsu province from one of China's
largest liquor companies. The purpose of the analysis is investigating factors that are associated with the monthly sales of liquor in Jiangsu province. The data set includes {\it monthly sales} $y_i$ and covariates information for $n = 280$ observations
in Jiangsu province  from 2011 to 2018. After data preprocessing, there are 249  covariates which include four parts: (a) the company's product information such as brand promotion and advertising investment, (b) brewing industry information such as monthly liquor yields and
monthly beer yields, (c) economic
information of related cities and towns such as per capita GDP, per capita disposable income and consumer price index,  and (d) geographic information such as monthly average temperature and monthly average relative humidity. Log transformation
is taken to response variable and all of covariates are standardized. A histogram of  {\it monthly sales} is shown in Figure \ref{fig:melambda}(a)  and the final response variable is took logarithm of  {\it monthly sales} (see Figure \ref{fig:melambda}(b)).  Then, we applied the TOSI method based on  high-dimensional regression models to explore  the influencing factors for sales of liquor.

First, we used lasso regression, implemented via the {\bf glmnet} package in R, to roughly separate the important variables and unimportant variables  based on the first 100 samples,
where the penalty parameter $\lambda_{opt}=0.1866$ is selected by ten-fold cross-validation; see  Figure \ref{fig:melambda}(c).
Then, we obtained {nine} important variables and 240 unimportant variables.

With some abuse of notation, we denote the index set with unimportant variables as $G^o$ and the index set with important variables as $G^{no}$.
Given $G^o$,  we used the rest samples to test whether there was nonzeros in $G^o$  using ToMax, {ZC1-17}, $p$-values correction method based on ridge
regression \citep{buhlmann2013statistical} ({B-13}), and $p$-values correction method based on Lasso regression \citep{zhang2014confidence,degeer2014on} ({ZZ-14}) and all methods failed to reject the null hypothesis.
The {TOSI} method could further test whether there were zeros in $G^{no}$, but the other methods could not.
The $p$-value of ToMin is 0.3018 and the adjusted $p$-values of ToMin$(5)$ is 0.7546, {indicating that} the null hypothesis could not be rejected; that is, there might be zeros in $G^{no}$. This is {consistent} with the simulation results that CV LASSO tended to over-select features.  Therefore, we increased the  value of penalty parameter by finely tuning within $\{0.2239, 0.3172, 0.3545\}$ until  the null hypothesis $\tilde H_{0,G^{no}}$ was rejected and the null hypothesis   $H_{0,G^{o}}$ was not rejected. Finally, we identified 6 important variables, as presented in Table \ref{tab:winetest}(a).
Furthermore, we  conducted another check whether the selected 243 variables were truly unimportant  via {ZC1-17, B-13 and ZZ-14}, and  none of these methods could reject the null hypothesis. Thus, {the analysis results obtained via TOSI were more interpretable than those from CV LASSO}.

Table \ref{tab:winetest}(b) presents the estimated coefficients for the 6 important covariates: the sales in the past months ({\it SL\underline{~}lag1} and {\it SL\underline{~}lag12}),  GDP from primary industry, mainly agriculture, in the last year ({\it gdp1\underline{~}lastyear}), brand promotion expenses for the past six months ({\it pptg\underline{~}six\underline{~}m}), the expense in giving products as gifts to customers in the past 12 months ({\it kq\underline{~}twelve\underline{~}m}), and the number of transactions of the stock 600779 in the past 10 months ({\it Stkcd600779\underline{~}lag10}), where 600779 is the stock code of another liquor company. From  Table \ref{tab:winetest}(b), we could draw following conclusions.
First, the positive coefficients of the sales in the past months ({\it SL\underline{~}lag1} and {\it SL\underline{~}lag12}) indicates that the larger the sales
in the past, the larger the sales in the current month. This is consistent with
intuition since the larger sales in the past could make customers trust this product more. Moreover, the coefficient of {\it SL\underline{~}lag1}  is much greater than that of {\it SL\underline{~}lag12}, which means {\it SL\underline{~}lag1} has greater influence on monthly sales.
Second,  GDP from primary industry, mainly agriculture, in last year
({\it gdp1\underline{~}lastyear})  had negative effect on
sales, which may be caused by the fact that areas with high agricultural output usually have low commercial operation ability.
Third,  the positive coefficient of brand promotion expenses for the past six months ({\it pptg\underline{~}six\underline{~}m}) shows that the higher the
brand promotion expenses, the larger the sales. \cite{porto2017multilevel} reported that promotional materials could generate a positive effect on product  sales.  Fourth, the positive coefficient of the expense in giving products as gifts to customers in the past 12 months ({\it kq\underline{~}twelve\underline{~}m}) suggests the expense in gift giving could improve the monthly sales since this gift-giving behavior can attract consumers' interest of the product. Lastly, the number of transactions of the stock 600779 in the past 10 months ({\it Stkcd600779\underline{~}lag10}) has a negative coefficient that indicates the transactions of stocks of the competitor could impair the monthly sales.
}

\subsection{Criminal data}
\begin{figure}[H]
\centering
 \subfigure[]{
	 \includegraphics[width=6cm, height=5.5cm]{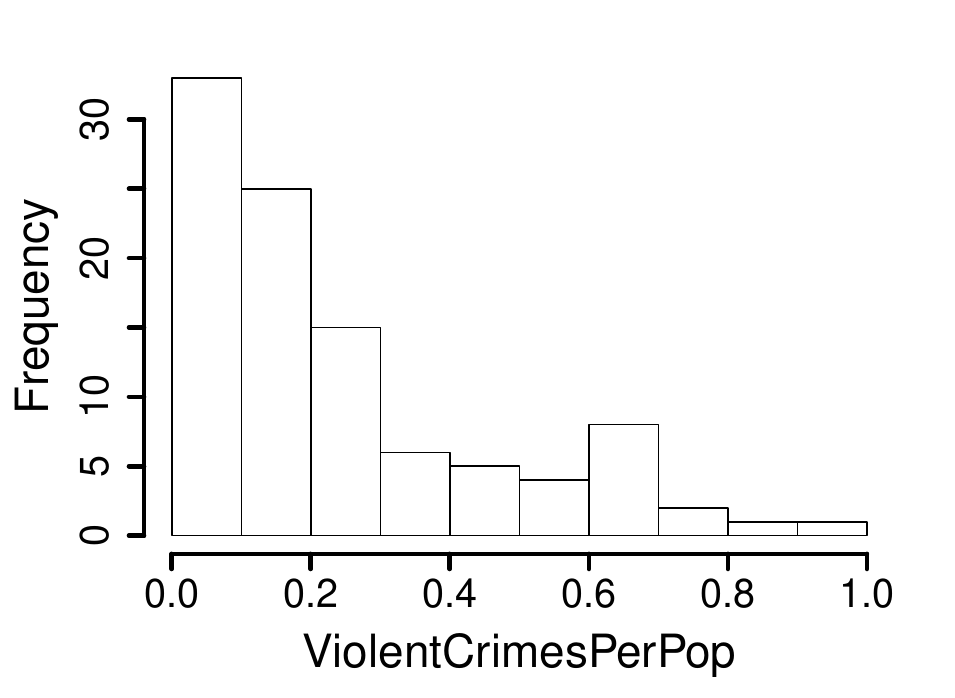}
	\label{fig:DBindex}
}
\subfigure[]{
	 \includegraphics[width=6cm, height=5.5cm]{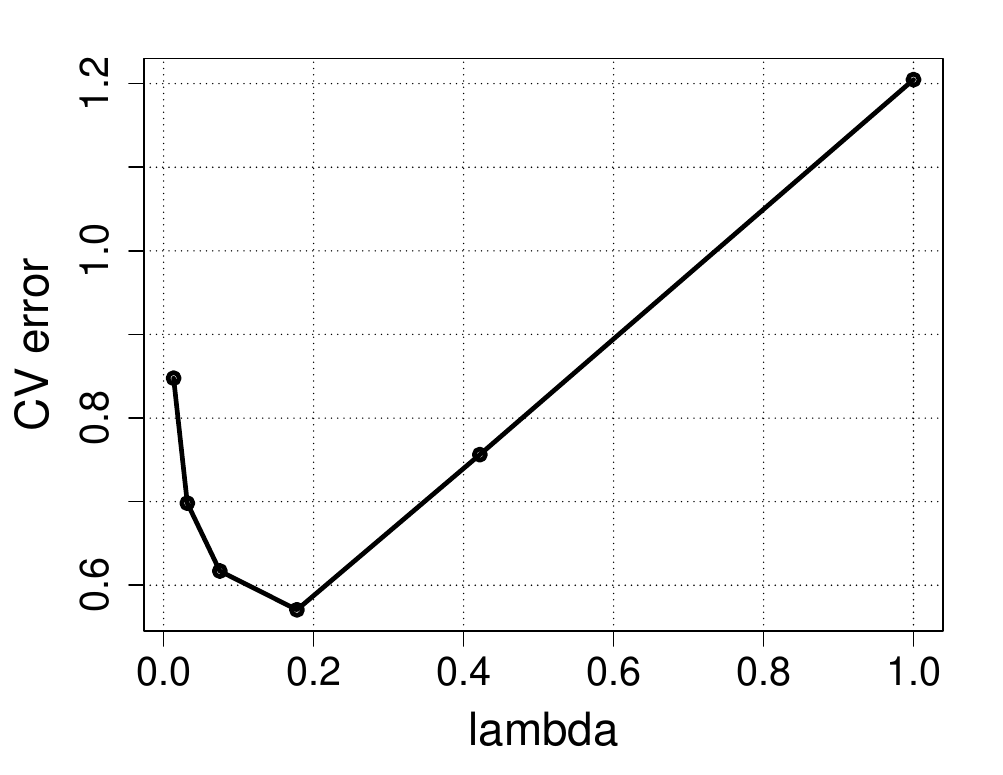}
	\label{fig:DBindex}
}
  \caption{(a) Histogram of the response variable {\it monthly sales}  in criminal data; (b) Ten-fold cross validation error VS lambda in  criminal data.}\label{fig:Crilambda}
\end{figure}

In {this} subsection, we {analyze}  a criminal dataset \citep{redmond2002data} to demonstrate the usefulness of TOSI.  {This} dataset is collected from 200 communities within the United States and combines socio-economic data
from the 1990 US Census, law enforcement data from the 1990 US LEMAS survey, and crime
data from the 1995 FBI UCR. After data preprocessing, {we removed variables with seriously missing values and  zero variance} and
obtained 101 variables for each community. The response variable of interest is the total number of violent crimes per 100K population ({\it ViolentCrimesPerPop}), which describes the severity of crime in the community. A histogram of {\it ViolentCrimesPerPop} is shown in Figure \ref{fig:Crilambda}(a)  and the final response variable is took logarithm of {\it ViolentCrimesPerPop}. The remaining 100 covariates{,  including}  population of the community, mean number of people per household,
percentage of males who are divorced, and percentage of kids aged 12-17 years in two-parent households, were used as predictors. We were interested
in which variables  had an impact on ViolentCrimesPerPop. To illustrate the application in  high-dimensional regression models, we applied our TOSI method to solve this problem.

Similarly, we used lasso regression to roughly determine the important variables and unimportant variables  based on the first 100 samples,
where the penalty parameter $\lambda_{opt}=0.1778$ is selected; see  Figure \ref{fig:Crilambda}(b).
Then, we obtained {nine} important variables and 91 unimportant variables.

We denote the index set with unimportant variables as $G^o$ and the index set with important variables as $G^{no}$.
Given $G^o$,  we used the last 100 samples to test whether there was nonzeros in $G^o$  using ToMax, {ZC1-17, B-13 and ZZ-14}.
All methods failed to reject the null hypothesis, so $G^o$ may contained all zeros.
In contrast to other methods, {TOSI} method could further test whether there were zeros in $G^{no}$.
The $p$-value of ToMin is 0.1432 and the adjusted $p$-values of ToMin$(5)$ is 0.2864, {indicating that} the null hypothesis could not be rejected; that is, there might be zeros in $G^{no}$. This is {consistent} with the simulation results that CV LASSO tended to over-select features.
 Therefore, we increased the  value of penalty parameter by finely tuning within $\{0.1828, 0.1878\}$ until  the null hypothesis $\tilde H_{0,G^{no}}$ was rejected and the null hypothesis   $H_{0,G^{o}}$ was not rejected at significance level 0.05. Finally, we  identified 6 important variables, as presented in Table \ref{tab:criminaltest}(a).
Furthermore, we  conducted another check whether the selected 94 variables were truly unimportant  via {ZC1-17, B-13 and ZZ-14}, and  none of these methods could reject the null hypothesis. Therefore, {the analysis results obtained via TOSI were more interpretable than those from CV LASSO}.

Table \ref{tab:criminaltest}(b) presents the estimated coefficients for the 6 important covariates: the percentage of the population that is Caucasian ({\it racePctWhite}),  the percentage of population that is of hispanic heritage ({\it racePctHisp}), the number of people under the poverty level ({\it NumUnderPov}), the percentage of females who are divorced ({\it FemalePctDiv}), the percentage of kids in family housing with two parents ({\it PctKids2Par}) and the percentage of people in owner occupied households ({\it PctPersOwnOccup}). From  Table \ref{tab:criminaltest}(b), we concluded that {\it racePctWhite, PctKids2Par} and  {\it PctPersOwnOccup} had negative effects on {\it ViolentCrimesPerPop}, while {\it racePctHisp, NumUnderPov} and
{\it FemalePctDiv} had positive effects on  {\it ViolentCrimesPerPop}.
 These results were easy to understand and explain.
First, {intuitively}, the greater the percentage of people below the poverty level is, the more serious the violent crime in the community is.
Second, divorce increases the likelihood of violent crime.  Moreover, having two parents at home has a substantial inhibitory effect on juvenile violent crime.

\section{Discussion}\label{sec:dis}
In this paper, we introduced a two-directional simultaneous inference framework for high-dimensional manifest  and latent  models.
With TOSI, we  can fully identify the zero and nonzero  parameters, resulting in  more interpretable and simpler  models.
The simultaneous inference procedure  achieves the prespecified significance level asymptotically and has power tending to one. Three typical models are considered as examples to illustrate the application of our TOSI framework, and the corresponding theoretical properties are established.
Simulation studies and two real high-dimensional data examples are used to verify the performance and effectiveness of the estimation and inference, and the results are satisfactory.

In this paper, we focus on the two testing problems for $G^o$ and $G^{no}$. In fact,  TOSI can be applied to any two  subsets $G_1$ and $G_2$ of $[p]$ and all the theoretical properties  hold. However, TOSI has several potential weakness. The single-split version of TOSI can control the Type I error at the prespecified significance level but lacks efficiency. Although the multi-split version of TOSI can mitigate this problem, it controls the Type I error tightly which leads to some loss of power. How to find a statistic in the multi-split method that can  control the Type I error exactly at the prespecified significance level,  is a potential direction for future research.

\section*{Acknowledgments}
The research were partially supported by National Natural Science Foundation of China (Nos. 11931014 and 11829101) and National Key R\&D Program of China (No. 2022YFA1003702).

\section*{Conflict of interest statement}
The authors report there are no competing interests to declare.

\begin{table}
            \scriptsize\centering\renewcommand{\arraystretch}{0.85}
            \caption{Results of Experiment 1: high-dimensional
            sparse linear regression models. Comparison of testing size and power for TOSI and other methods under the significance level $\alpha=0.05$, where ZC3-17($x$) represents that $x$ fraction of samples are used for screening step in the three-step procedure in \cite{zhang2017simultaneous} and we take $1/5$ and $1/3$ by following the setting in \cite{zhang2017simultaneous}.} \label{tab:regsplit}
            \begin{tabular}{ c l c c c c cc ccc}
           \cmidrule(lr){1-8}
           % &\multicolumn{7}{c}{{\bf (a) Testing results}} \\  \cmidrule(lr){1-8}
&Method&\multicolumn{3}{c}{$n=50$}& \multicolumn{3}{c}{$n=100$}  \\
\cmidrule(lr){3-5}\cmidrule(lr){6-8}
%&& \multicolumn{6}{c}{ToMax$(L)$ vs Others} \\
 \cmidrule(lr){1-8}
%\cmidrule(lr){3-5}\cmidrule(lr){6-8}
 &&  $G_{11}$&  $G_{12}$&  $G_{13}$&  $G_{11}$&  $G_{12}$&  $G_{13}$  \\
 \cmidrule(lr){3-5}\cmidrule(lr){6-8}
%\hline
Size&ToMax$(1)$&  0.030&  0.010&  0.015&  0.050&  0.045&  0.045  \\
&ToMax$(2)$ &  0.015&  0.030&  0.040&  0.030&  0.035&  0.060  \\
&ToMax$(5)$ &  0.015&  0.020&  0.045&  0.030&  0.045&  0.025  \\
&ToMax$(8)$ &  0.000&  0.020&  0.040&  0.025&  0.040&  0.015  \\
&{ZC1-17} &  0.070&  0.100&  0.130&  0.050&  0.065&  0.070  \\
&ZC3-17(1/5) & 0.029&  0.037&  0.039&  0.049&  0.041&  0.031 \\
&ZC3-17(1/3) & 0.031&  0.031&  0.034&  0.052&  0.026&  0.030\\
&ZZ-14 &  0.010&  0.010&  0.005&  0.025&  0.005&  0.000  \\
&B-13&  0.000&  0.002&  0.006& 0.004&  0.004&  0.006 \\
& MMB-09& 0.000&  0.000&  0.000 & 0.000&  0.000&  0.000 \\  \cmidrule(lr){1-8}
& &  $G_{14}$&  $G_{15}$&  $G_{16}$&  $G_{14}$&  $G_{15}$&  $G_{16}$  \\
\cmidrule(lr){3-5}\cmidrule(lr){6-8}
Power&ToMax$(1)$&  0.115&  0.115&  0.210&  0.215&  0.250&  0.400  \\
&ToMax$(2)$ &  0.135&  0.175&  0.275&  0.195&  0.350&  0.535  \\
&ToMax$(5)$ &  0.180&  0.200&  0.355&  0.250&  0.405&  0.620  \\
&ToMax$(8)$ &  0.205&  0.205&  0.400&  0.260&  0.440&  0.655  \\
&{ZC1-17}&  0.170&  0.200&  0.350&  0.200&  0.230&  0.585  \\
&ZC3-17(1/5) & 0.206&  0.146&  0.340 & 0.264&  0.140&  0.500\\
&ZC3-17(1/3) &  0.208&  0.170&  0.336& 0.248&  0.168&  0.446\\
&ZZ-14 &  0.150&  0.100&  0.255&  0.275&  0.160&  0.500  \\
&B-13&  0.000&  0.010&  0.036& 0.010&  0.022&  0.102 \\
&MMB-09&0.005&  0.120&  0.295   &0.035&  0.190&  0.465    \\  \cmidrule(lr){1-8}
%&& \multicolumn{6}{c}{ToMin$(L)$} \\
 \cmidrule(lr){1-8}
%\cmidrule(lr){3-5}\cmidrule(lr){6-8}
& &  $G_{21}$&  $G_{22}$&  $G_{23}$&  $G_{21}$&  $G_{22}$&  $G_{23}$  \\
\cmidrule(lr){3-5}\cmidrule(lr){6-8}
% L=1&  0.015&  0.030&  0.020&  0.055&  0.050&  0.045  \\
Size&ToMin$(1)$ &  0.015&  0.030&  0.020&  0.055&  0.050&  0.045  \\
&ToMin$(2)$&  0.010&  0.030&  0.025&  0.035&  0.020&  0.035  \\
&ToMin$(5)$ &  0.010&  0.010&  0.020&  0.025&  0.025&  0.020  \\
&ToMin$(8)$ &  0.005&  0.010&  0.020&  0.015&  0.035&  0.025  \\
%BY &  0.000&  0.000&  0.000&  0.000&  0.000&  0.000  \\\hline
 \cmidrule(lr){1-8}
& &  $G_{24}$&  $G_{25}$&  $G_{26}$&  $G_{24}$&  $G_{25}$&  $G_{26}$  \\
\cmidrule(lr){3-5}\cmidrule(lr){6-8}
Power& ToMin$(1)$&  0.105&  0.195&  0.130&  0.165&  0.345&  0.275  \\
&ToMin$(2)$&  0.095&  0.255&  0.185&  0.170&  0.410&  0.350  \\
&ToMin$(5)$&  0.160&  0.325&  0.255&  0.205&  0.540&  0.480  \\
&ToMin$(8)$&  0.195&  0.475&  0.380&  0.210&  0.640&  0.570  \\
%BY &  0.010&  0.000&  0.000&  0.005&  0.000&  0.000  \\
 \cmidrule(lr){1-8}
            \end{tabular}
            \end{table}

\begin{table}
            \scriptsize\centering\renewcommand{\arraystretch}{0.85}
            \caption{Results of Experiment 1: high-dimensional
            sparse linear regression models. Simulation results of identifying the structure of models for the proposed TOSI, cross-validation/AIC/BIC based on lasso regression (CV LASSO, AIC, BIC) and scaled LASSO, where $s=3$. NV, average number of the variables being selected; IN, percentage of occasions on which the correct variables are included in the selected model; CS, percentage of occasions on which correct variables are selected. %; Time, running time (seconds), averaged over 500 replications.
            } \label{tab:SelectLambda}
            \begin{tabular}{ c c c c cc c ccc c}
           \cmidrule(lr){1-11}
% &\multicolumn{7}{c}{{\bf (b) Results of variable selection and {\blue average running time!}}} \\ \cmidrule(lr){1-11}
&&\multicolumn{3}{c}{TOSI} & \multicolumn{3}{c}{CV LASSO}&\multicolumn{3}{c}{AIC} \\
\cmidrule(lr){1-2}\cmidrule(lr){3-5}\cmidrule(lr){6-8}\cmidrule(lr){9-11}
$(n,p)$& $\rho$ &   NV &  IN&  CS&  NV&    IN&  CS &  NV&    IN&  CS  \\ \cmidrule(lr){1-11}

$(50,50)$& 2 &  2.890&  0.864&  0.848 &     8.363&  1.000&  0.104 &    4.171& 1.000& 0.439  \\
%&&  SD& 5.6528&  0.0000&  0.3057&  1.3314&  0.3109&  0.3808  \\
&3 &  3.022&  0.998&  0.990  &  8.328&  1.000&  0.108   &   4.172& 1.000& 0.443  \\
% & &SD&   5.6629&  0.0000&  0.3110&  1.3669&  0.04477&  0.2304  \\ \hline
$(100,50)$& 2 &  3.042&  0.988&  0.966 &     7.864&  1.000&  0.112 &    5.166& 1.000& 0.264 \\
%&&  SD&    4.8395&  0.0000&  0.3157&  3.1490&  0.0996&  0.3742  \\  \cmidrule(lr){3-9}
&3 &  3.030&  1.000&  0.994  & 7.882&  1.000&  0.102 &  5.164  &  1.000& 0.278 \\
%& &SD&  4.8633&  0.0000&  0.3030&  2.0841&  0.0000&  0.3030  \\
\cmidrule(lr){1-11}
&&\multicolumn{3}{c}{BIC} & \multicolumn{3}{c}{scaled LASSO} \\
\cmidrule(lr){1-2}\cmidrule(lr){3-5}\cmidrule(lr){6-8}
$(n,p)$& $\rho$ &   NV &  IN&  CS&   NV&    IN&  CS  \\ \cmidrule(lr){1-11}
$(50,50)$& 2 &  3.269& 1.000& 0.778&   5.416& 1.000& 0.080 \\
&3 &  3.251& 1.000& 0.788  & 5.403& 1.000& 0.080 \\
$(100,50)$& 2 & 3.434& 1.000& 0.688&   5.430& 1.000& 0.078 \\
&3 &  3.430&    1.000& 0.692    & 5.420  &  1.000& 0.079 \\ \cmidrule(lr){1-11}
            \end{tabular}
            \end{table}

\begin{table}[H]
\scriptsize\centering \caption{Results of Experiment 2:  high-dimensional sparse factor models. Comparison of testing size and power for TOSI and other method under the significance level $\alpha=0.05$.}
\label{tab:Facsplit}
\begin{tabular}{c l c c c c c c }
\hline
%& \multicolumn{6}{c}{ToMax$(L)$} \\ \hline
& &\multicolumn{3}{c}{$n=200$}& \multicolumn{3}{c}{$n=400$}  \\ % $\sigma^2=1$
\cmidrule(lr){2-5}\cmidrule(lr){6-8}
& Method&  $G_{11}$&  $G_{12}$&  $G_{13}$&  $G_{11}$&  $G_{12}$&  $G_{13}$  \\
\cmidrule(lr){2-8}
&ToMax(1)&  0.060&  0.062&  0.072&  0.052&  0.034&  0.046  \\
%$L=2$&  0.062&  0.062&  0.066&  0.052&  0.036&  0.050  \\
&ToMax(2)&  0.072&  0.060&  0.066&  0.056&  0.052&  0.066  \\
&ToMax(8)&  0.052&  0.068&  0.064&  0.050&  0.048&  0.058  \\
&ToMax(15)&  0.060&  0.066&  0.070&  0.044&  0.044&  0.054  \\
&ToMax(20)&  0.054&  0.064&  0.062&  0.050&  0.040&  0.054  \\
&  UY-21 & 0.005&  0.0201&  0.045&  0.005&  0.035&  0.065  \\
% BC &  0.048&  0.058&  0.068& 0.048&  0.052&  0.066 \\
%BY&  0.030&  0.018&  0.012& 0.030&  0.020&  0.024  \\
%& \multicolumn{6}{c}{ToMin$(L)$} \\  \hline
%&\multicolumn{3}{c}{$n=200$}& \multicolumn{3}{c}{$n=400$}  \\
\cmidrule(lr){2-5}\cmidrule(lr){6-8}
Size& Method&  $G_{21}$&  $G_{22}$&  $G_{23}$&  $G_{21}$&  $G_{22}$&  $G_{23}$  \\
\cmidrule(lr){2-8}
&ToMin(1)&  0.062&  0.046&  0.046&  0.052&  0.038&  0.038  \\
%$L=2$&  0.070&  0.046&  0.046&  0.052&  0.034&  0.034  \\
&ToMin(5)&  0.060&  0.062&  0.062&  0.046&  0.040&  0.040  \\
&ToMin(8)&  0.062&  0.054&  0.054&  0.038&  0.048&  0.048  \\
&ToMin(15)&  0.046&  0.074&  0.074&  0.034&  0.052&  0.052  \\
&ToMin(20)&  0.046&  0.078&  0.078&  0.034&  0.046&  0.046  \\
%BY &  0.000&  0.000&  0.000&  0.000&  0.000&  0.000  \\
\hline
& &\multicolumn{3}{c}{$n=100$}& \multicolumn{3}{c}{$n=200$}  \\ % $\sigma^2=3$
\cmidrule(lr){2-5}\cmidrule(lr){6-8}
&Method&  $G_{14}$&  $G_{15}$&  $G_{16}$&  $G_{14}$&  $G_{15}$&  $G_{16}$  \\
\cmidrule(lr){2-8}
&ToMax(1)&  0.534&  0.394&  0.610&  0.848&  0.800&  0.932  \\
%$L=2$&  0.600&  0.440&  0.722&  0.910&  0.872&  0.980  \\
&ToMax(5)&  0.672&  0.504&  0.804&  0.946&  0.904&  0.994  \\
&ToMax(8)&  0.694&  0.530&  0.818&  0.954&  0.924&  1.000  \\
&ToMax(15)&  0.732&  0.520&  0.826&  0.958&  0.916&  0.998  \\
&ToMax(20)&  0.734&  0.522&  0.828&  0.960&  0.914&  0.998  \\
& UY-21 & 0.158&  0.224&  0.541&  0.531&  0.740&  0.980  \\
% BC & 0.842&  0.698&  0.958 & 0.984&  0.964&  1.000 \\
%BY&  0.800&  0.554&  0.892& 0.974&  0.920&  1.000  \\ \hline
%&\multicolumn{3}{c}{$n=100$}& \multicolumn{3}{c}{$n=200$}  \\ $\sigma^2=3$
\cmidrule(lr){2-5}\cmidrule(lr){6-8}
Power & Method&  $G_{24}$&  $G_{25}$&  $G_{26}$&  $G_{24}$&  $G_{25}$&  $G_{26}$  \\
\cmidrule(lr){2-8}
&ToMin(1)&  0.504&  0.528&  0.554&  0.782&  0.808&  0.778  \\
%$L=2$&  0.520&  0.660&  0.780&  0.852&  0.880&  0.902  \\
&ToMin(5)&  0.610&  0.750&  0.870&  0.928&  0.950&  0.980  \\
&ToMin(8)&  0.700&  0.810&  0.930&  0.928&  0.966&  0.990  \\
&ToMin(15)&  0.740&  0.850&  0.980&  0.954&  0.978&  1.000  \\
&ToMin(20)&  0.790&  0.890&  1.000&  0.964&  0.986&  1.000  \\
%BY & 0.470&  0.236&  0.000&  0.884&  0.778&  0.000  \\
\hline
\end{tabular}
\end{table}

\begin{table}[H]
	\centering\caption{Results for liquor sales data: (a) The results of the TOSI method  under different penalty parameters with $L=5$; (b) Estimates of coefficients for the {six} important variables, where {\it SL\underline{~}lag1}  and {\it SL\underline{~}lag12} are the sales in the past one month and 12 months, respectively, {\it gdp1\underline{~}lastyear} is the GDP from primary industry, mainly agriculture, in the last year, {\it pptg\underline{~}six\underline{~}m} is the brand promotion expense for the past six months, and  {\it kq\underline{~}twelve\underline{~}m} is the expense in giving products as gifts to customers in the past 12 months, and {\it Stkcd600779\underline{~}lag10} is the number of transactions of the stock 600779 in the past 10 months, where 600779 is the stock code of another liquor company.}\label{tab:winetest}
\renewcommand\tabcolsep{3.0pt} % 调整表格列间的宽度
	\begin{tabular}{l cc c c c c }
		\hline
&\multicolumn{5}{c}{\bf (a) The testing results}& \\ \hline
		$\lambda$&  $|G^o|$& $|G^{no}|$ & & p-value ($H_{0,G^{o}}$) &  &  p-value ($\tilde H_{0,G^{no}}$) \\ \hline
        $0.1866$& 240 & 9& ToMax&0.5832& ToMin& 0.3018 \\
       & & & ToMax$(L)$&0.5895& ToMin$(L)$& 0.7545 \\
       $0.2239$& 241 & 8& ToMax&0.2682& ToMin& 0.3699 \\
       & & & ToMax$(L)$&0.2009 & ToMin$(L)$& 0.3953 \\
       $0.3172$& 243 & 6& ToMax&0.1716& ToMin& 0.0431 \\
       & & & ToMax$(L)$&0.0684& ToMin$(L)$& 0.0137 \\
       $0.3545$& 244 & 5& ToMax&0.0253 & ToMin& 0.0035 \\
       & & & ToMax$(L)$&0.0001& ToMin$(L)$& 0.0002 \\
		\hline
&\multicolumn{5}{c}{\bf (b) The estimated coefficients of significant variables}& \\ \hline
var. name &  \multicolumn{2}{c}{{\it SL\underline{~}lag1}}&  \multicolumn{2}{c}{{\it SL\underline{~}lag12}}&  \multicolumn{2}{c}{{\it gdp1\underline{~}lastyear}}   \\

coef. est.& \multicolumn{2}{c}{0.1615}&  \multicolumn{2}{c}{0.0785}&  \multicolumn{2}{c}{-0.0574}    \\
var. name &  \multicolumn{2}{c}{{\it pptg\underline{~}six\underline{~}m}}&  \multicolumn{2}{c}{{\it kq\underline{~}twelve\underline{~}m}}&  \multicolumn{2}{c}{{\it Stkcd600779\underline{~}lag10}}    \\
coef. est.&  \multicolumn{2}{c}{0.0331}&  \multicolumn{2}{c}{ 0.1204}&  \multicolumn{2}{c}{-0.0236}     \\ \hline
	\end{tabular}
\end{table}

 \begin{table}[H]
	\centering\caption{Results for criminal data: (a) The results of the TOSI method  under different penalty parameters with $L=5$; (b) Estimates of coefficients for the {six} important variables, where {\it racePctWhite} is the percentage of the population that is Caucasian; {\it racePctHisp} is the percentage of population that is of hispanic heritage;
{\it NumUnderPov} is the number of people under the poverty level; {\it FemalePctDiv} is the percentage of females who are divorced;  {\it PctKids2Par} is the percentage of kids in family housing with two parents; and {\it PctPersOwnOccup} is the percentage of people in owner occupied households.}\label{tab:criminaltest}
\renewcommand\tabcolsep{3.0pt} % 调整表格列间的宽度
	\begin{tabular}{c cc c c c c }
		\hline
&\multicolumn{5}{c}{\bf (a) The testing results}& \\ \hline
		$\lambda$&  $|G^o|$& $|G^{no}|$ & & p-value ($H_{0,G^{o}}$) &  &  p-value ($\tilde H_{0,G^{no}}$) \\ \hline
        $0.1778$& 93 & 7& ToMax&0.9850& ToMin& 0.1432 \\
       & & & ToMax$(L)$&0.1589& ToMin$(L)$& 0.2864 \\
       $0.1828$& 94 & 6& ToMax&0.5806& ToMin& 0.0201 \\
       & & & ToMax$(L)$&0.1589& ToMin$(L)$& 0.0103 \\
       $0.1878$& 95 & 5& ToMax&0.0158& ToMin& 0.0201 \\
       & & & ToMax$(L)$&0.0468& ToMin$(L)$& 0.0095 \\
		\hline
&\multicolumn{5}{c}{\bf (b) The estimated coefficients of significant variables}& \\ \hline
var. name &  \multicolumn{2}{c}{{\it racePctWhite}}&  \multicolumn{2}{c}{{\it racePctHisp}}&  \multicolumn{2}{c}{{\it NumUnderPov}}   \\

coef. est.& \multicolumn{2}{c}{ -0.4248}&  \multicolumn{2}{c}{0.06946}&  \multicolumn{2}{c}{0.1366}    \\
var. name &  \multicolumn{2}{c}{{\it FemalePctDiv}}&  \multicolumn{2}{c}{{\it PctKids2Par}}&  \multicolumn{2}{c}{{\it PctPersOwnOccup}}    \\
coef. est.&  \multicolumn{2}{c}{1.5317}&  \multicolumn{2}{c}{-2.4265}&  \multicolumn{2}{c}{-0.009345}     \\ \hline
	\end{tabular}
\end{table}


\renewcommand{\thetable}{S\arabic{table}}
\renewcommand{\thefigure}{S\arabic{figure}}
\newcommand{\Appendix}{\def\thesection{Appendix~\Alph{section}}}
\baselineskip 18pt
\setcounter{section}{0}
\setcounter{table}{0}
\setcounter{figure}{0}
\Appendix

\section{Verify conditions (A1)--(A4)}

\subsection{Example 1. High-dimensional mean models}
 Recalling $G^o_{1} = \{j\in G: \btheta_j \neq 0\}, G^o_{0} = \{j\in G: \btheta_j = 0\}, G^{no}_{1} = \{j\in G^c: \btheta_j \neq 0\}$, $G^{no}_{0} = \{j\in G^c: \btheta_j = 0\}$,  and $\hat\btheta_j = \frac{1}{n} \sum_{i=1}^n x_{ij}$, we give the proofs by checking the Conditions (A1)--(A4) {in the main text.} %{\red $G_1= G^o_{1} \cup G^{no}_{1}$}

By Condition (B1), we know that $x_{ij}$ is the sub-exponential random variable, thus, it has {a} finite second moment. By {the} central limit theorem, we have $\sqrt{n}(\hat\btheta_j - \btheta_j) \stackrel{d}\rightarrow N(0, \sigma_j^2)$, where $\sigma_j^2$ can be estimated consistently by sample variance. Thus, Condition (A1) is satisfied.

Then, we verify the Conditions (A2) and (A3). First, by Conditions (B1), we have $\max_{j \in G^o_{0}} |\hat \btheta_j| = O_p(\sqrt{\frac{ln |G^o_{0}|}{n}})= O_p(\sqrt{\frac{ln (p)}{n}})$. Moreover, we know $\max_{j \in G^o_{1}} |\hat \btheta_j|\geq |\btheta_{j_0}|- |\hat\btheta_{j_0}-\btheta_{j_0}|\geq |\btheta_{j_0}|- \frac{c}{\sqrt{n}}\gg  O(\sqrt{\frac{ln (p)}{n}})$
%$max_{j \in G^o_{1}}(|\btheta_{j}| - |\hat \btheta_{j}- \btheta_{j}|) \geq max_{j \in G^o_{1}}(|\btheta_{j}| - max_{j \in G^o_{1}}|\hat \btheta_{j}- \btheta_{j}|) = max_{j \in G^o_{1}}|\btheta_{j}| -max_{j \in G^o_{1}}|\hat \btheta_{j}- \btheta_{j}| \gg max_{j \in G^o_{1}}|\btheta_{j}| - \frac{c\ln (s)}{\sqrt{n}} \gg  O(\sqrt{\frac{ln (p)}{n}})$
, where $j_0\in G^o_{1}$. Thus, we obtain $\max_{j \in G^o_{1}} |\hat \btheta_j| \gg \max_{j \in G^o_{0}} |\hat \btheta_j|$. {Therefore, Condition (A2) holds.}

Similarly, by Conditions  (B1) and (B2), we have $\min_{j  \in G^{no}_{1}}|\hat \btheta_j| \geq \min_{j  \in G^{no}_{1}}|\btheta_j| - \max_{j  \in G^{no}_{1}}$ $|\hat \btheta_j - \btheta_j| \gg O_p(\sqrt{\frac{ln s}{n}})$. As for $\min_{j \in G^{no}_{0}}|\hat \btheta_j|$, {we have, for $j_1 \in G^{no}_{0}$, $\min_{j \in G^{no}_{0}}|\hat \btheta_j| \leq |\hat \btheta_{j_1}| =  O_p(\sqrt{\frac{1}{n}}) \ll \min_{j  \in G^{no}_{1}}|\hat \btheta_j|$.  Thus, Condition  (A3) are satisfied.} And Condition (A4) is directly followed by $\inf_j var(z_{ij}^2) > c >0$.
Thus, we complete the verification of Example 1.

\subsection{Example 2. High-dimensional sparse linear regression models}
Before giving a formal verification, let's introduce a lemma  which will be used in the following.
\begin{Lem}\label{lem:reg}
  (Theorem 2.4 of \cite{degeer2014on}) {Under the conditions (C1)--(C4) in main text,} the nodewise lasso estimator $\wh\Theta$ and the de-biased estimator $\wh\bbeta$ satisfy
  $$\|\wh\Theta_{.j} - \Theta_{.j} \| = O_p(\frac{ \sqrt{s_j\ln p}}{\sqrt{n}})$$
  and
  $$\sqrt{n}(\wh\btheta - \btheta) = \wh\Theta \X\trans \bvarepsilon/ \sqrt{n} +o_p(1)$$
  by choosing the penalty parameter $\lambda_j$ and $\lambda$ such that $\lambda_j = O(\sqrt{\frac{\ln p}{n}})$  and $\lambda = O(\sqrt{\frac{\ln p}{n}})$, where $\Theta_{.j}$ is the $j$-th column of $\Theta$ and $\lambda_j$s are the penalty parameters used in nodewise lasso for estimating $\Theta_{.j}$s.
\end{Lem}

{\bf Proof. } By  Lemma \ref{lem:reg}, we only require to verify
\begin{equation}\label{eq:regmax}
 \max_{j\in G^o_{1}} |\hat\btheta_j| \gg \max_{j\in G^o_{0}}|\hat\btheta_j|
\end{equation}
and
\begin{equation}\label{eq:regmin}
 \min_{j\in G^{no}_{1}} |\hat\btheta_j| \gg \min_{j\in G^{no}_{0}}|\hat\btheta_j|.
\end{equation}
We denote $\xi_{ij}= \Theta_{.j}\trans \x_i \varepsilon_i$, $G$ is an arbitrary subset of $\{1, \cdots, p\}$, and $\u_i=(u_{i1}, \cdots, u_{ip})\trans$ is a sequence of independent normal vectors with mean zero and covariance matrix $\sigma^2\Theta$. Then by the proofs of Theorem 2.2 in \cite{zhang2017simultaneous}, we have
\begin{equation}\label{eq:bxi}
  \max_{j\in G} \sqrt{n}|\hat\btheta_j - \btheta_{j}| = \max_{j\in G}  |\sum_{i=1}^{n} \xi_{ij} /\sqrt{n}| + o_p(1).
\end{equation}
In addition, by Lemma 1.1 in \cite{zhang2017simultaneous}, we have
\begin{equation}\label{eq:xiz}
  \max_{j\in G}  |\sum_{i=1}^{n} \xi_{ij} /\sqrt{n}|  = \max_{j\in G}  |\sum_{i=1}^{n} u_{ij} /\sqrt{n}| + o_p(1).
\end{equation}
Using the tail probability inequality on i.i.d. normal sample, we obtain
\begin{equation}\label{eq:ztail}
  \max_{j\in G}  |\sum_{i=1}^{n} u_{ij} /\sqrt{n}| = O_p(\sqrt{\ln |G|}) = O_p(\sqrt{\ln (p)}).
\end{equation}
Combing \eqref{eq:bxi} -- \eqref{eq:ztail} and Condition (C3), we have
\begin{equation}\label{eq:maxbeta}
  \max_{j\in G}|\hat\btheta_j - \btheta_{j}| = O_p(\sqrt{\frac{\ln (p)}{n}}).
\end{equation}
Due to $G_{0}^o=\{j \in G^o: \btheta_{j} = 0\}$ and \eqref{eq:maxbeta}, we get
$$\max_{j\in G_{0}^o}|\hat\btheta_j| =  O_p(\sqrt{\frac{\ln (p)}{n}}),$$
which is dominated by $\max_{j\in G_{1}^o} |\hat\btheta_j|$ in order. Thus, we prove that equation \eqref{eq:regmax} holds.

In addition, by the triangular inequality, Condition (C5) and \eqref{eq:maxbeta}, we obtain
\begin{eqnarray} \label{eq:minb}
 \min_{j\in G_{1}^{no}} |\hat\btheta_j|&\geq & \min_{j\in G_{1}^{no}}|\btheta_{j}| - \max_{j\in G_{1}^{no}} |\hat\btheta_j - \btheta_{j}| \nonumber \\
  & \gg & \sqrt{\frac{\ln (p)}{n}}.
\end{eqnarray}
By Lemma \ref{lem:reg}, we have
\begin{equation}\label{eq:ming20}
  \min_{j\in G_{0}^{no}}|\hat\btheta_j| = O_p(\frac{1}{\sqrt{n}}).
\end{equation}
Coupling with \eqref{eq:minb} and \eqref{eq:ming20}, we complete the proof of \eqref{eq:regmin}.

%{\red
%{\noindent\bf Comparison of theoretical conditions with existing methods.}  Condition (C1) is the similar to Assumption 2.1 in \cite{zhang2017simultaneous} and (B1) in \cite{degeer2014on} to restrict the thin tail property of covariates. Condition (C2) is the same as Assumption 2.2 in \cite{zhang2017simultaneous}.  Condition (C3) is a standard sparsity assumption for the precision matrix of covariate vector and regression coefficients, which is also assumed in Theorem 2.4 in \cite{degeer2014on}. Condition (C4) restricts the good tail properties of error term, which  is also used in  the Assumption 2.3 (i) in \cite{zhang2017simultaneous}. Condition (C5) is a minimum signal strength assumption to  ensure the error rate in the variable selection step is ignorable and similar condition  is found in assumptions (A2) and (A3) in \cite{wasserman2009high} and assumption (A1) in \cite{meinshausen2009p} since they also adopted sample-splitting strategy for inference.
%}

\section{Proofs of Theorems 1--3}

{\noindent\bf Proof of Theorem 1}. The proofs include two parts, where the part one proves  the asymptotical $\chi^2(q)$ distribution of $\wh T_{{\max}}$ under $H_{0,G^o}$, and the part two proves the asymptotical unbiasedness of ToMax method.

{\bf Part 1. }
We show that  $\wh T_{{\max}}$ is asymptotically distributed as $\chi^2(q)$ under $H_{0,G^o}: \btheta_j = \bbo, \forall j \in G^o $.

Conditional on $\mathcal{D}_1$, $j_{\max}$ can be regarded as a constant, so $\hat \btheta^{(2)}_{j_{\max}}$ is asymptotically normal with mean zero and convariance $\Sigma_{j_{\max}}$ by Condition (A1) when $H_{0,G^o}$ is true. Since $\wh\Sigma_{2j}$ is a consistent estimator of $\Sigma_j$, $\wh T_{{\max}}$ is asymptotically distributed as $\chi^2(q)$ conditional on $\mathcal{D}_1$.
Furthermore,  for any $\epsilon > 0$ and $x \in R$, it holds that
\begin{equation}\label{eq:CDFcon}
F_{\chi^2(q)}(x) -\epsilon\leq P(\hat T_{\max} \leq x | \mathcal{D}_1) \leq F_{\chi^2(q)}(x) + \epsilon,
\end{equation}
where $F_{\chi^2(q)}(x)$ is the cumulative distribution function of $\chi^2(q)$. Taking expectation on $\mathcal{D}_1$ in \eqref{eq:CDFcon}, we get
$$F_{\chi^2(q)}(x) -\epsilon\leq P(\hat T_{\max} \leq x ) \leq F_{\chi^2(q)}(x) + \epsilon.$$
Then, letting $\epsilon \rightarrow 0$, we conclude
$$P(\hat T_{\max} \leq x )  \rightarrow F_{\chi^2(q)}(x),$$
which leads to the desired results.

{\bf Part 2.} We show that ToMax is unbiased.  Recall
$G^o_{1} = \{j\in G^o: \btheta_j \neq \bbo\}$, $G^o_{0} = \{j\in G^o: \btheta_j = \bbo\}$,  and $G_1= G^o_{1} \cup G^{no}_{1}$. If $H_{0,G^o}$ is false, then $G^o_{1}$ is nonempty.
Now, we have
\begin{eqnarray}
   \beta_{\hat T_{\max}}(\btheta_{G^o} )&\hat =&P(\hat T_{\max} > \chi^2_{1-\alpha}(q))  \nonumber \\
   &=& P(\hat T_{{\max}} > \chi^2_{1-\alpha}(q), j_{\max} \in G^o _{1}) + P(\hat T_{{\max}} > \chi^2_{1-\alpha}(q), j_{\max} \notin G^o _{1})  \nonumber \\
   &\geq & P(\hat T_{{\max}} > \chi^2_{1-\alpha}(q), j_{\max} \in G^o _{1})  \nonumber \\
   &=&P(\hat T_{{\max}} > \chi^2_{1-\alpha}(q)| j_{\max} \in G^o _{1})P(j_{\max} \in G^o _{1}). \label{eq:Powerfun}
\end{eqnarray}
Since $j_{\max}=\arg\max_{j\in G^o }\|\hat \Sigma_{2j}^{-1/2}\hat \btheta^{(2)}_{j}\|$, $$\max_{j\in G^o }\|\hat \Sigma_{2j}^{-1/2}\hat \btheta^{(2)}_{j}\|=\max\left\{\max_{j\in G^o _{1}}\|\hat \Sigma_{2j}^{-1/2}\hat \btheta^{(2)}_{j}\|, \max_{j\in G^o _{0}}\|\hat \Sigma_{2j}^{-1/2}\hat \btheta^{(2)}_{j}\| \right\}$$
and Condition (A2) holds, we have
\begin{equation}\label{eq:jmaxG}
P(j_{\max} \in G^o _{1}) \rightarrow 1.
\end{equation}

Under $H_{1,G^o }:\btheta_j \neq 0, \exists j\in G^o $, we denote $\hat L_1=\bar n \btheta^{\trans}_{j_{\max}} \wh\Sigma_{2j_{\max}}^{-1}(2 \hat \btheta^{(2)}_{j_{\max}}- \btheta_{j_{\max}} ), \tilde T_{\max} = \hat T_{\max}  - \hat L_1$.  Conditional on $\mathcal{D}_1$,  $j_{\max}$ can be regarded as a constant and $\tilde  T_{\max}$ is asymptotically $\chi^2(q)$ when $j_{\max} \in G^o _{1}$.  Hence,
\begin{eqnarray}
   && \hspace{-1cm} P(\hat T_{j_{\max}} > \chi^2_{1-\alpha}(q)| j_{\max} \in G^o _{1}, \mathcal{D}_1) \nonumber \\
   &=& P(\tilde  T_{\max} + \hat L_1 > \chi^2_{1-\alpha}(q)| j_{\max} \in G^o _{1}, \mathcal{D}_1 ). \label{eq:ThatL1}
\end{eqnarray}
Next, we consider to prove
\begin{equation}\label{eq:PowerL1}
P(\hat L_1> 0| j_{\max} \in G^o _{1},  \mathcal{D}_1) \rightarrow 1.
\end{equation}
By the definition of $\hat L_1$, {we known that there exists a sequence $a_{n,1}= o(1)$ ($\lim_{n\rightarrow \infty} a_{n,1} \rightarrow 0$) such that}
\begin{eqnarray}
   P(\hat L_1> 0| j_{\max} \in G^o _{1},  \mathcal{D}_1)& = &  P(\bar n \btheta^{\trans}_{j_{\max}}\Sigma_{j_{\max}}^{-1}\btheta^{\trans}_{j_{\max}} >0|j_{\max} \in G^o _{1} , \mathcal{D}_1 ) + a_{n,1}  \nonumber \\
   &> & P(\inf_{j\in G^o _{1}}\|\sqrt{\bar n}\Sigma_j^{-1/2}\btheta_j\|>0) + a_{n,1},  \nonumber
\end{eqnarray}
which indicates \eqref{eq:PowerL1} holds by Condition (A4).

Thus, by \eqref{eq:ThatL1}, {we known that there exist two sequences $a_{n,2}= o(1)$ and $a_{n,3}= o(1)$  such that}
\begin{eqnarray}
   && \hspace{-1cm} P(\hat T_{j_{\max}} > \chi^2_{1-\alpha}(q)| j_{\max} \in G^o _{1}, \mathcal{D}_1) \nonumber \\
   &> &P(\tilde T_{\max}> \chi^2_{1-\alpha}(q)| j_{\max} \in G^o _{1}, \mathcal{D}_1 ) + a_{n,2} \nonumber \\
   &=& \alpha + a_{n,3}, \nonumber
\end{eqnarray}
which implies
\begin{equation}\label{eq:Power2}
P(\hat T_{j_{\max}} > \chi^2_{1-\alpha}(q)| j_{\max} \in G^o _{11}) > \alpha + a_{n,3}.
\end{equation}
Therefore, combining \eqref{eq:Powerfun}, \eqref{eq:jmaxG} and \eqref{eq:Power2}, we conclude that
$$\beta_{\hat T_{\max}}(\btheta_{G^o} ) \geq \alpha,$$
when $n$ is sufficiently large.

In the following, we show that the power converges to 1 if  $\inf_{j\in G^o _{1}}\|\sqrt{n}\Sigma_j^{-1/2}\btheta_j\| \rightarrow \infty$.

By \eqref{eq:ThatL1} and following the proof of \eqref{eq:PowerL1}, we have, {there exists a sequence $a_{n,4}=o(1)$, such that}
\begin{eqnarray}
   && \hspace{-1cm} P(\hat T_{j_{\max}} > \chi^2_{1-\alpha}(q)| j_{\max} \in G^o _{1}, \mathcal{D}_1) \nonumber \\
   &\geq & P(\hat L_1 > \chi^2_{1-\alpha}(q)| j_{\max} \in G^o _{1}, \mathcal{D}_1 ) \nonumber \\
   & \geq & P((\inf_{j\in G^o _{1}}\|\sqrt{\bar n}\Sigma_j^{-1/2}\btheta_j\|)^2>\chi^2_{1-\alpha}(q)|j_{\max} \in G^o _{11},  \mathcal{D}_1 ) + a_{n,4}. \nonumber
\end{eqnarray}
By the condition that $\inf_{j\in G^o _{1}}\|\sqrt{n}\Sigma_j^{-1/2}\btheta_j\| \rightarrow \infty$, we obtain $ P(\hat T_{j_{\max}} > \chi^2_{1-\alpha}(q)| j_{\max} \in G^o _{1}, \mathcal{D}_1) \rightarrow 1$, which implies
\begin{equation}\label{eq:hTunCon}
P(\hat T_{j_{\max}} > \chi^2_{1-\alpha}(q)| j_{\max} \in G^o _{1}) \rightarrow 1.
\end{equation}
Coupling with \eqref{eq:Powerfun}, \eqref{eq:jmaxG} and \eqref{eq:hTunCon}, the desired results are proved.

{\noindent\bf Proof of Theorem 2}.

{\bf Part 1.} We show that  $\wh R_{j_{\min}}$ is asymptotically distributed as $\chi^2(q)$ under $\tilde H_{0,G^{no} }: \btheta_j = \bbo, \exists j \in G^{no}$.

Conditional on $\mathcal{D}_1$, $j_{\min}$ is a constant. By condition (A1) and the consistency of $\wh\Sigma_{j}$, we obtain $\hat R_{j_{\min}}\stackrel{d} \rightarrow \chi^2(q)$ by continuous mapping theorem if $\btheta_{j_{\min}}=\bbo$ conditional on $\mathcal{D}_1$. By Condition (A3) and following the proofs of Theorem 1, we have $P\{j_{\min} \in G^{no} _{0}\} \rightarrow 1.$
Thus, {there exists a sequence $b_{n,1}=o(1)$ such that }
\begin{eqnarray}
  P(\hat R_{j_{\min}} \leq x)&=&P(\hat R_{j_{\min}} \leq x, j_{\min} \in G^{no} _{0}) + P(\hat R_{j_{\min}} \leq x, j_{\min} \notin G^{no} _{0})  \nonumber \\
   &=& P(\hat R_{j_{\min}} \leq x, j_{\min} \in G^{no} _{0}) + b_{n,1}. \label{eq:hRjmin}
\end{eqnarray}
Conditional on $\mathcal{D}_1$, $\hat R_{j_{\min}}$ is asymptotically distributed as $\chi^2(q)$ if $j_{\min} \in G^{no} _{0}$, so
$$P(\hat R_{j_{\min}} \leq x, j_{\min} \in G^{no} _{0}|\mathcal{D}_1) \rightarrow F_{\chi^2(q)}(x),$$
{where $F_{\chi^2(q)}(x)$ is the cumulative distribution function of $\chi^2(q)$.}
Taking expectation on $\mathcal{D}_1$, we obtain
\begin{equation}\label{eq:RjFX2}
P(\hat R_{j_{\min}} \leq x, j_{\min} \in G^{no} _{0}) \rightarrow F_{\chi^2(q)}(x).
\end{equation}
Combing \eqref{eq:hRjmin} and \eqref{eq:RjFX2}, we conclude
$\hat R_{j_{\min}}\stackrel{d} \rightarrow \chi^2(q)$.

{\bf Part 2}. We show that the test is unbiased.

Under $ \tilde H_{1,G^{no} }: \btheta_j \neq \bbo, \forall j \in G^{no}$, we have
\begin{eqnarray}
   \beta_{\hat R_{\min}}(\btheta_{G^{no} })&\hat =&P(\hat  R_{\min} > \chi^2_{1-\alpha}(q))  \nonumber \\
   &=& P(\hat R_{\min} + \hat S_1 > \chi^2_{1-\alpha}(q)), \label{eq:RPower1}
\end{eqnarray}
where $\hat S_1=\bar n \btheta^{\trans}_{j_{\min}} \wh\Sigma_{2j_{\min}}^{-1}(2 \hat \btheta^{(2)}_{j_{\min}}- \btheta_{j_{\min}} )$ and $\hat R_{\min} = \hat R_{\min}  - \hat S_1$ is asymptotically distributed as $\chi^2(q)$ conditional on $\mathcal{D}_1$ by Condition (A1).

Next, we consider to prove
\begin{equation}\label{eq:PowerS1}
P(\hat S_1> 0| \mathcal{D}_1) \rightarrow 1.
\end{equation}
{By the definition of $\hat S_1$, we know that there exist two sequences $b_{n,2}=o(1)$ and $b_{n,3}=o(1)$ such that}
\begin{eqnarray}
   P(\hat S_1> 0| \mathcal{D}_1)& = &  P(\bar n \btheta^{\trans}_{j_{\min}}\Sigma_{j_{\min}}^{-1}\btheta^{\trans}_{j_{\min}} >0| \mathcal{D}_1 ) + b_{n,2}  \nonumber \\
   &> &P(\inf\limits_{j\in G^o _{1}}\|\sqrt{\bar n} \Sigma_j^{-1/2}\btheta_j\|>0| \mathcal{D}_1 ) + b_{n,3},
\end{eqnarray}
which indicates \eqref{eq:PowerS1} holds by Condition (A4).

Thus, {there exists a sequence $b_{n,4}=o(1)$ such that}
\begin{eqnarray}
   &&P(\hat R_{\min} + \hat S_1 > \chi^2_{1-\alpha}(q) | \mathcal{D}_1)  \nonumber \\
   & > & P(\hat R_{\min}  > \chi^2_{1-\alpha}(q) | \mathcal{D}_1) \nonumber \\
   &\geq & \alpha + b_{n,4}, \nonumber
\end{eqnarray}
which implies
\begin{equation}\label{eq:RPowerfun}
P(\hat R_{\min} + \hat S_1 > \chi^2_{1-\alpha}(q) ) > \alpha + b_{n,5},
\end{equation}
where $b_{n,5}=o(1)$. Coupling with \eqref{eq:RPower1} and \eqref{eq:RPowerfun}, the desired results are obtained.

Finally, we show that the power converges to 1 if  $\inf_{j\in G^{no} }\|\sqrt{n}\Sigma_j^{-1/2}\btheta_j\| \rightarrow \infty$.

 If $\inf_{j\in G^{no} }\|\sqrt{n}\Sigma_j^{-1/2}\btheta_j\| \rightarrow \infty$, we have $\hat R_{\min}$ is {asymptotically distributed as} $\chi^2(q)$ conditional on $\mathcal{D}_1$ by Condition (A1).

{Note that there exist two sequences $b_{n,6}=o(1)$ and $b_{n,7}=o(1)$ such that}
\begin{eqnarray*}
   &&P(\hat R_{\min} + \hat S_1 > \chi^2_{1-\alpha}(q) | \mathcal{D}_1) \\
   & \geq & P( \hat S_1 > \chi^2_{1-\alpha}(q)| \mathcal{D}_1) + b_{n,6} \\
   & \geq & P((\inf_{j\in G^{no} }\|\sqrt{\bar n}\Sigma_j^{-1/2} \btheta_j\|)^2 > \chi^2_{1-\alpha}(q)) + b_{n,7}.
\end{eqnarray*}
Since $\inf_{j\in G^{no} }\|\sqrt{n}\Sigma_j^{-1/2}\btheta_j\| \rightarrow \infty$, we have
\begin{equation}\label{eq:Rjmin1}
P(\hat R_{\min} + \hat S_1 > \chi^2_{1-\alpha}(q) | \mathcal{D}_1) \rightarrow 1.
\end{equation}
Taking expectation on $\mathcal{D}_1$ for \eqref{eq:Rjmin1}, we get
 \begin{equation}\label{eq:Rjmin11}
P(\hat R_{\min} + \hat S_1 > \chi^2_{1-\alpha}(q)) \rightarrow 1.
\end{equation}
Combing \eqref{eq:RPower1} and \eqref{eq:Rjmin11}, we complete the proofs.

%{\noindent\bf Proof of Theorem \ref{th:multitest}}.  By Theorem \ref{th:maxtest}, it asymptotically holds that
%\begin{equation}\label{eq:pmax}
%P(\wh p_{l,\max} \leq u) = u \mbox{ under $H_{0,G^o }$}.
%\end{equation}
%Thus, it asymptotically holds that
%\begin{eqnarray*}
%% \nonumber to remove numbering (before each equation)
%  P(\mbox{reject }H_{0,G^{o}}|H_{0,G^{o}}) &=& P(k_{\max}/L \geq r| H_{0,G^o } \mbox{ is true}) \\
%   &=&  P(k_{\max} \geq Lr| H_{0,G^o } \mbox{ is true})  \\
%   &=&  P(\sum_l 1_{\{\wh p_{l,\max} \leq \gamma\}} \geq k) \geq Lr| H_{0,G^o } \mbox{ is true})  \\
%   &\leq & \frac{E\{\sum_{l=1}^{L}1_{\{\wh p_{l,\max} \leq \gamma\}} \}}{Lr} \\
%   &\leq & \frac{\gamma}{r},
%\end{eqnarray*}
%where the first equality follows from the procedure of ToMax$(L, r, \gamma)$, the first inequality follows from the Markov inequality and the last inequality follows from \eqref{eq:pmax}. The proof of Theorem \ref{th:multitest} is completed.

{\noindent\bf Proof of Theorem 3}.
 By Theorem 1, it asymptotically holds that
\begin{equation}\label{eq:pmax}
P(\wh p_{l,\max} \leq u) = u \mbox{ under $H_{0,G^o }$}.
\end{equation}
{Let $\{\wh p_{(1),\max}, \wh p_{(2),\max}, \cdots, \wh p_{(L),\max}\}$ be the sorted sequence of $\{\wh p_{l,\max}, l\leq L\}$ in the increasing order.
By the definition of  Bonferroni-Holm procedure, rejecting at least one time in L times, i.e. $k_{\max} \geq 1$, % or $\min_l \wh p^{adj}_{l,\max} < \alpha$,
is equivalent to that $\wh p_{(1),\max} < \frac{\alpha}{L}$. Otherwise, there is no rejections.
Thus, it asymptotically holds that
\begin{eqnarray*}
  P(\mbox{reject }H_{0,G^{o}}|H_{0,G^{o}} \mbox{ is true}) &=& P(k_{\max} \geq 1| H_{0,G^o } \mbox{ is true}) \\
   &=&  P(\wh p_{(1),\max} < \frac{\alpha}{L}| H_{0,G^o } \mbox{ is true})  \\
   &\leq&  P(\exists \wh p_{l,\max} < \frac{\alpha}{L}, l\leq L| H_{0,G^o } \mbox{ is true})  \\
   &\leq &  \sum_{l=1}^L P(\wh p_{l,\max} < \frac{\alpha}{L} | H_{0,G^o } \mbox{ is true})  \\
   &\leq & \alpha,
\end{eqnarray*}
where  the first inequality follows from the fact $\{\wh p_{(1),\max} < \frac{\alpha}{L}\} \subseteq\{\exists \wh p_{l,\max} < \frac{\alpha}{L}, l\leq L\}$, the second inequality is from Bonferroni inequality and the third inequality follows from \eqref{eq:pmax}. The proof of Theorem 3 is completed.}

\section{Related materials of factor models}\label{proof:fac}
In this part, we present the proofs and other materials of the sparse latent factor model. Specifically, the NITS estimator is introduced in \ref{estalgo} and the parameters selection method of NITS estimator is given in \ref{sec:tuning}.
Then we give a proposition on the identifiability of model and a theorem on the Oracle properties of NITS estimator in \ref{app:pro1} and \ref{app:th1}, respectively. Next, we present the  proof of Theorem 4 in \ref{app:infer}. Finally, we show some simulation results on the performance of NITS estimator in \ref{sec:pvs}.
\subsection{Non-iterative two-step (NITS) estimation}\label{estalgo}

For ease of reading, here we may repeat some expression  in Section 4 of the main text. Considering the model (5) { with sparsity in} $\b_j$ in the main text,  we can estimate $\H$ and $\B$ by minimizing  the following least square error with an adaptive group-lasso penalty term on rows  and an adaptive lasso penalty on entries of $\B$,
\begin{eqnarray} \label{eq:spc}
&&\parallel \X-\H\B^{\trans}\parallel_F^2 + \lambda_1 \sum_{j=1}^{p} w_{1j}\|\b_j\| + \lambda_2 \sum_{j=1}^{p}\sum_{k=1}^{q} w_{2,jk}|b_{jk}|.
\end{eqnarray}

Since both $\H$ and $\B$ are unknown and   double penalties are presented,  estimating $\H$ and $\B$  by direct minimizing  (\ref{eq:spc}) is difficult. A possible method is  iteratively  estimating  $\H$ and $\B$. Since $\H$ and $\B$ are  matrix with $n\times q$ and $p\times q$, respectively, and both $n$ and $p$ are large, the iterative algorithm still requires  intensive computation. We solve the computational problem by using traditional
PCA without penalty on $\B$, which is the solution of the following objective function,
\begin{eqnarray} \label{eq:step1}
&&(\widetilde\H, \widetilde\B)=\mbox{argmin}_{\B,\H}\parallel \X-\H\B^{\trans}\parallel_F^2.
\end{eqnarray}
The solution satisfies
$\widetilde\h_i - \h_{i} = O_p(p^{-1/2} + n^{-1})$ \citep{bai2013principal,asz010}.
Since $p\gg n$, hence  $\widetilde\h_i - \h_{i} = O_p(n^{-1})$, which implies the estimator $\widetilde\H$ from the standard PCA is good enough to estimate $\b_j$ whose optimal rate is $O_p(n^{-1/2})$. {In other words},  we do not need to update $\B$
by replacing $\H$ with its improvements. Hence, the iterative computation between $\B$ and $\H$  is not necessary. The computation of $\widetilde\H$ and $\widetilde\B$ is simple and can be derived in closed forms. Particularly,
by \cite{bai2002determining}, $n^{-1/2}\widetilde\H$ is  the first $q$ eigenvectors of $n^{-1/2}p^{-1} \X\X\trans$ and $\widetilde\B= n^{-1} \X\trans \widetilde\H$. Obviously,  $n^{-1}\widetilde\H\trans \widetilde\H = \I_q$ and $\widetilde\B\trans\widetilde\B$ is diagonal matrix with decreasing diagonal entries, so the identification condition (E1) is satisfied. To adhere to (E2), we multiply  $1$ or $-1$ to each column of $\widetilde\B$ and $\widetilde\H$ so that the first nonzero element of each column of $\widetilde\B$ is  positive.

With fixing $\H$ at $\widetilde\H$ from (\ref{eq:step1}),
we  estimate  $\B$ by  minimizing a penalized least square error  with two adaptive penalty terms on $\B$,
\begin{eqnarray} \label{eq:step2}
&&\wh\B=\mbox{argmin}_{\B}\parallel \X- \widetilde\H\B^{\trans}\parallel_F^2 + \lambda_1 \sum_{j=1}^{p} w_{1j}\|\b_j\| + \lambda_2 \sum_{j=1}^{p}\sum_{k=1}^{q} w_{2,jk}|b_{jk}|,
\end{eqnarray}
where the adaptive weights can be taken as, e.g.,$w_{1j}= 1/\|\widetilde\b_j\|$ and $w_{2,jk}=1/|\tilde b_{jk}|$ from  (\ref{eq:step1}). The penalty parameters $\lambda_1$ and $\lambda_2$ are selected by cross validation described in \ref{sec:tuning}. With some calculations  in Appendix \ref{app:closeform}, $\wh\B$ has the following closed form
\begin{eqnarray} \label{eq:step21}
\wh\B = \left(\frac{\bzeta_1}{n}\left( 1 - \frac{\lambda_1 w_{11}}{2\|\bzeta_1\|} \right)_{+}, \cdots, \frac{\bzeta_p}{n}\left( 1 - \frac{\lambda_1 w_{1p}}{2\|\bzeta_p\|} \right)_{+} \right)\trans,
\end{eqnarray}
where $\bzeta_j = sign(\widetilde\H^{\trans}\x_{.j}) \times (|\widetilde\H^{\trans}\x_{.j}| -\frac{\lambda_2}{2} \w_{2j})_{+}$, $\x_{.j}$ is the $j$th column of $\X$, $\w_{2j}=(w_{2,j1}, w_{2,j2},$ $\cdots, w_{2,jq})\trans$, and $sign(\cdot), \times$, $|\cdot|$ and $(a)_{+}=max(0, a)$ represent entry-wise operation for a vector.
Obviously, if $\lambda_1 = \lambda_2 = 0$,  $\wh\B$ in (\ref{eq:step21}) degenerates to  the conventional PCA solution $\widetilde\B$.

With the  closed forms of the estimators $\widetilde\H$ and $\wh\B$,  the computational cost  is very low and the implementation is very simple.

\subsubsection{Selection of tuning parameters} \label{sec:tuning}
To estimate $\B$ and $\H$, we need to select  three tuning parameters including the dimension of latent factors $q$, and  penalty parameters $\lambda_1$ and $\lambda_2$. We use  eigenvalue ratio test  \citep{ly12, ah13} to select $q$. It has been shown that the eigenvalue ratio test
can be used to identify a consistent estimator for  the number of factors $q$ \citep{ly12} { with empirically good performance} %and empirically performs well
\citep{ma2015a,fan2020factor-adjusted}, and  is  computationally easy.

We  choose $\lambda_1$ and $\lambda_2$ by K-fold cross validation (CV) based on data $\Omega=\{(\X_{i}, \widetilde\h_i):  i=1,\cdots,n\}$. Particularly,
denote training and testing sets by $\Omega-\Omega_k$ and $\Omega_k$, respectively, for $k= 1,\cdots, K$. For each $\lambda_1$ from some grids of $(0,C_1]$ and $\lambda_2$ from some grids of $[0, C_2]$, where $C_1$ and $C_2$ are two constants,
we obtain the estimator $\hat{\B}^k(\lambda_1, \lambda_2)=(\hat{\b}_1^k(\lambda_1, \lambda_2),\cdots, \hat{\b}_p^k(\lambda_1, \lambda_2))^{\trans}$  of $\B$ using the training set $\Omega - \Omega_k$, and form the cross validation criterion by
$CV(\lambda_1, \lambda_2)= K^{-1}\sum_{k=1}^{K}[ p^{-1}\sumj \{|\Omega_k|^{-1}\sum_{i\in \Omega_k} (x_{ij}- \widetilde\h_i \trans \wh\b_j^{k}(\lambda_1,\lambda_2))^2\} ]$, where $|\Omega_k|$ is the cardinality.
We then find  $(\hat\lambda_1,\hat\lambda_2)$ that minimizes the criterion $CV(\lambda_1,\lambda_2)$. In the simulations and real data analysis, we {choose} $K = 5$.  The simulation studies and the real data analysis show that  { the cross-validation works well for choosing tuning parameters.} %the chosen tuning parameters work well.

\subsubsection{Derivation of the closed form of $\wh\B$}\label{app:closeform}
Noting the separability of $\b_j$'s and rewriting the objective function in \eqref{eq:step2},
$$c(\B) = \sumj\left\{\sumi (x_{ij}- \b_j^{\trans}\widetilde\h_i)^2+\lambda_1w_{1j}\|\b_j\| + \lambda_2 \sum_{k=1}^{q} w_{2,jk}|b_{jk}| \right\}.$$
Differentiating $c(\B)$ with respect to $\b_j$ and setting the derivatives to be zero, we obtain the following  equation,
$$n\b_j -  \widetilde\H^{\trans}\x_{.j} +\frac{\lambda_1 w_{1j} \b_j}{2\|\b_j\|} + \frac{\lambda_2}{2} \w_{2j}\times sign(\b_j)= \bbo,$$
where $\x_{.j}$ is the $j$th column of $\X$, $\w_{2j}=(w_{2,j1}, w_{2,j2}, \cdots, w_{2,jq})\trans$, and $\times$ represents entry-wise multiplication. Denote $a_j = n + \frac{\lambda_1 w_{1j}}{2\|\b_j\|}>0$,  we have
$$\b_j = \frac{\widetilde\H^{\trans}\x_{.j}}{a_j} - \frac{\lambda_2}{2a_j} \w_{2j}\times sign(\b_j).$$
Thus, the sign of each component of $\b_j$ must be same as the each component of $\widetilde\H^{\trans}\x_{.j}$, e.g., $sign(\b_j) = sign(\widetilde\H^{\trans}\x_{.j})$. Then we have
\begin{eqnarray}
\label{eq:sajbj}
a_j\b_j = sign(\widetilde\H^{\trans}\x_{.j}) \times (|\widetilde\H^{\trans}\x_{.j}| -\frac{\lambda_2}{2} \w_{2j})_{+},
\end{eqnarray}
where $|\widetilde\H^{\trans}\x_{.j}|$ is the entry-wise absolute value of $\widetilde\H^{\trans}\x_{.j}$ and $(a)_{+}= max\{0, a\}$. Denote $\bzeta_j = sign(\widetilde\H^{\trans}\x_{.j}) \times (|\widetilde\H^{\trans}\x_{.j}| -\frac{\lambda_2}{2} \w_{2j})_{+}$. Noting that $a_j = n + \frac{\lambda_1 w_{1j}}{2\|\b_j\|}$
and taking $l_2$ norm on both sides of \eqref{eq:sajbj}, we obtain the closed form of $\|\b_j\|$,
\begin{equation}\label{eq:l2bj}
\|\b_j\|=\left\{
\begin{aligned}
\frac{\|\bzeta_j\|}{n}- \frac{\lambda_1 w_{1j}}{2n}, & \mbox{ if } 2\|\bzeta_j\| \geq \lambda_1 w_{1j}, \\
0,& \mbox{ otherwise.}
\end{aligned}
\right.
\end{equation}
Substituting \eqref{eq:l2bj} into  \eqref{eq:sajbj}, we obtain
$$\b_j = \frac{\bzeta_j}{n}\left( 1 - \frac{\lambda_1 w_{1j}}{2\|\bzeta_j\|} \right)_{+}, j=1,\cdots,p.$$
Then we obtain the estimation for $\B$, which has the following closed form
$$\wh\B = \left(\frac{\bzeta_1}{n}\left( 1 - \frac{\lambda_1 w_{11}}{2\|\bzeta_1\|} \right)_{+}, \cdots, \frac{\bzeta_p}{n}\left( 1 - \frac{\lambda_1 w_{1p}}{2\|\bzeta_p\|} \right)_{+} \right)\trans.$$

\subsubsection{Proposition \ref{pro:01} and its proofs} \label{app:pro1}
{Let $\lambda_{min}(\M)$ and $\lambda_{max}(\M)$ be the  minimum and
maximum eigenvalues of a symmetric matrix $\M$, respectively,
and let $\Sigma_{\Lambda}=\lim\limits_{p \rightarrow
  \infty}p^{-1}\B^{\trans}\B$.
   Then, we  give the regularity Conditions (D1)-(D6) used for establishing the theoretical properties of the sparse latent factor model.

{\it \underline{{\bf (D1)}} there exists a constant $M$
    such that $E(||\h_{i}||_2^4) \leq M$, $cov(\h_{i})=\I_q$, and
    $\max_j\sigma_j^2\leq \tilde\sigma^2< \infty$;\
 \underline{{\bf (D2)}} $\sup_j||\b_{j}||_2\leq M$ and $\B$ satisfy  identifiability Condition (E2); \
\underline{{\bf (D3)}} $p^{-1/2}\sum\limits_{j=1}^{p}\b_{j}u_{ij} = O_p(1)$;\
\underline{{\bf (D4)}} there are two positive {constants} $c_1, c_2$  such that
    $c_1 < \lambda_{min}(\Sigma_{\Lambda})<\lambda_{max}(\Sigma_{\Lambda})<c_2$.\
 \underline{{\bf (D5)}} $n^{1/2} p^{-1} \rightarrow 0$ with $|J|=O(p)$. \
 \underline{{\bf (D6.1)}}   $\frac{\ln (p)}{n} \rightarrow 0$; \underline{{\bf (D6.2)}} there exist $r_1, r_2>0$ and $s_1,s_2>0$, such that for any $t>0,  k\leq q, j\leq p$,
     $ P(|h_{ik}|>t)\leq \exp(-(t/s_1)^{r_1})$ and $ P(|u_{ij}|>t)\leq \exp(-(t/s_2)^{r_2})$, where $h_{ik}$ is the $k$th element of $\h_{i}$; \underline{{\bf (D6.3)}}  $\min_{j\in J} \|\b_{j}\|\gg \sqrt{\frac{ln (p)}{n}}$.
}

Conditions (D1)--(D4) ensure model identifiability and are  similar to those  in \cite{bai2003inferential, bai2013principal} and \cite{asz010}. Specifically,  Condition (D1) gives the moment conditions of $\h_{i}$ and the upper bound of {the error variances}, where $cov(\h_{i})=\I_q$ is the population version of identifiability condition (E1), which indicates the lack of correlation of each component and determines the scale  of $\h_{i}$. Condition (D2) relates to the uniform upper bound of loading vectors. Condition (D3) ensures  the identifiability of $\H$ as $p$ goes to infinity. Condition (D4) guarantees that the factors signal and noise are {distinguishable}.
Condition (D5) requires $p$ to be sufficiently large so that the uncertainty from  estimating $\h_{i}$ based on $p$ variables  can be ignored, which ensures $\widetilde\b_j$ satisfies the asymptotical normality.
Condition (D6) is required to guarantee the validity of TOSI. Specifically, Condition (D6.1) specifies the relationship of $p$ and $n$.  Condition (D6.2) assumes a exponential tail of $\h_{i}$ and $\u_i$, which is a technical condition  to establish the uniform convergent rate of the loading estimator and can be found in \cite{bai2013statistical}. Condition (D6.3) requires the lower bound of the signals in the loadings to be separable from zero.}

Let $\| \M\| _1$ be the
1-norm of an arbitrary matrix $\M$, i.e. the maximum of the absolute column sums. Let
$\| \M\| _2$ be the 2-norm of an arbitrary matrix $\M$,
i.e. the maximum singular value of $\M$. Denote $\Sigma_\X = \var(\x_i)$ and  $\Sigma_\u=\var(\u_i)$. Then, we present Proposition \ref{pro:01} and its proofs.

\begin{Pro}\label{pro:01}
If Conditions (D1)--(D4)
hold,
then  $\B$ and $\H$ are unique when $p\rightarrow \infty$.
\end{Pro}

{\bf Proof.}  Based on the model (5) in Section 4 of the main text, we have
  $$p^{-1}\Sigma_\X=p^{-1}\B\B^{\trans} + p^{-1}\Sigma_\u.$$
  Note $\| \Sigma_\u\| _1\leq M$ by Condition (D2), so we obtain $p^{-1}\|
  \Sigma_\X-\B\B^{\trans}\| _1=p^{-1}\| \Sigma_\u\|
  _1
  \rightarrow 0$ when $p \rightarrow \infty$. Now let $\W\R^2\W\trans$
  be the singular value decomposition of $\Sigma_\X$,
  where $\W=(\w_1,\ldots, \w_p)$  and the first nonzero element of
  $\w_l$ is positive for $l=1, \ldots, p$, and $\R^2 =
  \mbox{diag}(r_1^2, \ldots, r_p^2)$ with $r_1^2\geq r_2^2\geq \ldots \geq r_p^2\geq 0$.
  We further define $\W_q=(\w_1, \ldots, \w_q) \mbox{ and
  }\R_q^2=\mbox{diag}(r_1^2, \ldots, r_q^2)$. Next, let the singular
  value
  decomposition of $\B$ be $\B=\A\Omega\V\trans$, where $\Omega$ is a
  $q\times q$ diagonal matrix with positive entries on
  the diagonal ordered in decreasing order, and $\A$ is a $q\times q$
  orthogonal matrix with $\w_l\trans\a_l\geq 0$, $l=1, \dots, q$, and
  $\V$ is a $q\times q$ orthogonal matrix. Then
  $\B\B^{\trans}=\V\Omega^2\V\trans$. According to Conditions (D1)--(D4)
  and following the same line in \cite{asz010}, we can
  show
  \begin{eqnarray}\label{MaConclu}
    &&\R_q^2, \W_q, \Omega, \A \mbox{ can be identified and } \\
    && \|
    \A-\W_q\| _2\rightarrow 0,p^{-1/2}\|
    \Omega-\R_q\| _2
  \rightarrow 0 \mbox{ when }p\rightarrow \infty. \label{eq:Aord}
  \end{eqnarray}

  Now, we show $\B$ can be identified when $p\rightarrow \infty$.
  By \eqref{eq:Aord} and  $\B=\A\Omega\V\trans$, we have
  \begin{eqnarray}\label{BWorder}
    p^{-1/2}\| \B-\W_q\R_q\V\trans\| _2 &=&
                                                          p^{-1/2}\| \A\Omega\V\trans-\W_q\R_q\V\trans\| _2 \nonumber\\
     &\leq& \| \A p^{-1/2}(\Omega-\R_q)\V\trans\| _2
            + \| (\A-\W_q)p^{-1/2}\R_q\V\trans\| _2
            \nonumber\\
     &\leq& \| p^{-1/2}(\Omega-\R_q)\| _2 + \|
            (\A-\W_q)\| _2 \| p^{-1/2}\R_q\| _2
            \nonumber\\
     &\rightarrow& 0.
  \end{eqnarray}
  Note that the first nonzero element in each column of $\W_q$ is
  positive, $\| \A-\W_q\| _2\rightarrow 0$
  implies that the first element in each column of $\A$ that has nonzero limit is also positive when $p$
  is sufficiently large. Hence by Condition (D1), we conclude
  that $\V$ is an identity matrix. This couples with
  (\ref{MaConclu}), so $\B$ can be
  identified.

  Now, we show $\H$ can be identified when $p\rightarrow \infty$. For any $i$ and a fixed $\B$, we have $\x_i = \B\h_{i} + \u_i$. Multiplying both sides by $p^{-1}\B\trans$ and letting $p\rightarrow \infty$, we have $p^{-1}\B\trans \x_i = p^{-1}\B\trans\B\h_{i} + p^{-1}\B\u_i = p^{-1}\B\trans\B\h_{i} + O_p(p^{-1/2})$ by Condition (D3). Therefore, $\|p^{-1}\B\trans \x_i - \Sigma_\Lambda\h_{i}\|=o_p(1)$. Note $\Sigma_\Lambda\h_{i}=O_p(1)$ by Conditions (D2) and (D4). Now since $\x_i$ and $\B$ are given, $p^{-1}\B\trans \x_i$ goes to a fixed value when $p\rightarrow \infty$. Thus, we have $\h_{i}= \Sigma_\Lambda^{-1}p^{-1}\B\trans \x_i$, hence is identifiable. $\hfill{} \Box$

\subsubsection{Theorem 6 and its Proofs}\label{app:th1}
Recall $J=\{j: \b_{j}\neq \bbo\}$, and denote $I_{j} = \{k: b_{jk} \neq 0\}$, $I=\{(j,k): b_{jk}\neq 0 \}$, $m_j = |{I}_{j}|, \hat{{I}_{j}}=\{k: \hat
b_{jk}\neq 0\}, \hat {{J}}= \{j: \hat \b_j \neq \bbo\}$ and $\hat {{I}}= \{(j,k): \hat  b_{jk} \neq 0\}$.  Then we present the oracle properties of the NITS estimators with an additional condition (D7) which is also used by  \cite{zou2006adaptive}.
\begin{description}
\item \underline{{\bf (D7)}} $\lambda_1n^{-1/2}\rightarrow 0, \lambda_2n^{-1/2}\rightarrow 0, \lambda_1 \rightarrow \infty$ and $\lambda_2 \rightarrow \infty$.
\end{description}
{{\bf Theorem 6}. {\it
(Oracle property) Under Conditions (D1)-(D5) and (D7),  we have\\
(1) for each $j \in [p]$,  $P(\hat {I_{j}}= I_{j})\rightarrow 1, P(\hat {J}= J)\rightarrow 1$ and $P(\hat {I}= I)\rightarrow 1$;\\
(2) For each $j \in J$, $\sqrt{n}(\hat  \b_{{I}_{j}} - \b_{I_{j}}) \stackrel{d}\rightarrow   N(\bbo, \sigma_j^2 \I_{m_j})$; for each $(j,k) \in I, \sqrt{n}(\hat  b_{jk} - b_{jk}) \stackrel{d}\rightarrow  N(0, \sigma_j^2)$, where $\b_{I_{j}}=\{b_{jk}, k\in I_j\}$.
}}

First of all, we present  Lemma \ref{lem:sqnbj} and Lemma \ref{lem:ujH}, which are the basis of proving Theorem 6.

\begin{Lem} \label{lem:sqnbj}
   Under Conditions (D1)--(D4),  we have
$\widetilde\h_i - \h_{i} = O_p(p^{-1/2} + n^{-1}), i = 1,\cdots, n,$ and $\widetilde\b_j - \b_{j} = O_p(n^{-1/2}+p^{-1}), j=1, \cdots, p.
 $
  Further, if Condition  (D5)  holds, then
  $\sqrt{n}(\widetilde\b_j - \b_{j}) \stackrel{d}\rightarrow N(\bbo, \sigma_j^2 \I_q).$
\end{Lem}
The rates of $\widetilde\h_i$s and $\widetilde\b_j$s  ensure the oracle properties of the proposed estimators.

{\bf Proof.} First, by lemma 3 in \cite{asz010}, under Conditions (D1)--(D4) we have
\begin{eqnarray*}
  \widetilde\h_i - \h_{i} = (p^{-1}\B\trans \B)^{-1} p^{-1}\sumj \b_{j}u_{ij} + O_p(n^{-1} + p^{-1} + p^{-1/2} n^{-1/2}).
\end{eqnarray*}
Moreover, by the Condition (D4), we have $(p^{-1}\B\trans \B)^{-1}=O_p(1)$. By Condition (D3), we obtain $p^{-1}\sumj \b_{j}u_{ij}=O_p(p^{-1/2})$. Thus, we have
\begin{eqnarray}\label{eq:wh-h}
  \widetilde\h_i - \h_{i} =O_p(p^{-1/2}+ n^{-1}).
\end{eqnarray}
Note that (B.2) in  \cite{bai2013principal}, we have
$$\widetilde\b_j - \K^{-1}\b_{j}= \K\trans n^{-1} \sumi \h_{i} u_{ij} + O_p(n^{-1} + p^{-1}),$$
where $ \K = p^{-1}\B\trans \B n^{-1} \H \trans\widetilde\H (p^{-1} \tilde \B\trans\tilde\B)^{-1}$. By the equation (2) in  \cite{bai2013principal}, we further have
$\K = \I_q + O_p(n^{-1} + p^{-1})$. Thus,
\begin{eqnarray}\label{eq:tib-b1}
  \widetilde\b_j - \b_{j} &=& n^{-1} \sumi \h_{i} u_{ij} + O_p(n^{-1} + p^{-1})\\
   &=& O_p(n^{-1/2} + p^{-1}). \label{eq:tb-b2}
\end{eqnarray}
By Condition (D5) and \eqref{eq:tib-b1}, we then have
$$\sqrt{n}(\widetilde\b_j - \b_{j}) \stackrel{d}\rightarrow N(\bbo, \sigma^2 \I_q).$$
Thus, we complete the proofs.

\begin{Lem} \label{lem:ujH}
  Under Conditions (D1)--(D5), for each $j$, we have
  \begin{eqnarray} \label{eq: ujh1}
   &&n^{-1/2}\u_{.j}\trans (\widetilde\H-\H)=O_p(n^{-1/2}+p^{-1/2}), \\
    &&n^{-1/2}\sumi (\widetilde\h_i-\h_{i})\trans \b_{j} \widetilde\h_i = O_p(p^{-1/2}+ p^{-1}n^{1/2} + n^{-3/2}). \label{eq:hibj}
  \end{eqnarray}
\end{Lem}
{\noindent\bf Proof.}
 By {the equation (7) in  \cite{bai2013principal},} we have
\begin{eqnarray}
&&n^{-1/2}\u_{.j}\trans (\widetilde\H-\H) \nonumber \\
&   =& n^{-1/2} p^{-1/2} \sumi \left\{ (\B\trans \B/ p)^{-1} p^{-1/2}\sum_{k=1}^{p} \b_{j}u_{ij}  + o_p(1)\right\}u_{ij} \nonumber \\
 &  =&  p^{-1/2}(\B\trans \B/ p)^{-1} n^{-1/2} \sumi (p^{-1/2}\sum_{k\neq j} \b_{k}u_{ik}u_{ij}) + (\B\trans \B/ p)^{-1}n^{-1/2} p^{-1} \sumi \b_{j} u_{ij}^2 \nonumber \\
 &\hat=  &  p^{-1/2}(\B\trans \B/ p)^{-1}I_1 + (\B\trans \B/ p)^{-1} I_2 \label{eq:ujH}.
\end{eqnarray}
 Noting $E I_1 = 0$ and
\begin{eqnarray*}
   E \|I_1\|^2 &=& n^{-1} \sum_{i_1=1}^n\sum_{i_2=1}^n E \{p^{-1}\sum_{j_1\neq j}\sum_{j_2\neq j} \b_{j_1}\trans\b_{j_2} u_{i_1j_1}u_{i_1j}u_{i_2j_2}u_{i_2j}\} \\
       &\leq & M n^{-1} \sumi \{p^{-1}\sum_{j_1\neq j}\sum_{j_2\neq j} E (u_{ij}^2 u_{ij_1}u_{ij_2})\} \\
       & = & M n^{-1} \sumi \{p^{-1}\sum_{k\neq j}E (u_{ij}^2 u_{ik}^2)\} \leq M\sigma^4,
\end{eqnarray*}
where the first inequality follows from Conditions (D2) and (D3), we have
\begin{equation}\label{eq:I1}
  I_1=O_p(1),
\end{equation}
by Chebyshev's inequality.

As for $I_2$, since $E\|n^{-1} \sumi \b_{j}u_{ij}^2\|\leq n^{-1}\sup_j\|\b_{j}\|\sumi E u_{ij}^2\leq M\sigma^2$ by Conditions (D2) and (D3), we obtain
$n^{-1} \sumi b_{j}u_{ij}^2=O_p(1)$ by Markov inequality, which implies
\begin{equation}\label{eq:I2}
  I_2=O_p(n^{1/2}p^{-1}).
\end{equation}

By Condition (D5), we have $(\B\trans \B/ p)^{-1}=O_p(1)$. Coupling with \eqref{eq:ujH}, \eqref{eq:I1} and \eqref{eq:I2}, we obtain
\begin{equation*}
  \u_{.j}\trans (\widetilde\H-\H)/ \sqrt{n}=O_p(p^{-1/2} + n^{1/2}p^{-1}).
\end{equation*}

Note
\begin{eqnarray}\label{eq:ujhdecomp}
 && n^{-1/2}\sumi (\widetilde\h_i-\h_{i})\trans \b_{j} \widetilde\h_i \nonumber \\
  &=&n^{-1/2}\sumi (\widetilde\h_i-\h_{i})\trans \b_{j} (\widetilde\h_i-\h_{i}) + n^{-1/2}\sumi (\widetilde\h_i-\h_{i})\trans \b_{j} \h_{i}= II_1 + II_2.
\end{eqnarray}
By Cauchy-Schwarz inequality, Condition (D4) and Lemma \ref{lem:sqnbj}, we have
\begin{equation}\label{eq:II1}
  II_1 \leq M n^{-1/2}\sumi \|\widetilde\h_i - \h_{i}\|^2 = O_p(n^{1/2}p^{-1} + n^{-3/2}).
\end{equation}
Denote the $k$th component of $II_2$ as $II_{2k}$. By the equation (7) in \cite{bai2013principal}, we have
\begin{eqnarray}
 II_{2k}& =& n^{-1/2} p^{-1/2} \sumi \left\{ \b_{j}\trans(\B\trans \B/ p)^{-1} p^{-1/2}\sum_{l=1}^{p} \b_{l}u_{il}  + o_p(1)\right\}h_{ik0} \nonumber \\
 &  =&  p^{-1/2}\b_{j}\trans(\B\trans \B/ p)^{-1} n^{-1/2} \sumi (p^{-1/2}\sum_{l=1}^p \b_{l}u_{il}h_{ik}) \nonumber \\
 &\hat= & p^{-1/2}\b_{j}\trans(\B\trans \B/ p)^{-1} II_{3k}. \label{eq:II2kdc}
\end{eqnarray}
By the central limit theory, we have $II_{3k} = O_p(1)$. By Condition (D2) and (D4), we have $\b_{j}\trans(\B\trans \B/ p)^{-1} = O_p(1)$. Thus, coupling with \eqref{eq:II2kdc}, we obtain
\begin{equation}\label{eq:II2}
  II_2 = O_p(p^{-1/2}).
\end{equation}
Finally, combing \eqref{eq:ujhdecomp}, \eqref{eq:II1} and \eqref{eq:II2}, we have
$$n^{-1/2}\sumi (\widetilde\h_i-\h_{i})\trans \b_{j} \widetilde\h_i = O_p(p^{-1/2}+ n^{1/2}p^{-1} + n^{-3/2}),$$
which completes the proof. $\hfill{} \Box$

Next, we give the proofs of Theorem 6 based on the above two Lemmas.

{\bf\noindent Proof of Theorem 6.} Previously, $\b_j$ represents both general symbol and true value in the absence of symbolic confusion. Here, for clarity of proofs of this Theorem, we use $\b_{j0}$ for the true value and $\b_j$ only for the general symbol.  Note that $\b_j$ is separatable in the objective function \eqref{eq:step2}, we can consider each $\b_j$ one by one. For each $j$, we let $\b_j = \b_{j0}+ \v_j/\sqrt{n}$ and $\phi_{nj}(\v_j)= \sumi \left\{x_{ij}- \widetilde\h_i\trans(\b_{j0}+ \v_j/\sqrt{n})\right\}^2 + \lambda_1  w_{1j}\|\b_{j0}+ \v_j/\sqrt{n}\| + \lambda_2 \sum_{k=1}^q w_{2,jk}|b_{jk0} + v_{jk}/\sqrt{n}|$. Let $\wh\v_j = \arg\min \phi_{nj}(\v_j)$; then $\wh\b_j = \b_{j0} + \wh\v_j / \sqrt{n}$ or $\wh\v_j = \sqrt{n}(\wh \b_j -\b_{j0})$.

Let $V_{nj}(\v_j)= \phi_{nj}(\v_j)- \phi_{nj}({\bf 0})$, then the specific form is $V_{nj}(\v_j)= \v_j\trans \widetilde\H\trans\widetilde\H/n \v_j - 2 n^{-1/2}\u_{.j}\trans \widetilde\H \v_j + 2 n^{-1/2}\sumi (\widetilde\h_i - \h_{i})\trans \b_{j0}\widetilde\h_i\trans \v_j + \lambda_1 n^{-1/2}  w_j\sqrt{n}(\|\b_{j0} + \v_j /\sqrt{n}\|- \|\b_{j0}\|) + \lambda_2 n^{-1/2} \sum_{k=1}^q w_{2,jk}n^{1/2} (|b_{jk0} + v_{jk}n^{-1/2}| - |b_{jk0}|)$.
By the identifiability condition (E1), we know $\widetilde\H\trans\widetilde\H/n= \I_q$. Let $s_j = \lambda_1 n^{-1/2}  w_j\sqrt{n}(\|\b_{j0} + \v_j /\sqrt{n}\|- \|\b_{j0}\|)$ and $r_{jk}= \lambda_2n^{-1/2} w_{2,jk}n^{1/2}\times$ $(|b_{jk0} + v_{jk}n^{-1/2}| - |b_{jk0}|)$, then we have the simplified form of $V_{nj}(\v_j)$,
\begin{eqnarray}\label{eq:Vnj}
V_{nj}(\v_j)&=& \v_j\trans \v_j - 2 \u_{.j}\trans\H/ \sqrt{n} \v_j -2 \u_{.j}\trans (\widetilde\H-\H)/ \sqrt{n} \v_j +  \nonumber \\
& & 2 n^{-1/2}\sumi (\widetilde\h_i - \h_{i})\trans \b_{j0}\widetilde\h_i\trans \v_j+ s_j + \sum_{k=1}^q r_{jk}.
\end{eqnarray}
By the central limit theorem, we have $\u_{.j}\trans\H/ \sqrt{n} \stackrel{d} \rightarrow \z_j \hat= N(0, \sigma_j^2 \I_q)$.
In addition, by Lemma \ref{lem:ujH}, we have $\u_{.j}\trans (\widetilde\H-\H)/ \sqrt{n} = o_p(1)$ and $n^{-1/2}\sumi (\widetilde\h_i - \h_{i})\trans \b_{j0}\widetilde\h_i\trans=o_p(1)$.

Now considering the limiting behaviour of $s_j$ in \eqref{eq:Vnj}. By Lemma \ref{lem:sqnbj}, we know if $\b_{j0} \neq {\bf 0}$, then $w_j\stackrel{p} \rightarrow \|\b_{j0}\|^{-1}$ and $\sqrt{n}(\|\b_{j0} + \v_{j}/\sqrt{n}\|-\|\b_{j0}\|)\stackrel{p} \rightarrow \b_{j0}\trans \v_j / \|\b_{j0}\|$.  By $\lambda_1 n^{-1/2}\rightarrow 0$ and Slutsky's theorem, we have $s_j \stackrel{p} \rightarrow 0$. By $\lambda_1 \rightarrow \infty$, we know if $\b_{j0} = {\bf 0}$, then $\sqrt{n}(\|\b_{j0} + \v_{j}/\sqrt{n}\|-\|\b_{j0}\|)= \|\v_j\|$ and $\lambda_1 n^{-1/2} w_j=\lambda_1 \|\sqrt{n}\widetilde\b_{j}\|^{-1} \rightarrow \infty$, which implies $s_j \rightarrow \infty$.

Then considering the limiting behaviour of $r_{jk}$ in \eqref{eq:Vnj}. If $b_{jk0} \neq 0$, then $w_{2,jk} \stackrel{p} \rightarrow |b_{jk0}|^{-1}$ and $n^{1/2} (|b_{jk0} + v_{jk}n^{-1/2}| - |b_{jk0}|)\stackrel{p} \rightarrow v_{jk}sign(b_{jk0})$. {By Slutsky's theorem}, we have $r_{jk} \stackrel{p} \rightarrow 0$. If $b_{jk0}=0$, then $n^{1/2} (|b_{jk0} + v_{jk}n^{-1/2}| - |b_{jk0}|)= |v_{jk}|$ and $\lambda_2 n^{-1/2} w_{2,jk} = \lambda_2 (|\sqrt{n} \tilde b_{jk}|)^{-1}\rightarrow \infty$, which implies $r_{jk} \rightarrow \infty$.
 Thus, again, by Slutsky's theorem, we obtain that $V_{nj}(\v_j) \stackrel{d} \rightarrow V_j(\v_j)$ for each $\v_j$, where
\begin{equation*}
V_j(\v_j)=\left\{
\begin{aligned}
\v_{j, I_{j}}\trans\v_{j, I_{j}} - 2\v_{j, I_{j}}\trans \z_{j,, I_{j}},& \mbox{ if $v_{jk} =0, \forall k \notin I_{j}$ },  \\
\infty, & \mbox{ otherwise.}
\end{aligned}
\right.
\end{equation*}
$V_j(\v_j)$ is convex, and the unique minimum of $V_j(\v_j)$ is $(\z_{j,, I_{j}},{\bf 0})\trans$. Following the epi-convergence results of \cite{geyer1994asymptotics} and \cite{knight2000asymptotics}, we have
\begin{equation}\label{eq:an}
  \wh\v_{j, I_{j}} \stackrel{d} \rightarrow \z_{j, I_{j}} \stackrel{d}= N(\bbo, \sigma_j^2 \I_{m_j}),~~\forall j \in J.
\end{equation}
Thus, we proved the part of the asymptotical normality.

Now we consider the part of the selection consistency. For any $k \in I_{j}$, the asymptotical result implies that $\hat b_{jk} \stackrel{p}
\rightarrow b_{jk0}$; thus $P\{k \in \hat I_j\} \rightarrow 1$. Then it suffices to show that for any $l \notin I_{j}$, $P\{l \in \hat I_{j}\} \rightarrow 0$. We consider the event $\{l \in \hat I_{j}\}$. By the KKT optimality conditions, we have
\begin{equation}\label{eq:kkt}
  n^{-1/2} \widetilde\h_{.l}\trans(\x_{.j} - \wh \H \wh \b_j)= \lambda_1 n^{-1/2} w_{1j} \hat b_{jl} / \|\hat\b_j\| + \lambda_2 n^{-1/2} w_{2,jl},
\end{equation}
where $\widetilde\h_{.l}$ is the $l$th column of $\widetilde\H$.
 By the conditions that $\lambda_1  \rightarrow \infty$ and $\lambda_2 \rightarrow \infty$, we obtain $\lambda_1 n^{-1/2} w_{1j} \hat b_{jl} / \|\hat\b_j\|= \lambda_1  (\|\sqrt{n}\|\widetilde\b_j\|\|)^{-1} \hat b_{jl} / \|\hat\b_j\|\rightarrow \infty$ and $\lambda_2 n^{-1/2} w_{2,jl}= \lambda_2 (|\sqrt{n}\tilde b_{jl}|)^{-1}\rightarrow \infty$. That is,
\begin{equation}\label{eq:right}
  \lambda_1 n^{-1/2} w_{1j} \hat b_{jl} / \|\hat\b_j\| + \lambda_2 n^{-1/2} w_{2,jl} \rightarrow \infty.
\end{equation}
  Moreover, we have
\begin{eqnarray} \label{eq:kkt1}
  &&n^{-1/2} \widetilde\h_{.l}\trans(\x_{.j} - \wh \H \wh \b_j) \nonumber\\
   &=&  n^{-1/2} \widetilde\h_{.l}\trans(\H\b_{j0}-\widetilde\H \wh\b_j) + n^{-1/2}\widetilde\h_{.l}\trans \u_{.j}= III_1 + III_2.
\end{eqnarray}
Then we consider the order of $III_1$.
\begin{eqnarray*}
  III_1& = & n^{-1/2} (\widetilde\h_{.l} - \h_{.l})\trans(\H-\widetilde\H) \wh\b_j + n^{-1/2}\h_{.l}\trans(\H-\widetilde\H)\wh\b_j \\
   &+&  n^{-1/2}(\widetilde\h_{.l} - \h_{.l})\trans\H (\b_{j0}- \wh\b_j) + n^{-1}\h_{.l}\trans \H n^{1/2}(\b_{j0}-\wh\b_j).
\end{eqnarray*}
Due to $\forall i, \widetilde\h_i - \h_{i}=o_p(1)$, $n^{-1}\h_{.l}\trans \H= O_p(1)$ and $n^{1/2}(\b_{j0}-\wh\b_j)=O_p(1)$, we obtain
\begin{equation}\label{eq:III1}
  III_1 = O_p(1).
\end{equation}
Next, we consider the order of $III_2$.
\begin{eqnarray}\label{eq:III2}
  III_2& = & n^{-1/2} \sumi (\hat h_{il}- h_{il})u_{ij} + n^{-1/2} \sumi h_{il} u_{ij} \nonumber \\
   &=& O_p(1).
\end{eqnarray}
Coupling with \eqref{eq:kkt}, \eqref{eq:kkt1}, \eqref{eq:III1} and \eqref{eq:III2}, we conclude that
$P\{l \in \hat{I_{j}}\}\leq P\{n^{-1/2} \widetilde\h_{.l}\trans(\x_{.j} - \wh \H \tilde \b_j)= \lambda_1 n^{-1/2} w_{1j} \tilde b_{jl} / \|\widetilde\b_j\| + \lambda_2 n^{-1/2} w_{2,jl}\} \rightarrow 0$. Thus, we obtain
$$P(\hat{I_{j}}= I_{j})\rightarrow 1.$$
Since it is separable for each $\b_j$, thus we obtain $P(\hat{J}= J)\rightarrow 1$ and $P(\hat I= I)\rightarrow 1$.
Therefore, we complete the proof of Theorem 6.
\subsection{Proof of Theorem 4 }\label{app:infer} % \ref{th:rowchiqmaxtest}
In this part, we  give the  proofs of Theorem  4 in the main text. Before presenting the proofs, % \ref{th:rowchiqmaxtest}
we first present a lemma that will be used for the followed proofs.

\begin{Lem}\label{lem:hsig-sig}
  Under  Conditions (D1)-(D4), we have
  $$|\hat\sigma^2_j- \sigma_j^2| = O_p(n^{-1/2} + p^{-1/2}).$$
\end{Lem}
{\bf\noindent Proof. } By triangular inequality and Cauchy-Schwarz inequality, we have
\begin{eqnarray}\label{eq:hsigdcomp}
  |\hat\sigma^2_j- \sigma_j^2| &=& \left|(\frac{1}{n}\sumi u_{ij}^2 -\sigma_j^2) + \frac{1}{n}\sumi(\b_{j}^{\trans}\h_{i}-\widetilde\b_{j}^{\trans}\widetilde\h_i)^2 + \frac{2}{n}\sumi(\b_{j}^{\trans}\h_{i}-\widetilde\b_{j}^{\trans}\widetilde\h_i)u_{ij}\right|  \nonumber\\
   & \leq & |\frac{1}{n}\sumi u_{ij}^2 -\sigma_j^2| + IV_{2j} + 2 \{\frac{1}{n}\sumi u_{ij}^2\}^{1/2} \{\frac{1}{n}\sumi (\b_{j}^{\trans}\h_{i}-\widetilde\b_{j}^{\trans}\widetilde\h_i)^2\}^{1/2} \nonumber \\
   &=& IV_{1j} + IV_{2j} +2 \{\frac{1}{n}\sumi u_{ij}^2\}^{1/2} IV_{2j}^{1/2}.
\end{eqnarray}
By central limit theorem, we have
\begin{equation}\label{eq:uij2}
 IV_{1j} = O_p(n^{-1/2}).
\end{equation}
Next, we consider the order of $IV_{2j}$.  By the fact that $(a +b )^2 \leq 2a^2 + 2b^2$ and $\a^{\trans}\b \leq \|\a\|\|\b\|$, we have
\begin{eqnarray*}
   IV_{2j} &\leq &  \|\b_{j}-\widetilde\b_{j}\|^2  \frac{2}{n}\sumi \|\h_{i}\|^2+ 4\|\b_{j}\|^2 \frac{1}{n}\sumi \|\h_{i} - \widetilde\h_i\|^2 \\
   &&  + 4\|\b_{j}-\widetilde\b_{j}\| \frac{1}{n}\sumi \|\h_{i}- \widetilde\h_i\|^2. \\
\end{eqnarray*}
According to Condition (D2) and the results of Lemma \ref{lem:sqnbj}, we obtain
\begin{equation}\label{eq:unifbh}
 |IV_{2j}| = O_p(n^{-1}+ p^{-1}).
\end{equation}

Then, combing \eqref{eq:hsigdcomp}, \eqref{eq:uij2} and \eqref{eq:unifbh} we have
$$ |\hat\sigma^2_j- \sigma_j^2| = O_p(n^{-1/2} + p^{-1/2}).$$
$\hfill{} \Box$

{\noindent\bf Proof of Theorem 4}.
By Theorems 1 -- 3 in the main text and Lemma \ref{lem:sqnbj}, we only require to verify
\begin{equation}\label{eq:Rmax}
 \max_{j\in G^o_1} \|\widetilde\b_j\| \gg \max_{j\in G^o_0}\|\widetilde\b_j\|
\end{equation}
and
\begin{equation}\label{eq:Rmin}
 \min_{j\in G^{no}_1} \|\widetilde\b_j\| \gg \min_{j\in G^{no}_0}\|\widetilde\b_j\|.
\end{equation}

For any $j_0\in G^o_1$, we have $\max_{j\in G^o_1} \|\tilde \b_{j}\| \geq \|\b_{j_0}\| - \|\tilde \b_{j_0} - \b_{j_0}\| \gg \sqrt{\frac{ln (p)}{n}}$ by Condition (D6.3) and Lemma 2, where $j_0 \in G^o_1$.
Under Conditions (D1) -- (D6), and by
Theorem 3.2 in \cite{bai2013statistical}, we have $\max_{j\in G^o_0} \|\tilde \b_j\|= O_p(\sqrt{\frac{\ln (|G^o_0|)}{n}}) = O_p(\sqrt{\frac{\ln (p)}{n}})$. Thus, the order of $\max_{j\in G^o_1} \|\tilde \b_j\|$   dominates the order of $\max_{j\in G^o_0} \|\tilde \b_j\|$, which implies equation \eqref{eq:Rmax} holds.

 In addition, by the triangular inequality and Condition (D6.3), we obtain
\begin{eqnarray} \label{eq:Rminb}
 \min_{j\in G^{no}_1} \|\widetilde\b_j\|&\geq & \min_{j\in G^{no}_1}\|\b_{j}\| - \max_{j\in G^{no}_1} \|\widetilde\b_j - \b_{j}\| \nonumber \\
  & \gg & \sqrt{\frac{\ln (p)}{n}}.
\end{eqnarray}
By Lemma \ref{lem:sqnbj}, we have
\begin{equation}\label{eq:Rming20}
  \min_{j\in G^{no}_0}|\widetilde\b_j| = O_p(\frac{1}{\sqrt{n}}).
\end{equation}
Coupling with \eqref{eq:Rminb} and \eqref{eq:Rming20}, we complete the proof of \eqref{eq:Rmin}.
Moreover, we complete the proof of Theorem 4. % \ref{th:rowchiqmaxtest}

\subsection{Performance of variable selection from NITS estimation}\label{sec:pvs}
We investigated the performance of variable selection for the entries or rows of $\B$  using six examples with $q=6$ and $\sigma^2=1$, in order to see  the effect of the sample size $n$, the dimension $p$, the size of true model ($s_1=p-s$), the heteroscedasticity and  signal-to-noise ratio (SNR). The SNR is determined by  $\sigma^2$ and $\rho$. The settings of  six examples {were} listed in Table
\ref{tab:facset}. In Ex.4, we set $\gamma=0.5$ to ensure the average of $\sigma_j^2$'s  equals to $\sigma^2=1$. In all simulations, we selected the tuning parameters $(q,\lambda_1, \lambda_2)$ by using the method in Appendix \ref{sec:tuning}. All results were  based on 500 repetitions.
\begin{table}
\centering
\caption{The setting of six examples.}\label{tab:facset}
\begin{tabular}{ c c c c c c c c }
\hline
Example&  $n$&  $p$&  $s_1$&  $\rho$&  $\sigma^2$&  $\u_i$&    \\ \hline

Ex.1&  50 or 100&  300&  $\lfloor p/4\rfloor$&  1&  1&  $N(0, \sigma^2\I_p)$&    \\
Ex.2&  100&  100 or 500&  $\lfloor p/4\rfloor$&  1&  1&  $N(0, \sigma^2\I_p)$& \\
Ex.3& 100&  500&  10 or $\lfloor p/4\rfloor$&  1&  1&  $N(0, \sigma^2\I_p)$&    \\
Ex.4&  100&  500&  $\lfloor p/4\rfloor$&  1&  1&  $N(0, \Sigma_\u)$&    \\
Ex.5&  100&  500&  $\lfloor p/4\rfloor$&  1&  1 or 2& $N(0, \sigma^2\I_p)$ &    \\
Ex.6&  100&  500&  $\lfloor p/4\rfloor$& 1 or 2&  1& $N(0, \sigma^2\I_p)$&  \\ \hline
\end{tabular}
\end{table}

\begin{figure}
  \centering
  \subfigure[Ex.1]{
	 \includegraphics[width=4cm, height=4cm]{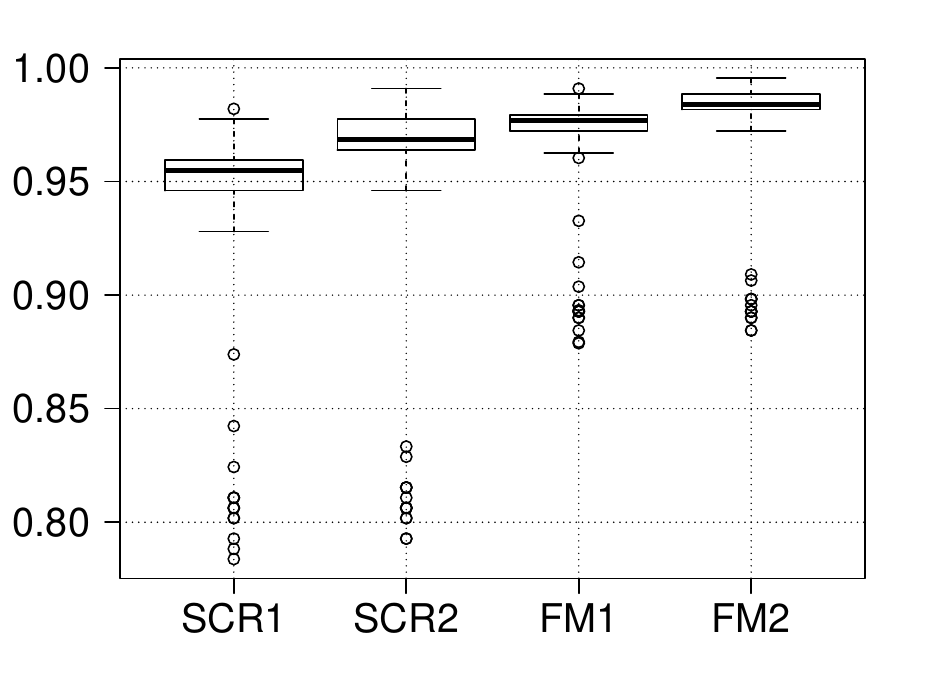}
	\label{fig:ex1-row}
}
  \subfigure[Ex.2]{
  \includegraphics[width=4cm, height=4cm]{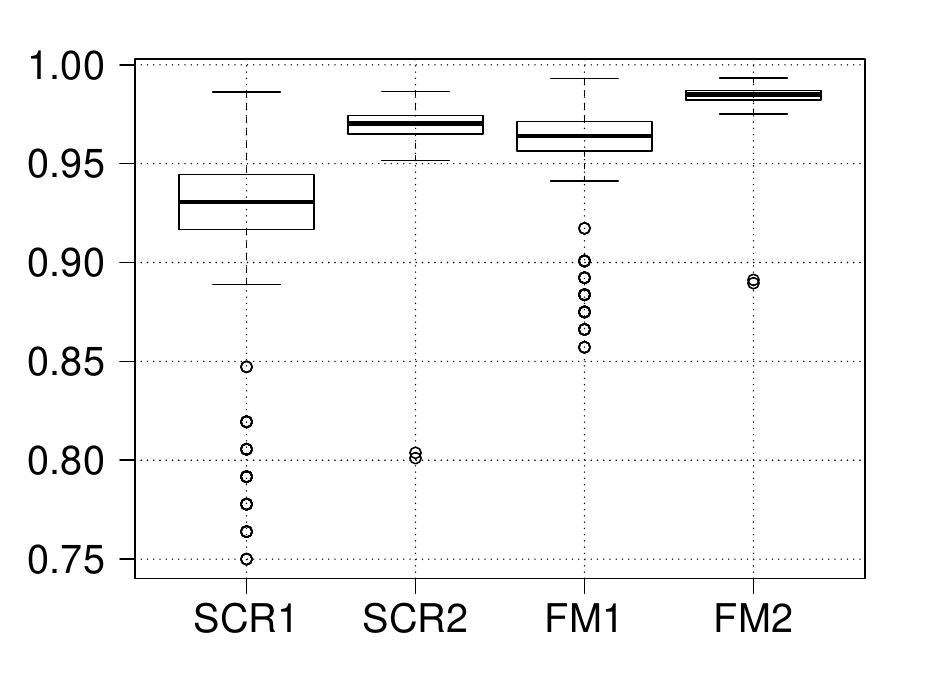}
	\label{fig:ex2-row}
}
\subfigure[Ex.3]{
  \includegraphics[width=4cm, height=4cm]{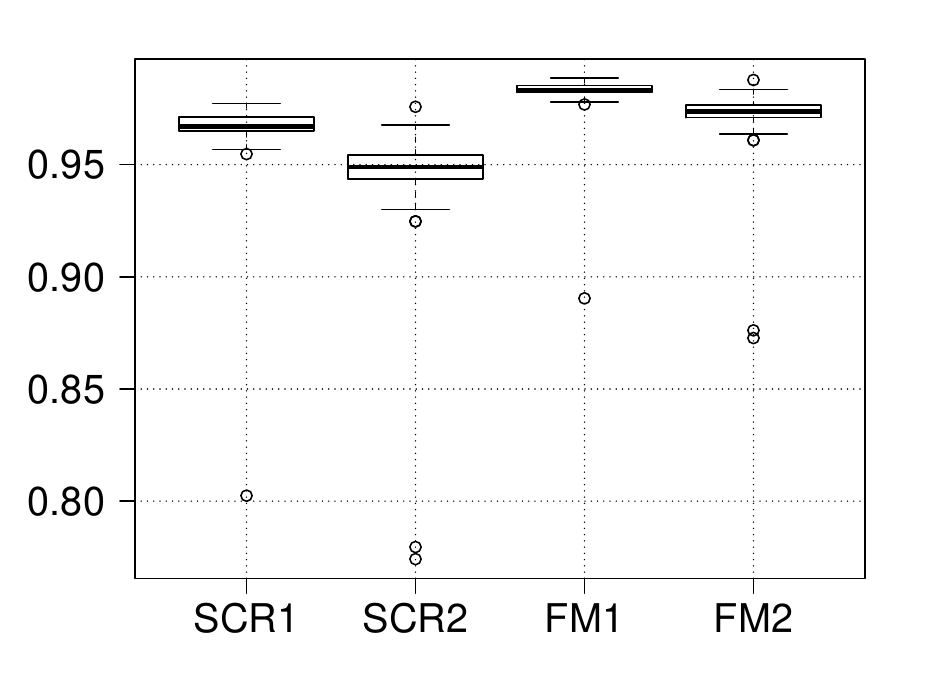}
	\label{fig:ex3-row}
}
\subfigure[Ex.4]{
	 \includegraphics[width=4cm, height=4cm]{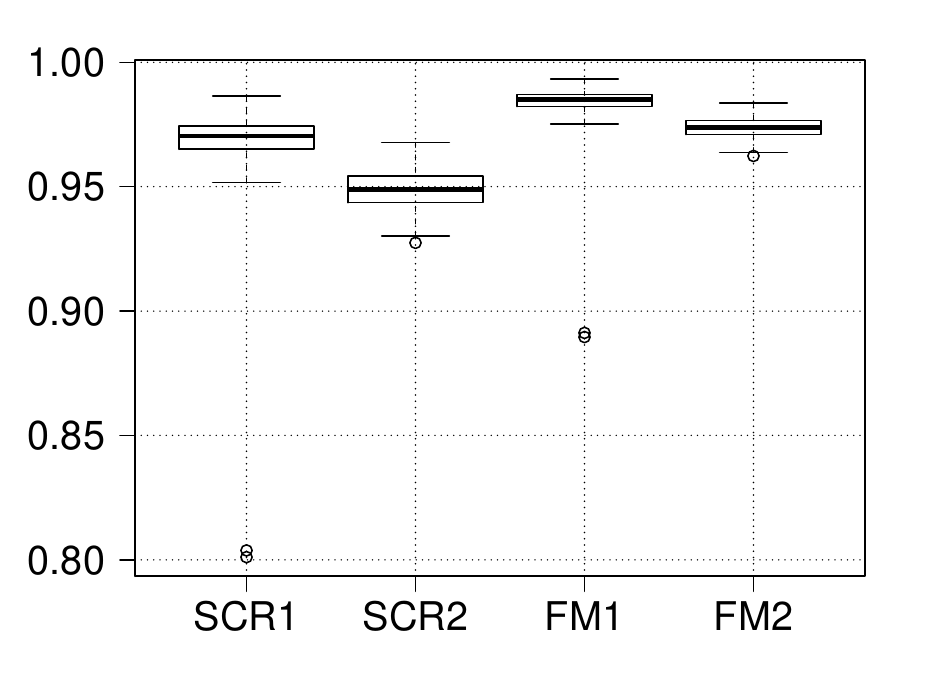}
	\label{fig:ex4-row}
}
  \subfigure[Ex.5]{
  \includegraphics[width=4cm, height=4cm]{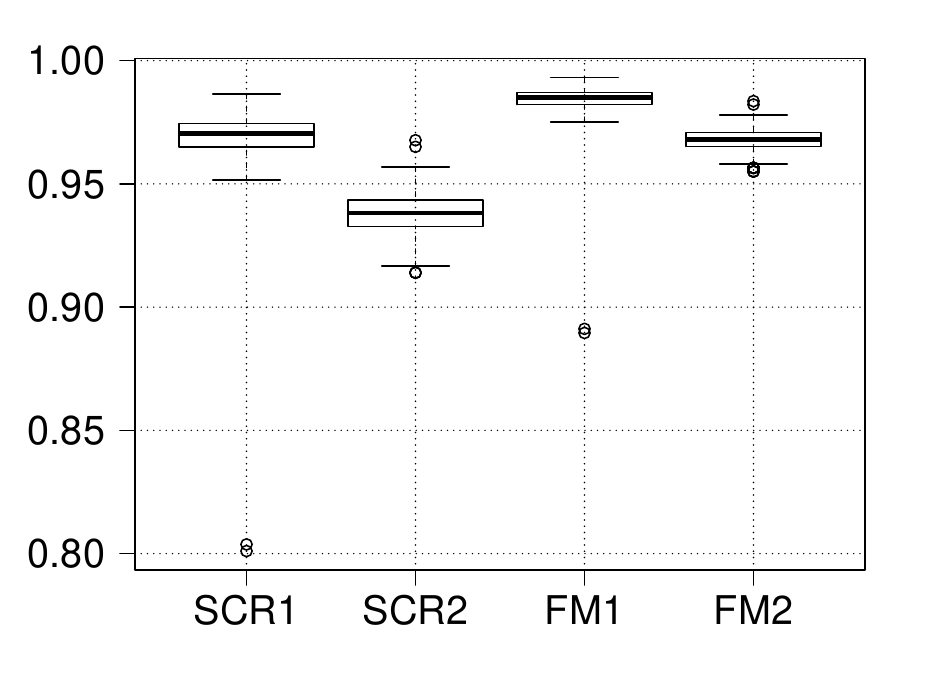}
	\label{fig:ex5-row}
}
\subfigure[Ex.6]{
  \includegraphics[width=4cm, height=4cm]{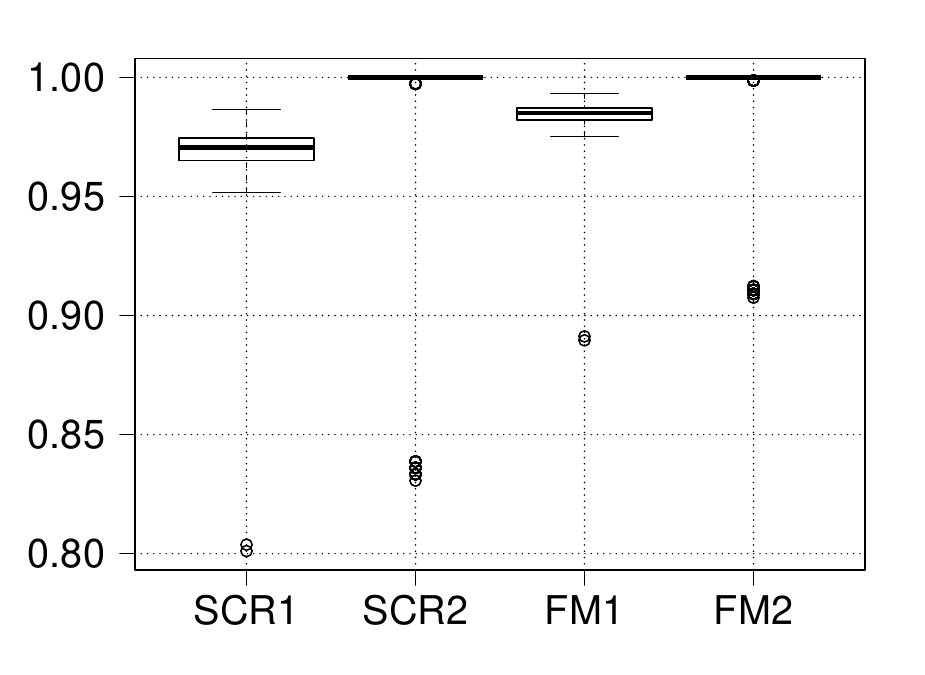}
	\label{fig:ex6-row}
}
  \caption{Performance of variable selection for entries and rows of $\B$ with 500 repeats. (a) : SCR1 and FM1 for $n=50$, SCR2 and FM2 for $n=100$; (b): SCR1 and FM1 for $p=100$, SCR2 and FM2 for $p=500$; (c): SCR1 and FM1 for $s_1=10$, SCR2 and FM2 for $s_1 = \lfloor p/4\rfloor$; (d): SCR1 and FM1 for homoscedasticity with $\sigma^2=1$, SCR2 and FM2 for heteroscedasticity with $\gamma=0.5$; (e): SCR1 and FM1 for $\sigma^2=1$, SCR2 and FM2 for $\sigma^2=2$; (f): SCR1 and FM1 for $\rho=1$, SCR2 and FM2 for $\rho=2$.}\label{fig:fmea-row}
\end{figure}

\begin{figure}[H]
  \centering
  \subfigure[Ex.1]{
	 \includegraphics[width=4cm, height=4cm]{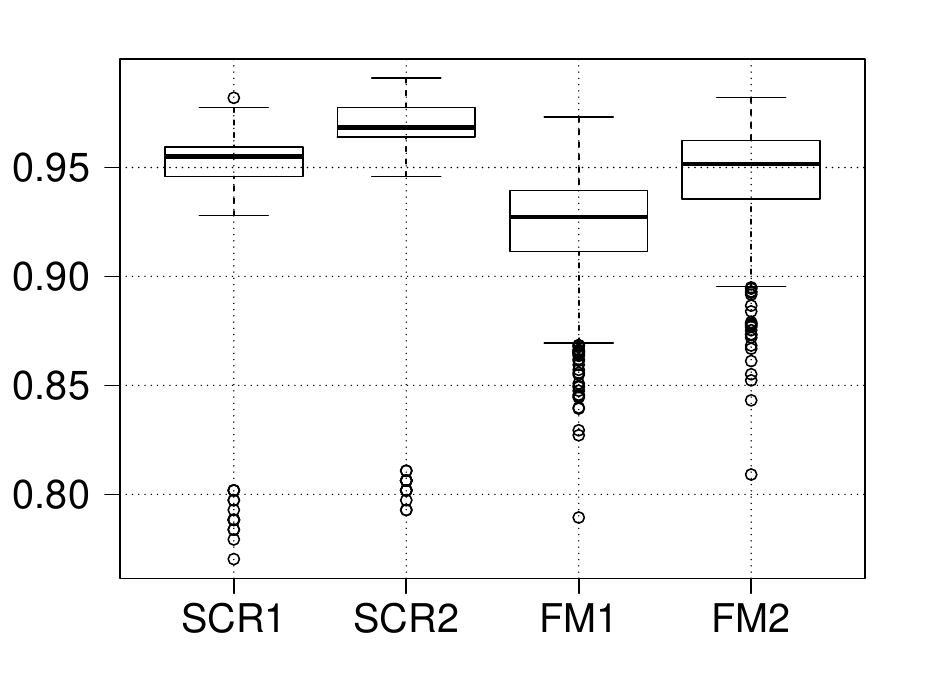}
	\label{fig:ex1-entry}
}
  \subfigure[Ex.2]{
  \includegraphics[width=4cm, height=4cm]{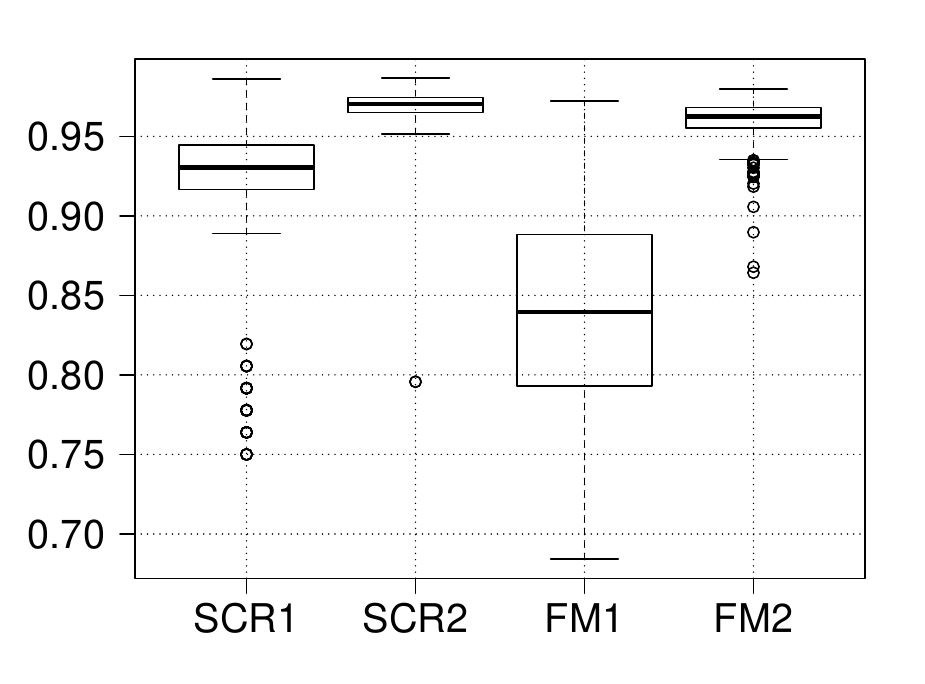}
	\label{fig:ex2-entry}
}
\subfigure[Ex.3]{
  \includegraphics[width=4cm, height=4cm]{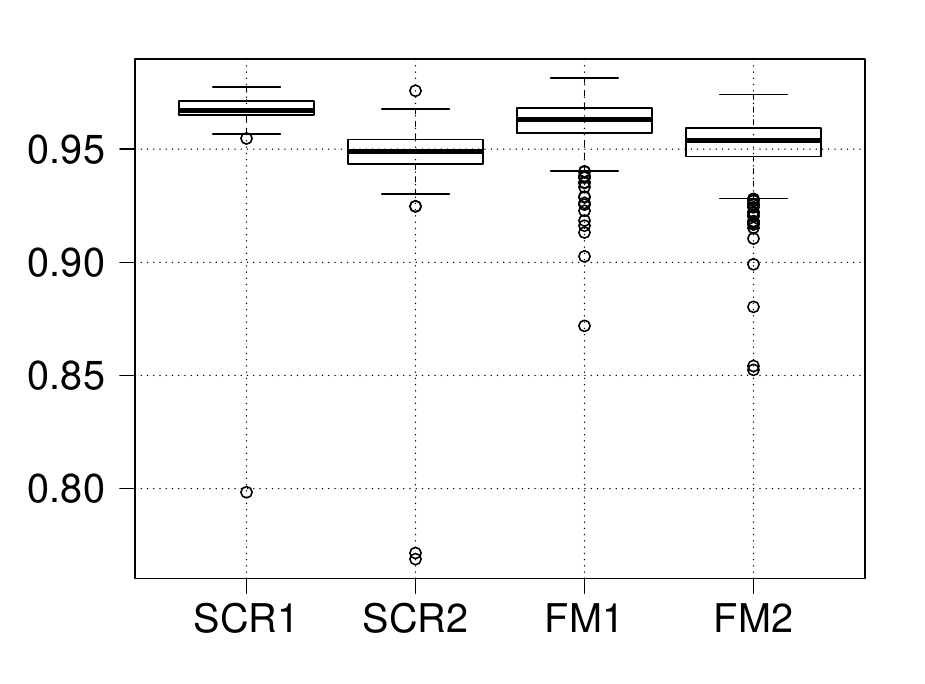}
	\label{fig:ex3-entry}
}
\subfigure[Ex.4]{
	 \includegraphics[width=4cm, height=4cm]{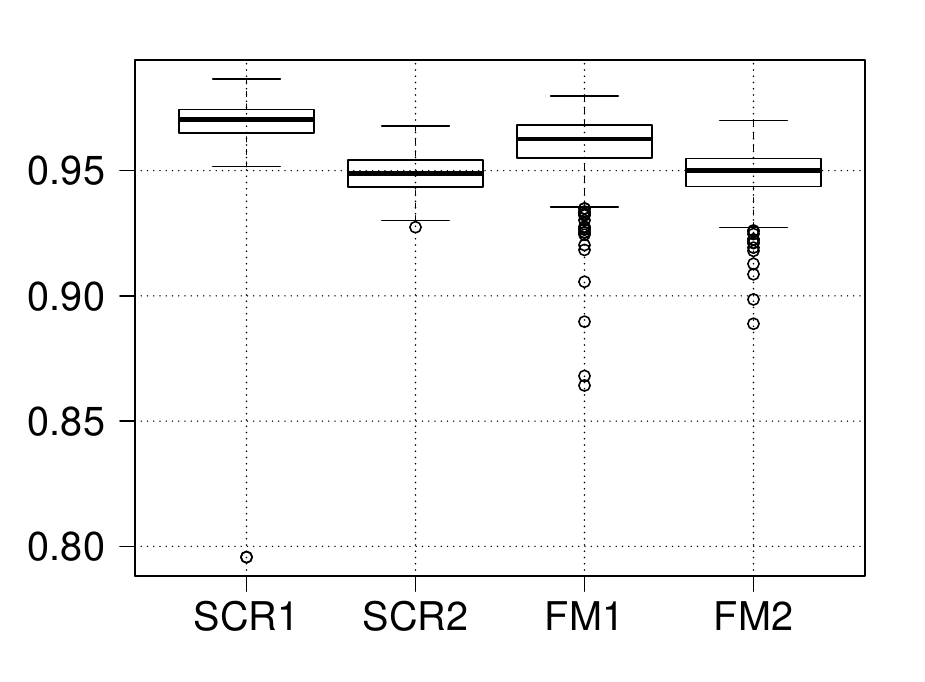}
	\label{fig:ex4-entry}
}
  \subfigure[Ex.5]{
  \includegraphics[width=4cm, height=4cm]{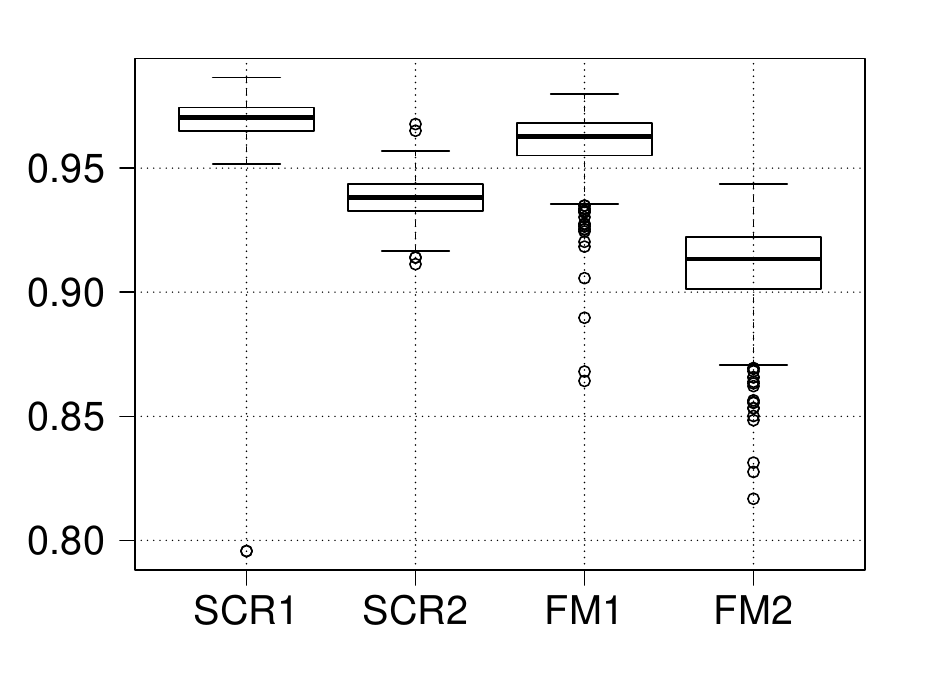}
	\label{fig:ex5-entry}
}
\subfigure[Ex.6]{
  \includegraphics[width=4cm, height=4cm]{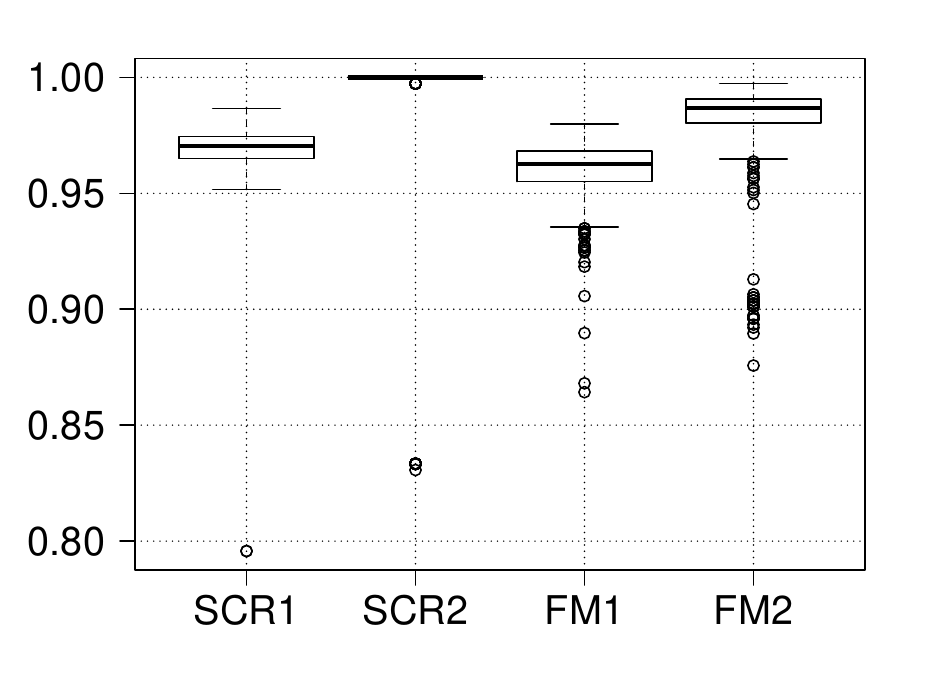}
	\label{fig:ex6-entry}
}
  \caption{Performance of variable selection for entries of $\B$ with 500 repeats.  (a) : SCR1 and FM1 for $n=50$, SCR2 and FM2 for $n=100$; (b): SCR1 and FM1 for $p=100$, SCR2 and FM2 for $p=500$; (c): SCR1 and FM1 for $s_1=10$, SCR2 and FM2 for $s_1 = \lfloor p/4\rfloor$; (d): SCR1 and FM1 for homoscedasticity with $\sigma^2=1$, SCR2 and FM2 for heteroscedasticity with $\gamma=0.5$; (e): SCR1 and FM1 for $\sigma^2=1$, SCR2 and FM2 for $\sigma^2=2$; (f): SCR1 and FM1 for $\rho=1$, SCR2 and FM2 for $\rho=2$.}\label{fig:fmea-entry}
\end{figure}

 We evaluated the performance in terms of model selection consistency rate (SCR) and F-measure (FM,\cite{p11}).
The results were summarized in Figures \ref{fig:fmea-row} and \ref{fig:fmea-entry}. From Figures \ref{fig:fmea-row} and \ref{fig:fmea-entry}, we {could} see both SCR and FM  approach to  $1$, suggesting the proposed method {could} select $J$ and $J^c$ in high accuracy. The proposed method {worked} better  as $n$,  $p$ or SNR increased, or $s_1$ was fixed, or $\u_i$ is homogeneous.  Particularly, since $\h_i$ is estimated based on $p$ variables, hence  larger $p$ provide more information on $\h_i$ then $\b_j$, which is also confirmed by Lemma \ref{lem:sqnbj}. In addition, larger $s_1$ implies more noise, which  had very significant effect on the resulting estimators.


\section{Additional simulation results}
{
To evaluate the performance of TOSI in selecting the penalty parameter in Section 6.1 of the main text, we compare it with CV LASSO, AIC, BIC and scaled LASSO. %Here, we present the implementation details about  CV LASSO, AIC, BIC and scaled LASSO.
CV LASSO is implemented by the function {\it cv.glmnet} with default settings in the R package {\bf glmnet}; AIC and BIC  using the formula in \cite{cherkassky2003comparison} based on LASSO are implemented  by the function {\it glmnet} in the R package {\bf glmnet}; and scaled LASSO is implemented by the function {\it scalreg} with default settings in the R package {\bf scalreg}.

To demonstrate the generality of the proposed method, We apply TOSI to a nonlinear logistic model setting. We generate data from the model $y_i|\x_i \sim Bernoulli(\pi_i), \pi_i = \frac{1}{1+ \exp(-\x_i^{\trans}\bbeta)}, i=1,\cdots, n=100,$
where $\x_i \sim N(0, \Sigma^x)$ with $\sigma^x_{jk} = 0.2^{|j-k|}$ and $\bbeta=(\beta_1, \cdots, \beta_p)^{\trans}$ with $p=30$. $\beta_{j} = 0.5 z $ for $j\leq 5$ and $\beta_{j} = 0$ for $j>5$, where $z$ is a random variable following uniform distribution $U[0,2]$. $\bbeta$ is fixed after being generated. Similar to that in linear regression models, we require to construct debiased estimators as well as standard errors for regression coefficients of the logistic regression model. Here, we base on a newly  developed  approach in \cite{cai2021statistical}. To benchmark the testing performance of TOSI in the logistic regression, we compare TOSI with three other methods: (1) Benjamini-Yekutieli $p$-value-adjusted method~\citep{benjamini2001control} based on the $p$-values obtained from \cite{cai2021statistical}, denoted as CZM-21; (2) Holm $p$-value-adjusted method~\citep{holm1979simple} based on the $p$-values obtained from \cite{degeer2014on} on logistic regression, denoted as DBRD-14; and (3) Holm $p$-value-adjusted method~\citep{holm1979simple} based on the $p$-values obtained from \cite{buhlmann2013statistical} on logistic regression, denoted as B-13. In implementation, CZM-21 is implemented by using R function {\it GLM\_binary} in the R package {\bf SIHR}; DBRD-14 is conducted using R function {\it lasso.proj} with {\it family='binomial'} in the R package {\bf hdi}; and B-13 is implemented using the function {\it ridge.proj} with {\it family='binomial'} in the R package {\bf hdi}. The simulation results are summarized in Table \ref{tab:logitregsplit}. We observe ToMax$(L)$ outperforms the three $p$-values-adjusted methods in terms of testing size and power. The $p$-values-adjusted methods control Type I error too conservative (0.00 for $G_{11}$) and have lower testing powers (0.02-0.05 for $G_{14}$). In contrast, ToMax$(L)$/ToMin$(L)$ controls the Type I error around the minimal level $0.05$ and with larger $L$ tends to have higher powers.

}
\begin{table}[H]
            \scriptsize\centering\caption{{Results of Experiment 1: high-dimensional
            sparse linear regression models with two other signal-noise-ratio settings, $\rho=0.5$ and $0.8$}. Comparison of testing size and  power of ToMax$(L)$ with other methods under the significance level $\alpha=0.05$.} \label{tab:regsplit}
            \begin{tabular}{ c c c c c c cc }
            \hline
& Method&\multicolumn{3}{c}{$n=50$}& \multicolumn{3}{c}{$n=100$}  \\
%\cmidrule(lr){3-5}\cmidrule(lr){6-8}
\hline
%& & \multicolumn{6}{c}{ToMax$(L)$ vs Others} \\
%\hline
%\cmidrule(lr){3-5}\cmidrule(lr){6-8}
$\rho=0.5$ & &  $G_{11}$&  $G_{12}$&  $G_{13}$&  $G_{11}$&  $G_{12}$&  $G_{13}$  \\
 \cmidrule(lr){3-5}\cmidrule(lr){6-8}
%\hline
Size&ToMax(1)&  0.025&  0.025&  0.050&  0.030&  0.030&  0.065  \\
&ToMax(2) &  0.015&  0.015&  0.035&  0.035&  0.030&  0.065  \\
&ToMax(5) &  0.005&  0.020&  0.040&  0.030&  0.040&  0.040  \\
&ToMax(8) &  0.005&  0.030&  0.040&  0.025&  0.040&  0.020  \\
&ZC1-17 &  0.070&  0.110&  0.130&  0.055&  0.070&  0.080  \\
&ZC3-17(1/5)& 0.035&  0.038&  0.031&  0.048&  0.032&  0.032  \\
& ZC3-17(1/3)& 0.024&  0.026&  0.024&  0.046&  0.022&  0.032 \\
&ZZ-14&  0.010&  0.010&  0.000&  0.025&  0.000&  0.000  \\
&B-13& 0.000&  0.004&  0.008&  0.004&  0.004&  0.006  \\
& MMB-09& 0.000&  0.000&  0.000 & 0.000&  0.000&  0.000 \\
 \hline
& &  $G_{14}$&  $G_{15}$&  $G_{16}$&  $G_{14}$&  $G_{15}$&  $G_{16}$  \\
\cmidrule(lr){3-5}\cmidrule(lr){6-8}
Power&ToMax(1)&  0.332&  0.452&  0.664&  0.458&  0.630&  0.868  \\
&ToMax(2) &  0.416&  0.548&  0.782&  0.518&  0.720&  0.950  \\
&ToMax(5) & 0.508&  0.648&  0.890&  0.600&  0.782&  0.982  \\
&ToMax(8) &  0.580&  0.672&  0.928&  0.664&  0.806&  0.986  \\
&ZC1-17&  0.336&  0.404&  0.872&  0.518&  0.518&  0.988  \\
&ZC3-17(1/5)& 0.348&  0.316&  0.784&  0.488&  0.460&  0.950  \\
& ZC3-17(1/3)&0.346&  0.346&  0.778&  0.450&  0.404&  0.888  \\
&ZZ-14&  0.402&  0.350&  0.790&  0.600&  0.518&  0.974  \\
&B-13&  0.012&  0.060&  0.212&  0.022&  0.120&  0.540   \\
& MMB-09&0.022&  0.064&  0.280 &0.098&  0.114&  0.768   \\
\hline
%& & \multicolumn{6}{c}{ToMax$(L)$ vs Others} \\ \hline
%\cmidrule(lr){3-5}\cmidrule(lr){6-8}
 $\rho=0.8$&&  $G_{11}$&  $G_{12}$&  $G_{13}$&  $G_{11}$&  $G_{12}$&  $G_{13}$  \\
 \cmidrule(lr){3-5}\cmidrule(lr){6-8}
%\hline
Size&ToMax(1)&  0.040&  0.020&  0.040&  0.034&  0.034&  0.050  \\
&ToMax(2) &  0.026&  0.018&  0.046&  0.032&  0.032&  0.052  \\
&ToMax(5) &  0.006&  0.018&  0.046&  0.022&  0.024&  0.040  \\
&ToMax(8)&   0.006&  0.020&  0.044&  0.028&  0.016&  0.024  \\
&ZC1-17 &   0.056&  0.092&  0.098&  0.050&  0.052&  0.066  \\
&ZC3-17(1/5)&  0.040&  0.024&  0.022&  0.048&  0.030&  0.032  \\
&ZC3-17(1/3)& 0.020&  0.020&  0.03&  0.044&  0.020&  0.034  \\
&ZZ-14 &  0.014&  0.006&  0.004&  0.030&  0.002&  0.000  \\
&B-13&  0.000&  0.002&  0.006&  0.004&  0.004&  0.006  \\
& MMB-09& 0.000&  0.000&  0.000 & 0.000&  0.000&  0.000 \\
\hline
& &  $G_{14}$&  $G_{15}$&  $G_{16}$&  $G_{14}$&  $G_{15}$&  $G_{16}$  \\
\cmidrule(lr){3-5}\cmidrule(lr){6-8}
Power&ToMax(1)&  0.512&  0.746&  0.882&  0.728&  0.916&  0.990  \\
&ToMax(2) &  0.648&  0.842&  0.962&  0.806&  0.950&  1.000  \\
&ToMax(5) &  0.778&  0.892&  0.992&  0.890&  0.976&  1.000  \\
&ToMax(8) &  0.838&  0.924&  0.994&  0.904&  0.984&  1.000  \\
&ZC1-17& 0.636&  0.750&  1.000&  0.846&  0.934&  1.000  \\
&ZC3-17(1/5)&  0.618&  0.680&  0.976&  0.806&  0.856&  1.000   \\
& ZC3-17(1/3)& 0.560&  0.632&  0.968&  0.742&  0.784&  0.998  \\
&ZZ-14 &  0.714&  0.716&  0.978&  0.900&  0.912&  1.000  \\
&B-13&  0.062&  0.300&  0.658&  0.122&  0.520&  0.962  \\
& MMB-09& 0.024&  0.036&  0.498  &0.246&  0.254&  0.970  \\
\hline
            \end{tabular}
            \end{table}

\begin{table}[H]
\centering \caption{{Results of Experiment 1: sparse logistic regression model}. Comparison of testing size and power for TOSI with other methods under the significance level $\alpha=0.05$.} \label{tab:logitregsplit}
\begin{tabular}{ c c c c c c c c }
\hline
&&\multicolumn{3}{c}{Size}& \multicolumn{3}{c}{Power}  \\
\cmidrule(lr){3-5}\cmidrule(lr){6-8}
%&& \multicolumn{6}{c}{ToMax$(L)$ vs Others} \\
% \cmidrule(lr){1-5} \cmidrule(lr){6-8}
 &Method & $G_{11}$&  $G_{12}$&  $G_{13}$& $G_{14}$&  $G_{15}$&  $G_{16}$  \\ \hline

& ToMax(1) &  0.050&  0.050&  0.045&  0.150&  0.225&  0.530  \\
& ToMax(2) &  0.030&  0.030&  0.035&  0.210&  0.260&  0.575  \\
& ToMax(5) &  0.045&  0.025&  0.030&  0.280&  0.330&  0.720  \\
& ToMax(8) &  0.045&  0.030&  0.035&  0.270&  0.360&  0.750  \\
& CZM-21 &  0.000&  0.010&  0.040&  0.040&  0.140&  0.710  \\
& DBRD-14 &  0.000&  0.020&  0.060&  0.020&  0.170&  0.660  \\
& B-13 &  0.000&  0.020&  0.050&  0.050&  0.310&  0.870  \\ \hline
%&& \multicolumn{6}{c}{ToMin$(L)$} \\
%\cmidrule(lr){1-5} \cmidrule(lr){6-8}
 &Method & $G_{21}$&  $G_{22}$&  $G_{23}$& $G_{24}$&  $G_{25}$&  $G_{26}$  \\ \hline
&  ToMin(1)&  0.035&  0.025&  0.040&  0.165&  0.335&  0.265  \\
&  ToMin(2)&  0.025&  0.050&  0.045&  0.205&  0.465&  0.370  \\
&  ToMin(5)&  0.040&  0.065&  0.075&  0.250&  0.620&  0.575  \\
&  ToMin(8)&  0.035&  0.060&  0.055&  0.330&  0.725&  0.675  \\ \hline
\end{tabular}
\end{table}

\begin{table}[H]
            \scriptsize\centering\renewcommand{\arraystretch}{1}
            \caption{Results of Experiment 1: high-dimensional
            sparse linear regression models. Sensitivity analysis of the proposed TOSI in identifying the model structure  by setting three different  nominal levels ($\alpha=0.1, 0.05$ and $0.01$), where $s=3$. NV, average number of the variables being selected; IN, percentage of occasions on which the correct variables are included in the selected model; CS, percentage of occasions on which correct variables are selected. %; Time, running time (seconds), averaged over 500 replications.
            } \label{tab:SelectLambdaAlpha}
            \begin{tabular}{ c c c c cc c ccc c}
           \cmidrule(lr){1-11}
% &\multicolumn{7}{c}{{\bf (b) Results of variable selection and {\blue average running time!}}} \\ \cmidrule(lr){1-11}
&&\multicolumn{3}{c}{$\alpha=0.1$} & \multicolumn{3}{c}{$\alpha=0.05$}&\multicolumn{3}{c}{$\alpha=0.01$} \\
\cmidrule(lr){1-2}\cmidrule(lr){3-5}\cmidrule(lr){6-8}\cmidrule(lr){9-11}
$(n,p)$& $\rho$ &   NV &  IN&  CS&  NV&    IN&  CS &  NV&    IN&  CS  \\ \cmidrule(lr){1-11}

$(50,50)$& 2 & 2.920&  0.884&  0.858 &    2.890&  0.864&  0.848 &  2.826&  0.826&  0.826  \\
%&&  SD& 5.6528&  0.0000&  0.3057&  1.3314&  0.3109&  0.3808  \\
&3 &  3.028&  0.998&  0.984 & 3.022&  0.998&  0.990  &  2.990&  0.990&  0.990  \\
% & &SD&   5.6629&  0.0000&  0.3110&  1.3669&  0.04477&  0.2304  \\ \hline
$(100,50)$& 2 & 3.124&  0.990&  0.940   &    3.042&  0.988&  0.966 &    2.982&  0.964&  0.960 \\
%&&  SD&    4.8395&  0.0000&  0.3157&  3.1490&  0.0996&  0.3742  \\  \cmidrule(lr){3-9}
&3 &  3.072&  1.000&  0.988   & 3.030&  1.000&  0.994  & 3.018&  1.000&  0.998\\
%& &SD&  4.8633&  0.0000&  0.3030&  2.0841&  0.0000&  0.3030  \\
\cmidrule(lr){1-11}
\end{tabular}
\end{table}

\begin{figure}[H]
  \centering
  \subfigure[ToMax]{
  \includegraphics[width=6.5cm, height=4cm]{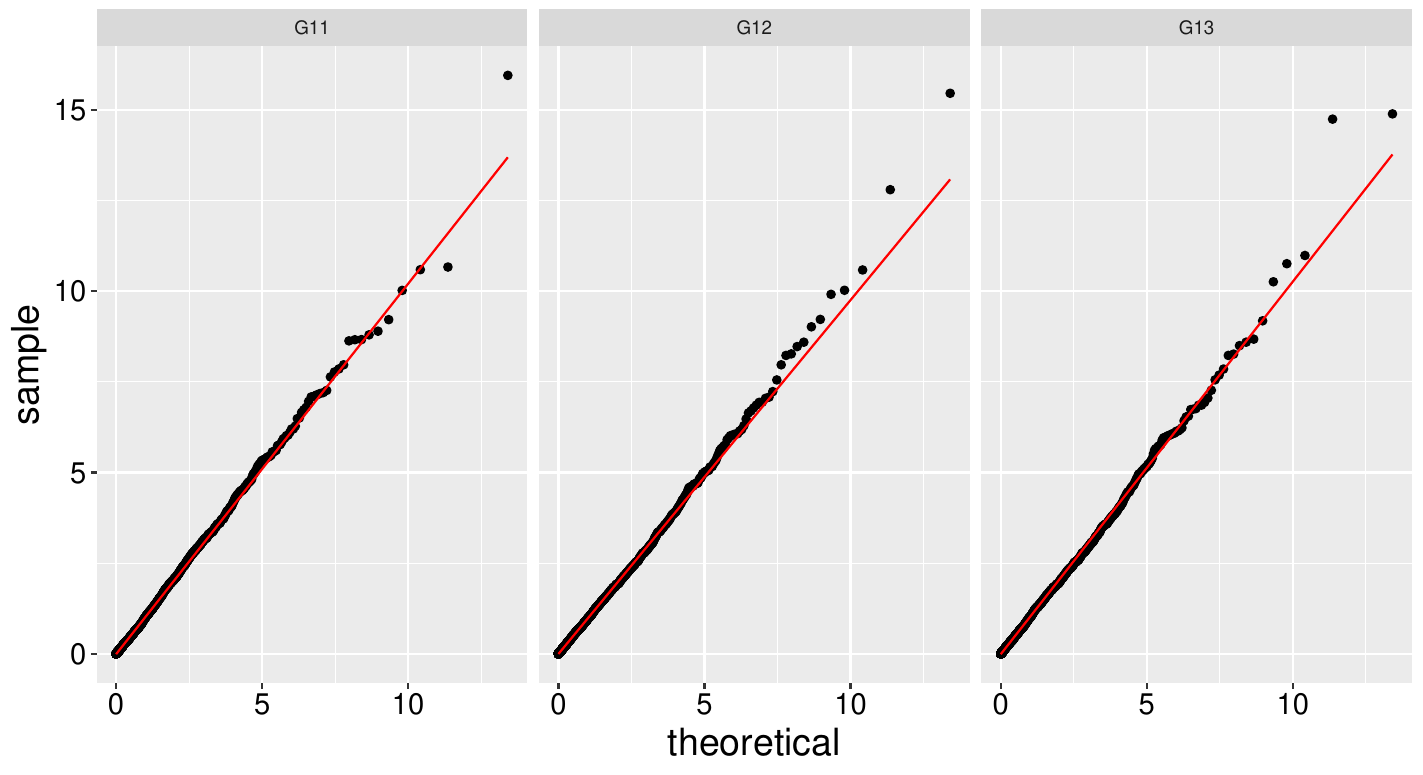}
}
\subfigure[ToMin]{
  \includegraphics[width=6.5cm, height=4cm]{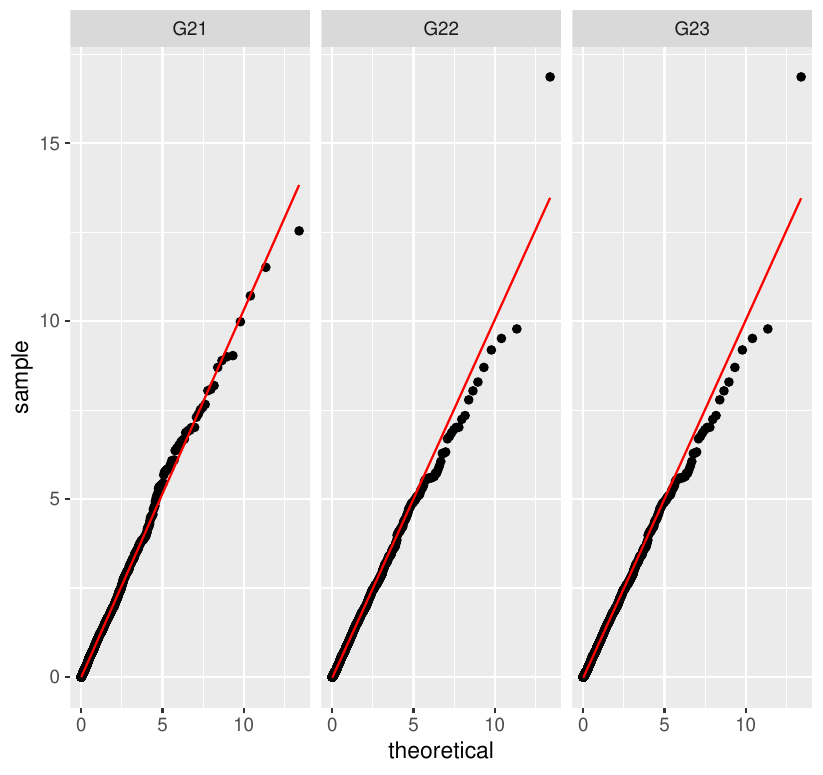}
}
  \caption{QQ plot with 2000 repeats for ToMax and ToMin  with $\chi^2_{(1)}$ distribution under high-dimensional mean models under $n=200$.}\label{fig:MeanQQplot}
\end{figure}

{\bf\noindent Experiment 3: High-dimensional mean models.} We consider the high-dimensional mean model,
$\x_i = \bmu + \bvarepsilon_i, i = 1,\cdots,n=200,$
where $\x_i \in R^p, \bvarepsilon_i \sim N(0, \Sigma^e)$ with $p=100, \Sigma^e=(\sigma^e_{jk})$ and $\sigma^e_{jk} =  0.5^{|j-k|}$. We set $\mu_{j} = z$ for $j\leq 5$ and $\mu_{j} = 0$ for $j>5$, where $z$ is a random variable following uniform distribution $U[0,1]$.
In this experiment, we {compare our approach } with the {test statistic $\max_j|\hat \mu_j - \mu_{j}|$ in \cite{chernozhukov2013gaussian} for $H_{0, G^{o}}$}, denoted as MaxMu. As far as we know, no existing literature on testing $\tilde H_{0, G^{no}}$ is available.
Figure \ref{fig:MeanQQplot} shows that the asymptotical $\chi^2_{(1)}$ distribution is accordance with the theoretical distribution, which implies the correction of our Theorems 1 and 2. Table \ref{tab:meansplit} shows testing size and power of TOSI under various settings. Under high-dimensional mean models, ToMax$(L)$ has lower testing powers than MaxMu even if we increase $L$ from $5$ to $20$. This point is different from that conclusion in sparse regression models since the regression model is more complex than mean model.

\begin{table}[H]
            \centering\caption{Results of Experiment 3. Testing size and power under the significant level $\alpha=0.05$: ToMax$(L)$ and ToMin$(L)$ with different splits for high-dimensional mean models.}\label{tab:meansplit}
            \begin{tabular}{ l c c c c c c }
      \hline
&\multicolumn{3}{c}{$n=100$}& \multicolumn{3}{c}{$n=200$}  \\
\cmidrule(lr){2-4}\cmidrule(lr){5-7}
& \multicolumn{6}{c}{$\rho=0.5$} \\
\cmidrule(lr){1-7}
 &  $G_{11}$&  $G_{12}$&  $G_{13}$&  $G_{11}$&  $G_{12}$&  $G_{13}$  \\ \hline
 ToMax(1)&  0.068&  0.052&  0.054&  0.056&  0.052&  0.038  \\
%$L=2$&  0.072&  0.046&  0.044&  0.044&  0.038&  0.030  \\
ToMax(5)&  0.056&  0.040&  0.038&  0.034&  0.038&  0.036  \\
ToMax(8)&  0.060&  0.042&  0.032&  0.032&  0.050&  0.042  \\
ToMax(15)&  0.066&  0.038&  0.040&  0.036&  0.042&  0.030  \\
ToMax(20)&  0.066&  0.046&  0.030&  0.028&  0.040&  0.032  \\
MaxMu&  0.082&  0.074&  0.056& 0.032&  0.050&  0.044 \\
%BY &  0.046&  0.024&  0.020& 0.036&  0.008&  0.010  \\
\hline
&  $G_{14}$&  $G_{15}$&  $G_{16}$&  $G_{14}$&  $G_{15}$&  $G_{16}$  \\ \hline
ToMax(1)&  0.190&  0.176&  0.580&  0.374&  0.484&  0.926  \\
% $L=2$& 0.208&  0.178&  0.688 &  0.398&  0.540&  0.982  \\
ToMax(5)&  0.222&  0.194&  0.760&  0.470&  0.596&  1.000  \\
ToMax(8)&  0.234&  0.184&  0.764&  0.480&  0.598&  1.000  \\
ToMax(15)&  0.224&  0.198&  0.798&  0.490&  0.598&  0.996  \\
ToMax(20)&   0.236&  0.182&  0.786&  0.468&  0.592&  0.998  \\
MaxMu&   0.344&  0.312&  0.868&  0.654&  0.732&  1.000  \\
% BY & 0.298&  0.162&  0.754& 0.600&  0.576&  0.996  \\
\hline
& \multicolumn{6}{c}{$\rho=1$} \\
\cmidrule(lr){1-7}
Method&  $G_{21}$&  $G_{22}$&  $G_{23}$&  $G_{21}$&  $G_{22}$&  $G_{23}$  \\ \hline
ToMin(1)&  0.036&  0.044&  0.044& 0.046&  0.048&  0.048  \\
% $L=2$&  0.048&  0.058&  0.056&  0.036&  0.054&  0.054  \\
ToMin(5)&  0.064&  0.040&  0.044& 0.048&  0.042&  0.042  \\
ToMin(8)&  0.060&  0.046&  0.054&   0.036&  0.046&  0.040  \\
ToMin(15)&  0.062&  0.038&  0.046&  0.036&  0.036&  0.036  \\
ToMin(20)&  0.052&  0.040&  0.046&   0.040&  0.026&  0.028  \\ \hline
 &  $G_{24}$&  $G_{25}$&  $G_{26}$&  $G_{24}$&  $G_{25}$&  $G_{26}$  \\ \hline
ToMin(1)&   0.536&  0.546&  0.398&  0.796&  0.798&  0.626  \\
% $L=2$&  0.598&  0.596&  0.450&  0.884&  0.852&  0.658  \\
ToMin(5)&  0.680&  0.684&  0.538&  0.914&  0.902&  0.756  \\
ToMin(8)&  0.730&  0.716&  0.578&  0.930&  0.924&  0.800  \\
ToMin(15)&  0.762&  0.740&  0.646&  0.938&  0.928&  0.826  \\
ToMin(20)&  0.778&  0.764&  0.682&  0.956&  0.942&  0.848  \\  \hline
            \end{tabular}
            \end{table}

%\bibliographystyle{apalike}
%
%\bibliography{ref_library}
\bibliographystyle{apalike}

\bibliography{ref_library}

\end{document}